\documentclass[]{jfm}
\usepackage{graphicx}
\usepackage{multirow}
\usepackage{newtxtext}
\usepackage{newtxmath}
\usepackage{natbib}
\usepackage{hyperref}
\hypersetup{
    colorlinks = true,
    urlcolor   = blue,
    citecolor  = blue,
    linkcolor  = blue  
}

\newcommand{\RomanNumeralCaps}[1]
\linenumbers

\title{Turbulent Pipe Flow of Thixotropic Fluids}

\author{Noman Yousuf\aff{1},
  Daniel Lester\aff{1}
  \corresp{\email{daniel.lester@rmit.edu.au}},
  Murray Rudman\aff{2},
  Marco Dentz \aff{3},
 \and Nicky Eshtiaghi \aff{1}}

\affiliation{\aff{1}Chemical and Environmental Engineering, RMIT University, VIC 3000, Australia
\aff{2}Department of Mechanical and Aerospace Engineering, Monash University, VIC 3800, Australia
\aff{3}Spanish National Research Council (IDAEA-CSIC), 08034 Barcelona, Spain}
\begin{document}
\maketitle

\begin{abstract}
Complex materials with internal microstructure such as suspensions and emulsions exhibit time-dependent rheology characterized by viscoelasticity and thixotropy. In many large-scale applications such as turbulent pipe flow, the elastic response occurs on a much shorter timescale than the thixotropy, hence these flows are purely thixotropic. The fundamental dynamics of thixotropic turbulence is poorly understood, particularly the interplay between microstructural state, rheology, and turbulence structure. To address this gap, we conduct direct numerical simulations (DNS) of fully developed turbulent pipe flow of a model thixotropic (Moore) fluid over a range of thixoviscous numbers $\Lambda$ from slow ($\Lambda\ll 1$) to fast ($\Lambda\gg 1$) thixotropic kinetics relative to the eddy turnover time. Analysis of DNS results in the Lagrangian frame shows that, as expected, in the limits of slow and fast kinetics, these time-dependent flows behave as time-independent purely viscous (generalized Newtonian) analogues. For intermediate kinetics ($\Lambda\sim 1$), the rheology is governed by a \emph{path integral} of the thixotropic fading memory kernel over the distribution of Lagrangian shear history, the latter of which is modelled via a simple stochastic model for the radially non-stationary pipe flow. DNS computations based on this effective viscosity closure exhibit excellent agreement (within 2.4\% error) with the fully thixotropic model for $\Lambda=1$, indicating that the purely viscous (generalized Newtonian) analogue persists for arbitrary values of $\Lambda\in[0,\infty^+)$ and across nonlinear rheology models. These results 
uncover the feedback mechanisms between microstructure, rheology, and turbulence and
offer fundamental insights into the structure of thixotropic turbulence.
\end{abstract}

\begin{keywords}
Turbulence, Thixotropy, Non-Newtonian Flow
\end{keywords}

\section{Introduction}\label{sec:Introduction}

Complex fluids such as colloidal suspensions, emulsions, biological assemblies, polymer solutions and melts~\citep{Larson1999} are all characterized by the presence of an internal microstructure that imparts non-Newtonian flow behaviors such as yield-pseudo plasticity~\citep{Barnes1999,Bonn2017}, nonlinear viscoelasticity~\citep{Ewoldt2009,Mckinley1996}, thixotropy~\citep{Larson2019,Mewis2009} and thixo-elasto-visco-plasticity~\citep{Ewoldt2017}. These fluids arise in large-scale process equipment such as pumps, heat exchangers, mixing tanks and pipelines across industrial applications spanning minerals processing, paints and coatings, energy resources, food processing and wastewater treatment. From a process perspective, turbulent flow of these complex fluids is desirable due to improved heat and mass transfer characteristics, mitigation of sedimentation and optimal pumping efficiency. Despite widespread application, there remain significant fundamental questions regarding the turbulent flow of complex fluids.

The onset of rheological complexity in complex fluids arises from their internal microstructure, which exhibits time-dependent properties characterized by nonlinear viscoelasticity and thixotropy. Viscoelasticity involves the partial recovery of prior elastic deformation of the microstructure over small timescales~\citep{Pipkin1972}, whereas thixotropy manifests as a reversible, time-dependent change in viscosity under flow conditions, driven by the shear degradation (breakdown) and the thermal recovery (rebuild) of the microstructure~\citep{Larson2019}. In this sense, viscoelasticity involves the impact of \emph{strain history} on the current stress state, whereas thixotropy involves the impact of \emph{strain rate history} on the current stress state. Many complex fluids can be broadly classified as thixotropic elasto-visco-plastic (TEVP) materials~\citep{Ewoldt2017}, where the interplay of thixotropic and elastic effects, along with a plastic yield criterion gives rise to complex rheological responses that can be difficult to deconvolve~\citep{Ewoldt2022}. 

However, for many large-scale applications that involve continuous flow over longer timescales, some TEVP materials can be validly approximated as thixotropic fluids with shear-dependent viscosity as the elastic relaxation timescale of the fluid is several orders of magnitude shorter than both the flow and thixotropic timescales~\citep{Dullaert2005,Mewis2012}. One such practical example is turbulent pipe flow, where TEVP materials exhibit drag reduction primarily due to thixotropic breakdown of the microstructure near pipe walls~\citep{Escudier1996,Pereira1999,Pereira2002}. Although there certainly exist industrial applications where elastic effects are important, many large-scale flows of complex fluids can be treated as purely thixotropic fluids~\cite{Larson2019}.

Despite extensive applications in the process industries~\citep{Cayeux2020,Escudier1999,Pereira1999,Pereira2002}, thixotropic turbulence remains poorly understood. Turbulent thixotropic flows are complex in that they involve an interplay~\citep{Thompson2020} between microstructural evolution due to thixotropy, macroscopic fluid rheology that arises from the microstructural state, and the turbulence structure that informs fluid shear and transport which drive thixotropy. The evolution of microstructure due to thixotropy is typically characterized by the structural parameter $\lambda$~\citep{Goodeve1939,Cheng1965,Barnes1997} that represents the state of the microstructure, ranging from fully structured ($\lambda = 1$) to an unstructured ($\lambda = 0$) material. Typically, the viscosity of the fluid decreases with decreasing $\lambda$, and in a spatially homogeneous system the structural parameter decreases with shear exposure due to shear-induced structure breakdown, while simultaneously rebuilding due to Brownian motion~\citep{Nguyen1985}. 

In turbulent flows, the ubiquity of coherent structures~\citep{Lee1990} can generate large spatial variations in shear rate, which can generate heterogeneous spatial distributions of the structural parameter $\lambda$. Hence, in these flows the structural parameter evolves through a balance between shear-induced microstructure breakdown, thermal rebuild and advective transport. Diffusion is typically negligible as this is governed by the very slow self-diffusion of the microstructure~\citep{Eckstein1977}. This feedback loop remains poorly understood, particularly when the timescale of the thixotropic kinetics are similar to that of the eddy turnover time, leading to highly nonlinear flow behavior.

In this study, we explore the following questions regarding the fundamentals of thixotropic turbulence: (i) What mechanisms govern the interplay between rheology, microstructural state, and turbulence structure? Understanding this relationship is essential for capturing the non-equilibrium dynamics of thixotropic turbulent pipe flows. 
(ii) How do these interactions evolve as the thixo-viscous timescale changes? Variations in timescales can lead to qualitatively different behaviour, influencing both microstructural evolution and turbulence dynamics. (iii) Can a simplified rheological model be developed to accurately describe these interactions in fully developed turbulent flow? Development of such a model would significantly improve our ability to predict and control flow of thixotropic fluids in engineering applications.

We address these questions in this study by simulating and analyzing fully developed turbulent pipe flow of a model thixotropic fluid at a moderate Reynolds number. We employ a high-resolution spectral element code~\citep{Blackburn2019} to conduct the first direct numerical simulation (DNS) of fully developed turbulent of a thixotropic fluid. We consider a range of thixotropic kinetics relative to the eddy turnover time, as characterized by the \emph{thixoviscous number} $\Lambda$~\citep{Ewoldt2017}, spanning from essentially instantaneous breakdown and rebuild $(\Lambda\gg 1)$ to very slow thixotropic kinetics $(\Lambda\ll 1)$. We analyze the Eulerian characteristics of these flows for the range of $\Lambda$ and compare results with turbulent pipe flow of Newtonian and generalised Newtonian (GN) fluids. The Lagrangian frame is then used to explore the relationship between Lagrangian shear history, viscosity and flow structure across the range of $\Lambda$, and a stochastic model is developed for the effective viscosity, enabling elucidation of the mechanisms governing thixotropic turbulence. Despite the complex coupling between rheology, microstructure and turbulence, our findings indicate that the turbulent flow of time-dependent thixotropic fluids can be accurately captured by a time-independent (generalised Newtonian) effective viscosity over the entire spectrum of $\Lambda$. These effective viscosity models are verified through DNS simulations, thus providing valuable insights into the structure of thixotropic turbulence.

The remainder of this paper is structured as follows. In \S\ref{sec:Governing Equations and Numerical Method} the governing equations and numerical methods used are summarised. Results from the numerical simulations are analysed in the Eulerian frame in \S\ref{sec:Eulerian Characteristics of Thixotropic Turbulence}, with a focus on flow behavior and structural evolution over the range of thixotropic kinetics. A Lagrangian frame for analysis of thixotropic turbulence is developed in \S\ref{sec:Lagrangian Thixotropy}, including derivation and validation of analytic closures for effective viscosity in the limits of fast ($\Lambda \gg 1$) and slow ($\Lambda \ll 1$) thixotropic kinetics. This frame is then used in \S\ref{sec:stochastic} to develop and validate stochastic models for effective viscosity in the intermediate kinetics range ($\Lambda \sim 1$), and conclusions are made in \S\ref{sec:Conclusions}.

\section{Governing Equations and Numerical Method}\label{sec:Governing Equations and Numerical Method}

\subsection{Governing equations}\label{subsec:Governing equations}

In this study we consider the simplest representative scenario of fully developed turbulent flow of a single-phase incompressible and inelastic thixotropic fluid through a smooth long pipe. The flow domain consists of an axially-periodic straight pipe section of diameter $D$ and length $L_z=4 \pi D$ with no-slip wall boundary conditions that is driven by a constant axial body force~\citep{Singh2018}. Flow within the pipe is governed by the incompressible Navier-Stokes equations
\begin{equation}\label{eq:gvn}
 \rho \left( \frac{\partial \boldsymbol{v}}{\partial t} + \boldsymbol{v} \bcdot \bnabla \boldsymbol{v} \right) = \bnabla \bcdot \mathsfbi{\tau} -\bnabla p + \boldsymbol{f}
,\quad \bnabla\bcdot\boldsymbol{v} = 0,
\end{equation}
where $\boldsymbol{v}$ is the fluid velocity, $\rho$ is density and $p$ static pressure, and the flow is driven by the constant axial body force
\begin{equation}
    \boldsymbol{f} =  -\frac{d\langle p\rangle}{dz}\hat{\mathbf{e}}_z.
\end{equation}
The shear stress tensor $\mathsfbi{\tau} = 2 \eta\mathsfbi{S}$ is the product of the rate of strain tensor $\mathsfbi{S}=1/2\left(\bnabla \boldsymbol{v} + \bnabla \boldsymbol{v}^\top\right)$ and the apparent viscosity
\begin{equation}\label{eqn:eta}
    \eta=\eta\left(\dot\gamma, \lambda \right),
\end{equation}
which is a function of local shear rate $\dot\gamma =\sqrt{2 (\mathsfbi{S}:\mathsfbi{S})}$ and the structural parameter $\lambda$ that quantifies the impact of thixotropy upon the fluid rheology. A range of models for $\eta\left(\dot\gamma, \lambda \right)$ have been proposed in the literature that span a wide range of materials~\citep{Burgos2001,Toorman1997,Farno2020,Mewis2009,Barnes1999}. As the fluid is inelastic, quantities such as the Deborah number and the thixoelastic number~\citep{Ewoldt2017} are not relevant.

For the thixotropic evolution of $\lambda$ in a homogeneous flow, various kinetic models~\citep{Mujumdar2002,Mewis2009,Larson2015,Barnes1997} have been proposed, most of which are of the form
\begin{equation}\label{eq:kinetics}
\frac{\partial \lambda}{\partial t}=-g(\dot\gamma,\lambda) + f(\dot\gamma,\lambda),\quad \lambda \in [0,1]
\end{equation}
where the kinetic functions $g(\dot\gamma,\lambda)$ and $f(\dot\gamma,\lambda)$ respectively represent the shear-induced breakdown and thermal and/or shear rebuild of the microstructure. Along with the shear viscosity $\eta$, determination of the kinetic functions form an integral part of the rheological characterization of thixotropic fluids. Under steady shear conditions ($\dot\gamma=\text{const.}$), these processes come into equilibrium and the structural parameter approaches its equilibrium value $\lambda\rightarrow\lambda_{\text{eq}}(\dot\gamma)$ given by
\begin{equation}\label{eqn:thixo_homog}
    g(\dot\gamma,\lambda_{\text{eq}}(\dot\gamma)) = f(\dot\gamma,\lambda_{\text{eq}}(\dot\gamma)),
\end{equation}
where typically $\lambda_{\text{eq}}(\dot\gamma)\rightarrow 1$ as $\dot\gamma\rightarrow 0$, $\lambda_{\text{eq}}(\dot\gamma)\rightarrow 0$ as $\dot\gamma\rightarrow\infty$. In general, the viscosity $\eta(\dot\gamma,\lambda)$ of thixotropic fluids varies strongly (typically decreasing) with both shear rate $\dot\gamma$ (shear thinning) and the structural parameter $\lambda$ (thixotropy) as the microstructure responds to imposed shear. 

As this study aims to understand the fundamental mechanisms that govern thixotropic turbulence, we consider the simplest possible non-trivial thixotropic kinetic model, where the shear breakdown $g=k\,m\,\dot\gamma\,\lambda$ and thermal rebuild $f=m(1-\lambda)$ terms are simple linear functions of $\dot\gamma$ and $\lambda$, yet still capture the essential physics of more complex thixotropic models. 
Extension of (\ref{eqn:thixo_homog}) to heterogeneous flow such as turbulent pipe flow yields an advection-diffusion-reaction equation (ADRE) for the evolution of $\lambda$ as
\begin{equation} \label{eq:transport}
\frac{\partial \lambda}{\partial t} +\boldsymbol{v} \bcdot \bnabla \lambda= m\left[-k\dot\gamma \lambda + (1- \lambda)\right] +  \alpha \bnabla^2 \lambda
,\quad 0 \leq \lambda \leq 1,
\end{equation}
where $\alpha$ characterizes the self-diffusivity of the microstructure which is typically very small, corresponding to P\'{e}clet number $\Pen \sim 10^{12}$~\citep{Morris1996} for e.g. colloidal suspensions. However, a much larger value of $\alpha$ is typically used in  simulations for numerical stability~\citep{Billingham1993}.
The thixotropic rate parameter $m$ governs the rate of convergence of $\lambda$ to the equilibrium value $\lambda_{\text{eq}}(\dot\gamma)$. The parameter $k$ governs the relative rates of breakdown and rebuild, and impacts the equilibrium state as
\begin{equation} \label{eq:equil}
\lambda_{\text{eq}} (\dot\gamma) = \frac{1}{1+k \dot\gamma},
\end{equation}
which also corresponds to the limit of instantaneous thixotropic kinetics $m\rightarrow\infty$, where $\lambda\rightarrow\lambda(\dot\gamma)$. Indeed, shear thinning behavior in non-Newtonian fluids may be conceptualized as thixotropic fluids with very rapid kinetics~\citep{Scott1951,Barnes1997}.

Following the ``eagle and rat'' metaphor~\citep{Thompson2020} for homogeneous flows with time-varying shear rate, the parameter $k$ defines $\lambda_{eq}(\dot\gamma)$ (position of the ``rat''), while the parameter $m$ governs the rate at which $\lambda\rightarrow\lambda_{\text{eq}}$, analogous to the speed at which the ``eagle'' approaches the ``rat''. However, for heterogeneous flows such as fully developed turbulence, this picture is more complicated as the shear rate varies spatially as well as temporally, in addition to the feedback mechanism that exists between the microstructure, rheology and turbulence. 

Similar to the thixotropic model, for simplicity we also consider the simplest non-trivial rheological model given a purely viscous (non-elastic) Moore fluid~\citep{Moore1959}
\begin{equation} \label{eq:rheology}
\eta(\lambda)=\eta_{\infty}+(\eta_{0}-\eta_{\infty})\lambda 
,\quad 
\eta_{\infty} \leq \eta(\lambda) \leq \eta_{0}.
\end{equation}
where $\eta_{0}$ and $\eta_{\infty}$ represent the limiting viscosities for the structured $(\lambda=1)$ and unstructured $(\lambda=0)$ material. The coupling between the governing equations (\ref{eq:gvn}), (\ref{eq:transport}), (\ref{eq:rheology}) directly quantify the feedback loop between the turbulence structure, microstructural evolution and the bulk rheology. 

\subsection{Non-dimensionalisation}\label{subsec:Non-dimensionalisation}

The governing equations (\ref{eq:gvn}), (\ref{eq:transport}), (\ref{eq:rheology}) are non-dimensionalised with respect to the characteristic lengthscale $D$, advective timescale $\tau_v\equiv D/U_b$ and the wall viscosity
\begin{equation}
    \eta_w = \dfrac{\tau_w}{\dot{\gamma}_w}, 
\end{equation}
is chosen as a viscosity scale~\citep{Rudman2004}, where subscript $w$ denotes values at the pipes walls. Henceforth all variables are non-dimensional and the same notation is used for simplicity. The resultant non-dimensional governing equations are then
\begin{align}
\label{eq:gvn_nD}
 \frac{\partial \boldsymbol{v}}{\partial t} + \boldsymbol{v} \bcdot \bnabla \boldsymbol{v} &= \frac{1}{\Rey_G} \bnabla \bcdot \left[\eta(\lambda) \bnabla \boldsymbol{v} \right] -\bnabla p + F\boldsymbol{f}
,\quad 
\bnabla\bcdot\boldsymbol{v} = 0,\\
\label{eq:transport_nD}
\frac{\partial \lambda}{\partial t} +\boldsymbol{v} \bcdot \bnabla \lambda &= \frac{1}{\Pen} \bnabla^2 \lambda + \Lambda \left[1-\lambda \left(1+K\dot\gamma\right) \right].\\
\label{eq:rheology_nD}
\eta(\lambda)&= \eta_{\infty} \left[ 1 + \lambda(\eta_r -1) \right].
\end{align}
In (\ref{eq:gvn_nD})-(\ref{eq:rheology_nD}), the generalised Reynolds number
\begin{equation}
    Re_G\equiv \frac{\rho U_b D}{\eta_w}, 
\end{equation}
characterizes the ratio of inertial to viscous forces, the viscosity ratio $\eta_{r}\equiv\eta_{0}/\eta_{\infty}=2$ characterizes the ratio between the fully structured and fully unstructured viscosity, and the non-dimensional body force magnitude $F\equiv\vert d\langle{p}\rangle/dz \vert  D/ \rho U_b^2=1.8\times10^{-2}$ characterizes the ratio of inertial to body forces. With respect to evolution of $\lambda$, the diffusion timescale is defined as $\tau_{\alpha} \equiv D^2/\alpha$ and the thixotropic reaction timescale is $\tau_r \equiv 1/m$, which characterize the P\'{e}clet and Damkh\"{o}ler numbers as
\begin{align}
    \Pen \equiv \frac{\tau_\alpha}{\tau_v}=\frac{D U_b}{\alpha} && Da \equiv \frac{\tau_\alpha}{\tau_r}=\frac{D^2m}{\alpha}
\end{align}
which respectively characterise the timescales of advection and thixotropy to the diffusive timescale. A related dimensionless parameter known as the thixoviscous number~\citep{Ewoldt2017}
\begin{equation}
    \Lambda\equiv \frac{\tau_v}{\tau_r}=\frac{Da}{Pe}=\frac{m D}{U_b},
\end{equation}
which characterises the timescale of thixotropic kinetics relative to that of advection. Here $\Lambda \gg 1$ represents thixotropic fluids that evolve on timescales much faster than the flow, whereas $\Lambda \ll 1$ corresponds to thixotropic fluids that evolve over several advection times. The non-dimensional equilibrium parameter $K\equiv k \dot\gamma_{\text{nom}}$ in (\ref{eq:transport_nD}) characterises the equilibrium value of $\lambda$ under the nominal shear rate $\dot{\gamma}_{\text{nom}}$ as (\ref{eq:equil}). This parameter although is defined as $K\equiv k(U_b/D)$, however, we have characterised it as $K\equiv k(8U_b/D)=0.4$ to ensure that the breakdown and the rebuild occur at roughly similar rates.
Representative values of the key dimensionless variables used in the simulations are summarised in Table~\ref{tab:dmless_par}. The generalised Reynolds number $Re_G$ is in the moderately turbulent regime, and varies by roughly a factor of $\eta_r$ due to change in viscosity with fixed $F$. The thixoviscous number $\Lambda$ is varied from slow $(\Lambda\ll 1)$ to fast $(\Lambda\gg 1)$ kinetics. Although the P\'{e}clet number $Pe\gg 1$ indicates the transport of $\lambda$ is always advection dominated, the broad range of $\Lambda$ means that the Damkh\'{o}ler number $Da$ is order unity for slow kinetics, meaning diffusion occurs on a similar timescale to thixotropy in this case.

\begin{table}
\centering
\begin{tabular}{lcc}
Parameter			& Symbol					& Representative Value \\
\hline
Reynolds number	 & 	$Re_G$ 
& $6-14 \times 10^3$                     \\
P\'{e}clet Number 	 &   	$Pe$ 
&  $10^3$                     \\
Damk\"{o}hler Number    &    	$Da$ 
&    $10^1 - 10^5$                    \\
Thixoviscous Number&     	$\Lambda$ 
&   $10^{-2} - 10^{2}$                   \\
\end{tabular}
\caption{Summary of key dimensionless parameters}
\label{tab:dmless_par}
\end{table}

\subsection{Numerical Method}\label{sec:Details of the numerical simulations}

The DNS method utilized in this study is based on an generalised Newtonian (GN) extension \emph{nnewt}~\citep{Rudman2006} of the spectral element code \emph{Semtex}~\citep{Blackburn2019}, which has been previously validated for the turbulent pipe flow simulations of GN fluids~\citep{Rudman2006,Singh2016,Yousuf2024}. This code is further extended to simulate turbulent flow of thixotropic fluids, as is described below. The DNS method is well suited for simulation of thixotropic turbulence as it resolves the turbulent flow across all spatio-temporal scales, eliminating the need for sub-grid scale turbulent closures, and facilitating direct solution of the transport equation (\ref{eq:transport_nD}) for the evolution of $\lambda$.

The pipe cross-section is divided into discrete two-dimensional quadrilateral elements representing standard tensor-product Lagrange interpolants with Gauss-Lobatto-Legendre collocation points. The axial direction is effectively managed through the application of Fourier expansion, which naturally imposes periodic boundary conditions~\citep{Temperton1992}. The temporal resolution is handled by the second-order time integration method~\citep{Karniadakis1991} to maintain numerical stability. The code strictly monitors the Courant-Friedrichs-Lewy (CFL) criterion as well as the average divergence of the numerical solution for diagnostics.

To simulate thixotropy, the advective non-linear terms in (\ref{eq:gvn_nD}) and (\ref{eq:transport_nD}) are discretized explicitly, while the diffusive terms are handled implicitly to ensure numerical stability and convergence, following the methodology outlined by \citet{Rudman2006}. For the breakdown and rebuild terms in (\ref{eq:transport_nD}), a fully implicit novel numerical scheme has been integrated into the code to enhance numerical robustness, see Appendix~\ref{app:Numericals} for more details.

In the present study, the thixoviscous number $\Lambda$ is systematically varied in DNS computations from fast ($\Lambda = 10^2$) to slow kinetics ($\Lambda = 10^{-2}$). Simulations at higher $\Lambda$ values were found to be infeasible, as sharp gradients in the $\lambda$ field led to numerical instabilities, even with the implicit solver detailed in Appendix~\ref{app:Numericals} for the thixotropic kinetics.  Similarly, computations for very small values of $\Lambda$ were also found infeasible, as the timescale to reach thixotropic stationarity becomes arbitrarily greater than the eddy turnover time, thus requiring extensive computational overhead. However, as shall be shown in \S\S\ref{subsec:Fast Kinetics} - \ref{subsec:Slow Kinetics}, these computational limits do not restrict understanding of thixotropic turbulence as the dynamics in the limits $\Lambda\rightarrow\infty$ and $\Lambda\rightarrow 0$ can be easily understood and the behaviour at endpoints of this finite range $\Lambda\in[10^{-2},10^2]$ is close to that given by these limits. To provide reference cases, we also compute two Newtonian turbulent flow simulations corresponding to fluids with fully structured ($\lambda=1$, $\eta=\eta_0$) and unstructured ($\lambda=0$, $\eta=\eta_\infty$) microstructures. 

The mesh design for these DNS cases adhere to the established guidelines for Newtonian~\citep{Piomelli1997} and non-Newtonian turbulence~\citep{Rudman2004,Rudman2006}. The pipe cross-sectional mesh has 300 spectral elements with ninth-order tensor-product shape functions, and 384 axial data planes, corresponding to 11.52 million nodal points. Note that the structural parameter diffusivity $\alpha$ was chosen to yield a moderate P\'{e}clet number ($\Pen = 10^3$) to ensure numerical stability, but this value is much smaller than characteristic values ($\Pen \sim 10^{12}$) based on self-diffusivity of e.g. colloidal suspensions~\citep{Morris1996}. A summary of the computational parameters is given in Table~\ref{tab:computational_par}. The DNS code monitors the Courant-Friedrichs-Lewy (CFL) criterion which was kept at $~10^{-1}$ to maintain numerical stability and convergence. Each case was run on the Gadi resource on the National Computational Infrastructure (NCI), Australia, which is a HPC cluster comprised of 8 x 24-core Intel Xeon Platinum 8274 (Cascade Lake) with 3.2 GHz CPUs and 192 GB RAM per node. 

The DNS computations for the thixotropic and the structured ($\lambda=1$) Newtonian cases were initialised by introducing isotropic Gaussian-distributed white noise (with variance 0.01) to the velocity field of a fully developed laminar pipe flow, and the $\lambda$ field is set to unity. For the unstructured ($\lambda=0$) Newtonian case, the $\lambda$ field is set to zero during initialisation. Under fully developed conditions the thixotropic turbulent pipe flow is symmetric and thus stationary in time, axial, and azimuthal directions but non-stationary in the radial coordinate. Each DNS computation was run until the velocity and $\lambda$ fields both achieved statistical stationarity (typically 30 wash-through times), after which Eulerian turbulent flow statistics were accumulated and analyzed for another 30 wash-through times. In addition, Lagrangian data (velocity, velocity gradient and $\lambda$) was also collected for each run for $10^4$ randomly placed tracer particles over a period of around 8 wash-through times.

\begin{table}
  \begin{center}
\def~{\hphantom{0}}
  \begin{tabular}{lccccccccccc}
DNS Case & $\Lambda$ & $Pe$ & $Da$ & $K$ & $U_b/u_{\tau}$ & $Re_G$ & $\Delta y^+$ & $\Delta (r\theta)^+$ & $\Delta z^+$ & $\Delta t / [\eta_{w} /\rho u_{\tau}^2]$ \\
\hline
Structured    & -                     & -                     & -          & -      & 14.74   & 5928         & 0.69        & 3.14              & 13.16        & $2.70\times 10^{-2}$ \\
Slow          & $9.9 \times 10^{-3}$  & $1.01\times 10^{3}$   & $10^{1}$   & 0.40   & 15.13   & 7147         & 0.81        & 3.68              & 15.46        & $3.17\times 10^{-2}$ \\
Intermediate  & $9.4\times 10^{-1}$   & $1.07\times 10^{3}$   & $10^{3}$   & 0.43   & 15.90   & 8896         & 0.96        & 4.36              & 18.31        & $3.75\times 10^{-2}$ \\
Fast          & $9.3\times 10^{1}$   & $1.08\times 10^{3}$   & $10^{5}$   & 0.43   & 16.10   & 9623         & 1.02        & 4.66              & 19.56        & $4.01\times 10^{-2}$ \\
Unstructured  & -                     & -                     & -          & -      & 16.47   & 13 246~~     & 1.37        & 6.27              & 26.32        & $5.39\times 10^{-2}$ \\
  \end{tabular}
  \caption{Summary of computational parameters.}
  \label{tab:computational_par}
  \end{center}
\end{table}

\section{Eulerian Characteristics of Thixotropic Turbulence}\label{sec:Eulerian Characteristics of Thixotropic Turbulence}

\subsection{Instantaneous Flow}\label{subsec:Instantaneous Flow}

\begin{figure}
\centering
\begin{tabular}{ccccc}

\multicolumn{5}{c}{\includegraphics[width=0.8\textwidth]{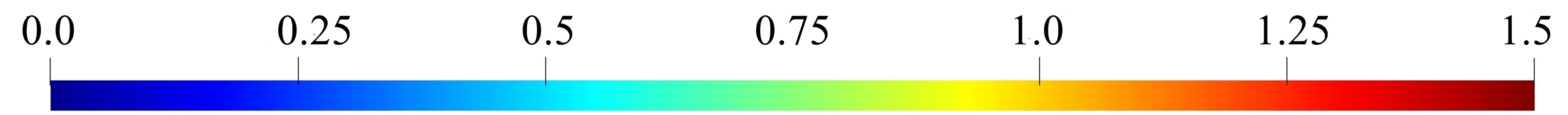}} \\
\includegraphics[width=0.19\textwidth]{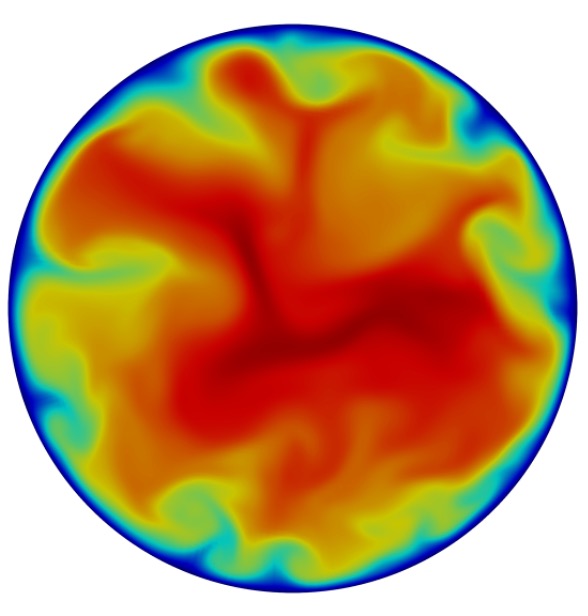}&
\includegraphics[width=0.19\textwidth]{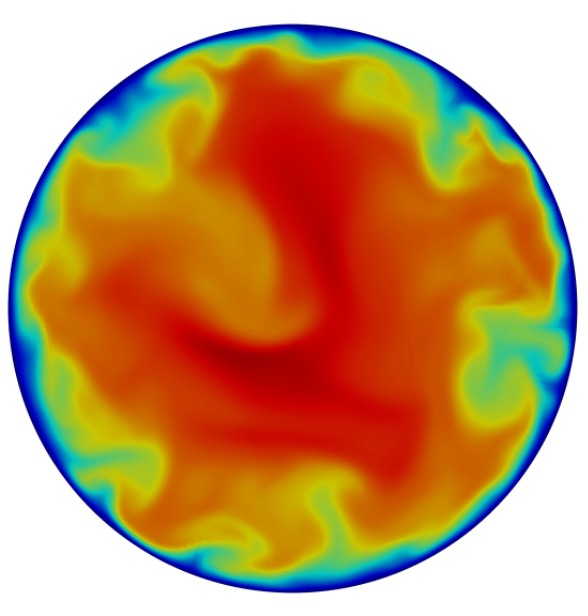}&
\includegraphics[width=0.19\textwidth]{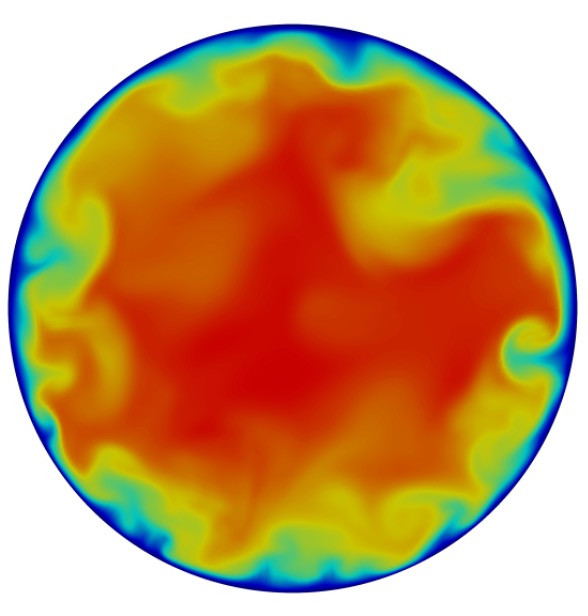}&
\includegraphics[width=0.19\textwidth]{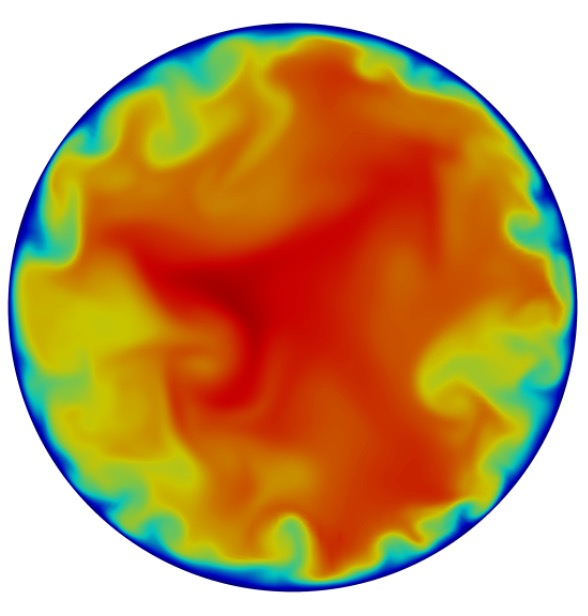}&
\includegraphics[width=0.19\textwidth]{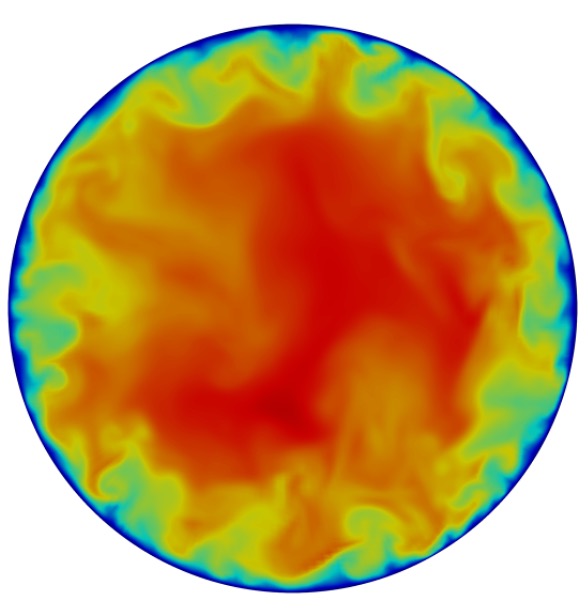}
\\
(a) $\lambda=1$ & (b) $\Lambda=10^{-2}$ & (c) $\Lambda=1$ & (d) $\Lambda=10^{2}$ & (e) $\lambda=0$
\end{tabular}
\caption{Typical cross-sectional contour plots of the instantaneous axial velocity $u_z(\mathbf{x},t)$ for (a) Newtonian flow with $\lambda=0$, (b) thixotropic flow with $\Lambda=10^{-2}$, (c) thixotropic flow with $\Lambda=1$, (d) thixotropic flow with $\Lambda=10^{2}$, (e) Newtonian flow with $\lambda=0$.}
\label{fig:vel_contours}
\end{figure}

\begin{figure}
\centering
\begin{tabular}{ccccc}
\multicolumn{5}{c}{\includegraphics[width=0.8\textwidth]{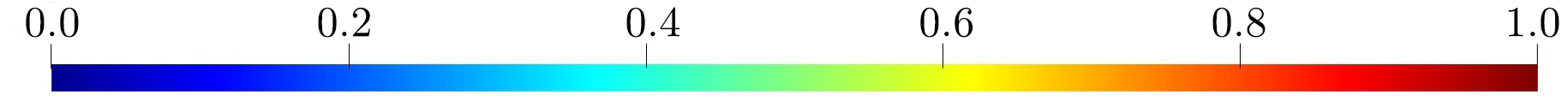}}\\
\includegraphics[width=0.19\textwidth]{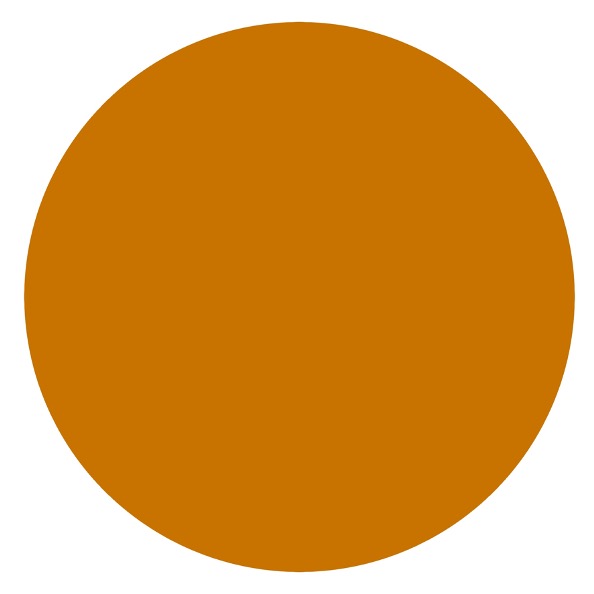}&
\includegraphics[width=0.19\textwidth]{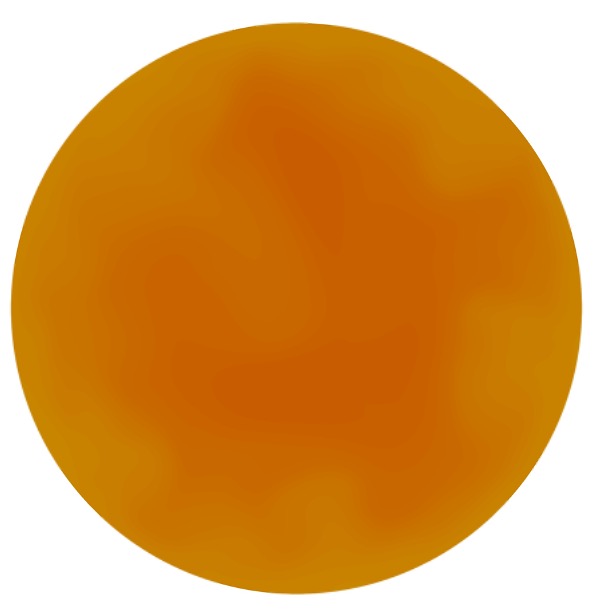}&
\includegraphics[width=0.19\textwidth]{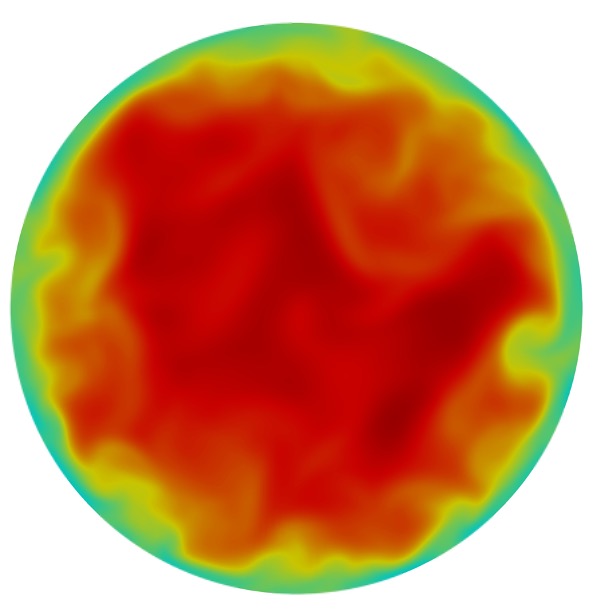}&
\includegraphics[width=0.19\textwidth]{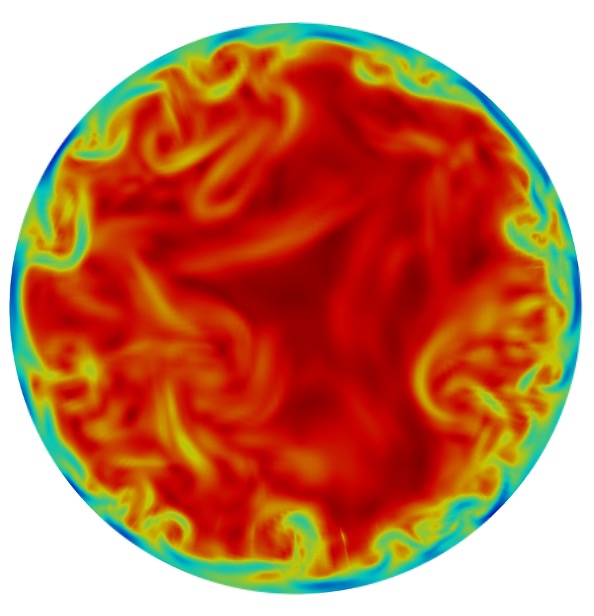}&
\includegraphics[width=0.19\textwidth]{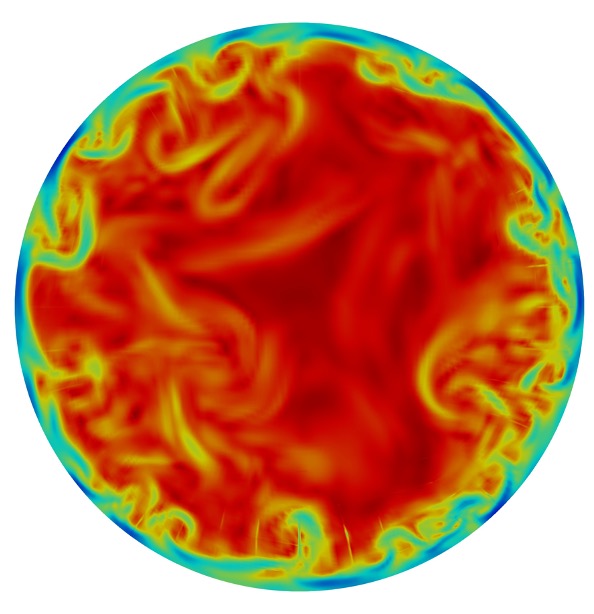}\\   
(a) $\Lambda=10^{-2}$ & (b) $\Lambda=10^{-2}$ & (c) $\Lambda=1$ & (d) $\Lambda=10^{2}$ & (e) $\Lambda=10^{2}$
\end{tabular}
\caption{Typical cross-sectional contour plots of structural parameter $\lambda(\mathbf{x},t)$
for (a) closure (\ref{eq:lag_lambda_slow}) computed from thixotropic flow with $\Lambda=10^{-2}$, (b) thixotropic flow with $\Lambda=10^{-2}$, (c) thixotropic flow with $\Lambda=1$, (b) thixotropic flow with $\Lambda=10^{2}$, (b)  closure (\ref{eq:lag_lambda_fast}) computed from thixotropic flow with $\Lambda=10^{2}$.}
\label{fig:lambda_contours}
\end{figure}

In this section we examine the characteristics of fully developed thixotropic turbulent pipe flow from an Eulerian perspective. Figures~\ref{fig:vel_contours} and \ref{fig:lambda_contours} respectively show representative cross-sectional plots of axial velocity $u_z$ and structural parameter $\lambda$ fields for for the two Newtonian and three thixotropic cases.
At larger values of $\Lambda$, sharper and more fine-scale structures in the $\lambda$ field (and thus viscosity) are observed due to strong coupling between $\lambda$ and the instantaneous shear field, and from (\ref{eq:equil}), in the limit $\Lambda\rightarrow\infty$ the structural parameter is solely dependent upon the local shear rate $\dot\gamma$, as demonstrated in Figure~\ref{fig:vel_contours}e. In this fast kinetics regime, the flow behavior effectively reduces to that of a shear-thinning rheology and the turbulence structure exhibits characteristics similar to that of shear-thinning turbulence~\citep{Rudman2004,Singh2017a}.

For smaller values of $\Lambda$, the structural parameter evolves more slowly along pathlines and the relevant shear rate history that governs $\lambda$ is longer. This leads to a reduction in fine-scale structures observed in the $\lambda$ field due to an effective averaging process backwards along pathlines. Although this results in more diffuse $\lambda$ distributions, this behavior persists in the limit of vanishing diffusivity ($\Pen\rightarrow\infty)$ due to the averaging process. For these computations, diffusion of $\lambda$ plays a minor role due to the moderate P\'{e}clet number ($Pe=10^3$) required for numerical stability. As the Damkh\"{o}ler number scales with $\Lambda$ as $Da=\Lambda \Pen$, and $Da = 10$ for $\Lambda = 10^{-2}$, diffusion acts on a similar timescale as the thixotropic kinetics in this case, leading to the discrepancies observed in Figure~\ref{fig:FDM_all_compare}c. Note that for complex fluids $Pe\gg 10^{3}$, and so for these flows smoothing of the $\lambda$ field is solely due to averaging of the Lagrangian shear rate. In the limit of $\Lambda\rightarrow 0$, the $\lambda$ field is almost uniform, as demonstrated in Figure~\ref{fig:lambda_contours}, meaning that the viscosity is likewise and so the flow resembles that of Newtonian turbulence~\citep{Toonder1997,ElKhoury2013}.

\subsection{Turbulence Statistics}\label{subsec:Turbulence Statistics}

\begin{figure}
\centering
\begin{tabular}{cc}
\includegraphics[width=0.45\textwidth]{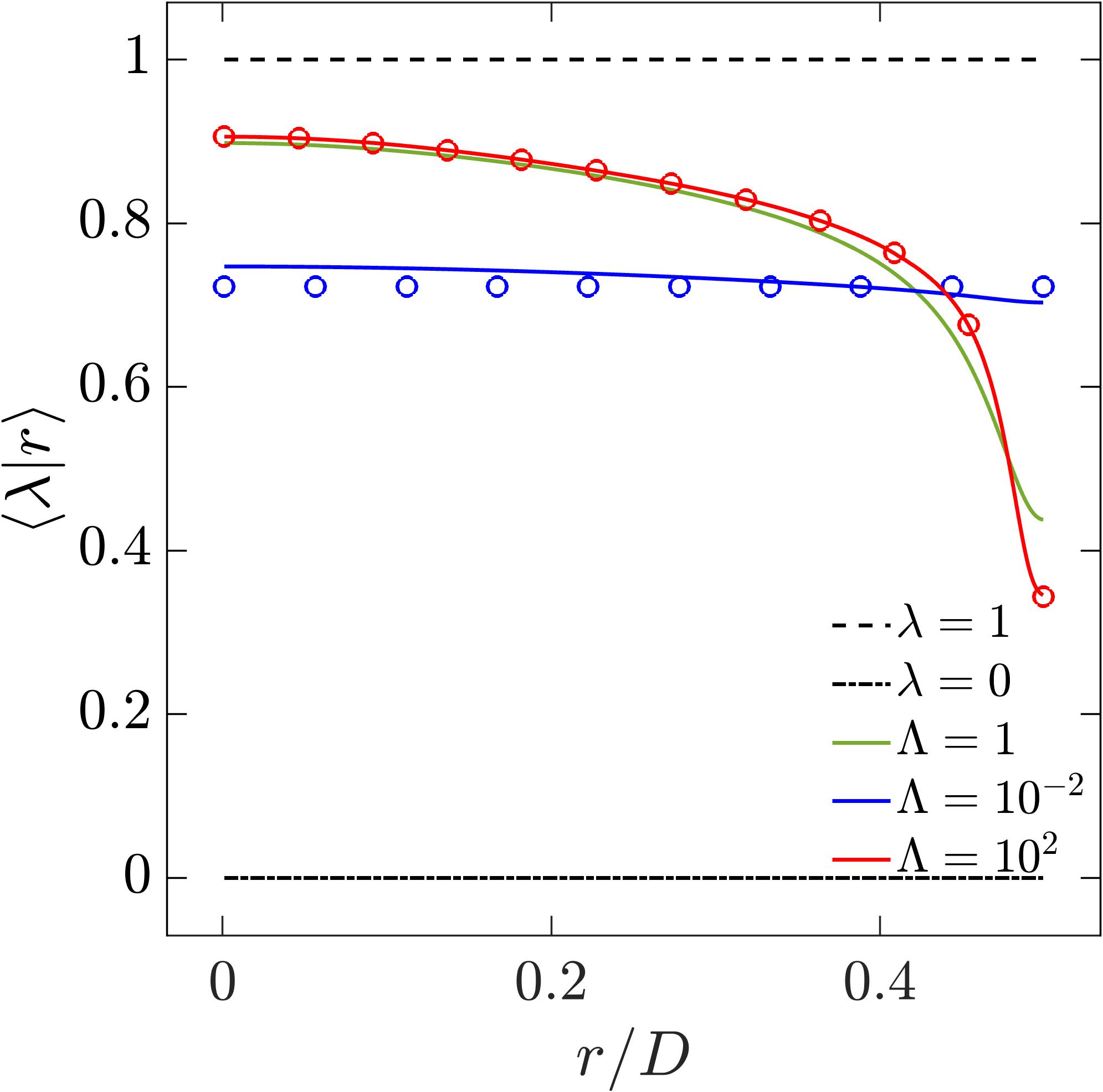}&
\includegraphics[width=0.45\textwidth]{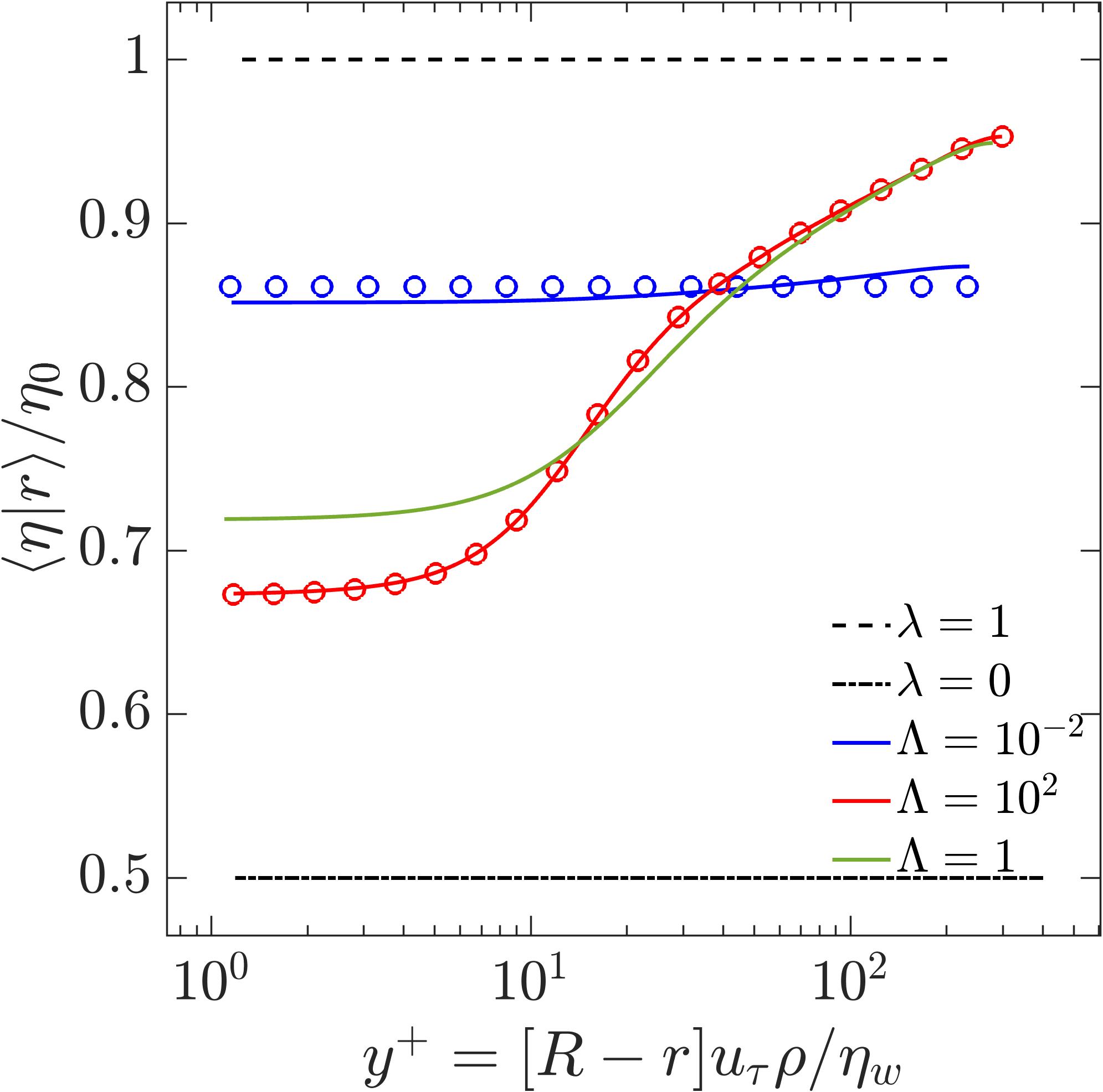}
\\
(a) & (b)\\
\end{tabular}
\caption{Radial profiles of (a) conditionally averaged structural parameter $\langle\lambda|r\rangle$ and (b) viscosity $\langle\eta|r\rangle$, for thixotropic flows and Newtonian reference cases. Lines indicates results from DNS computations of thixotropic and Newtonian models and circles indicate results from the analytic closures (\ref{eq:lag_lambda_slow}) and (\ref{eq:lag_lambda_fast}).}
\label{fig:Mprofiles_1}
\end{figure}

Mean field and second order statistics also provide insights into the impact of thixotropy on turbulent pipe flow. The mean radial viscosity and $\lambda$ profiles for the three thixotropic cases are illustrated in Figure~\ref{fig:Mprofiles_1}, and the Newtonian reference cases provide upper and lower bounds for these quantities.

For all thixotropic cases, shear degradation in the viscous sub-layer ($y^+ \leq 5$) is more significant than in the pipe core due to increased shear near the pipe walls. This is more pronounced for the faster thixotropic kinetics, as the structural parameter $\lambda$ exhibits a stronger correlation with local shear rate $\dot{\gamma}$, leading to a monotone decreasing radial viscosity profile similar to that of turbulent pipe flow of shear thinning fluids~\citep{Rudman2004,Singh2017a}. Similar to the more diffusive profiles shown in Figure~\ref{fig:FDM_all_compare}, this effect is less pronounced for the slower thixotropic kinetics, where the radial profiles of $\lambda$ and $\eta$ are significantly flatter, and the viscosity profile approaches a constant in the limit $\Lambda\rightarrow 0$.

\begin{figure}
\centering
\begin{tabular}{cc}
\includegraphics[width=0.45\textwidth]{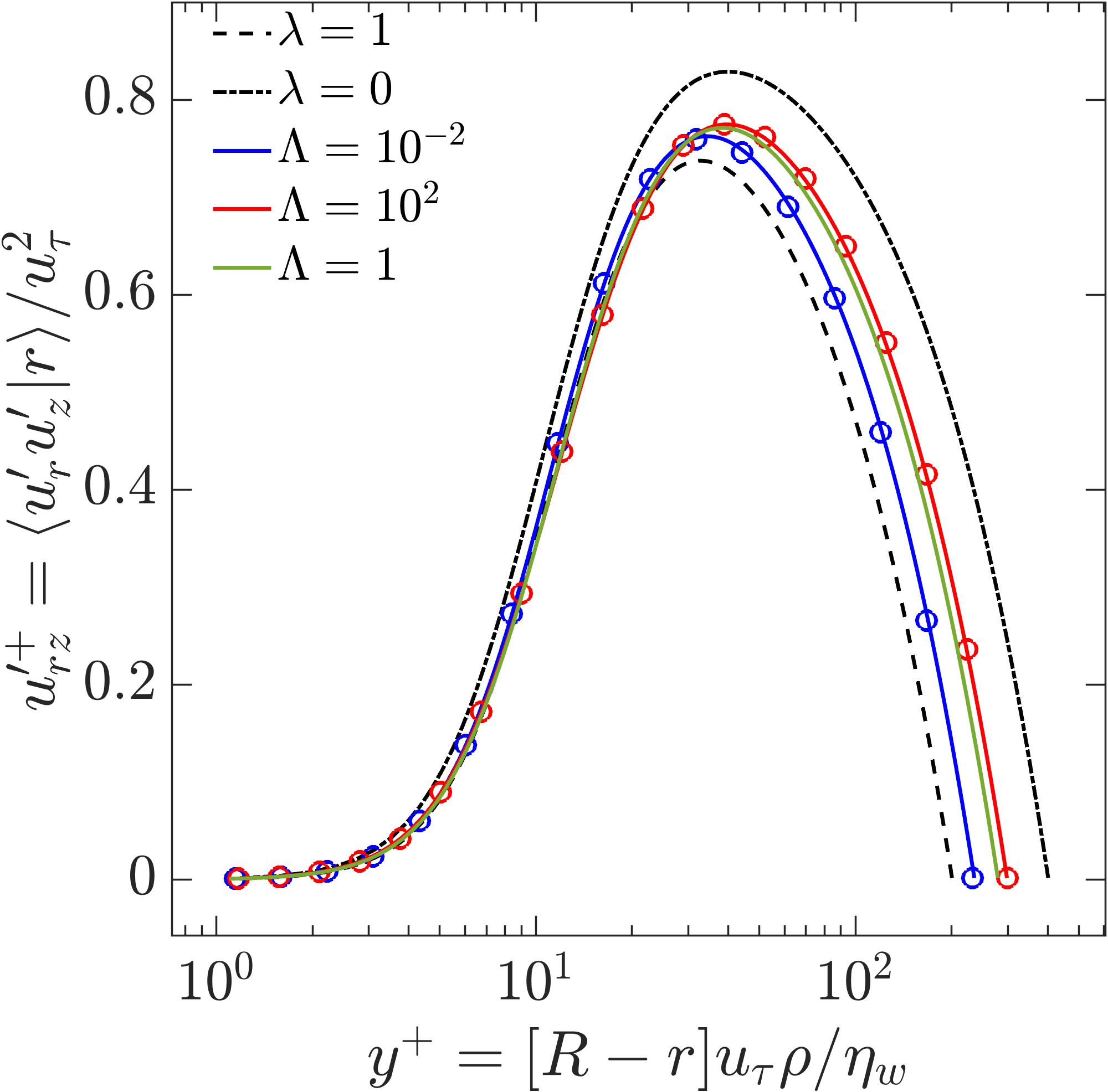}&
\includegraphics[width=0.45\textwidth]{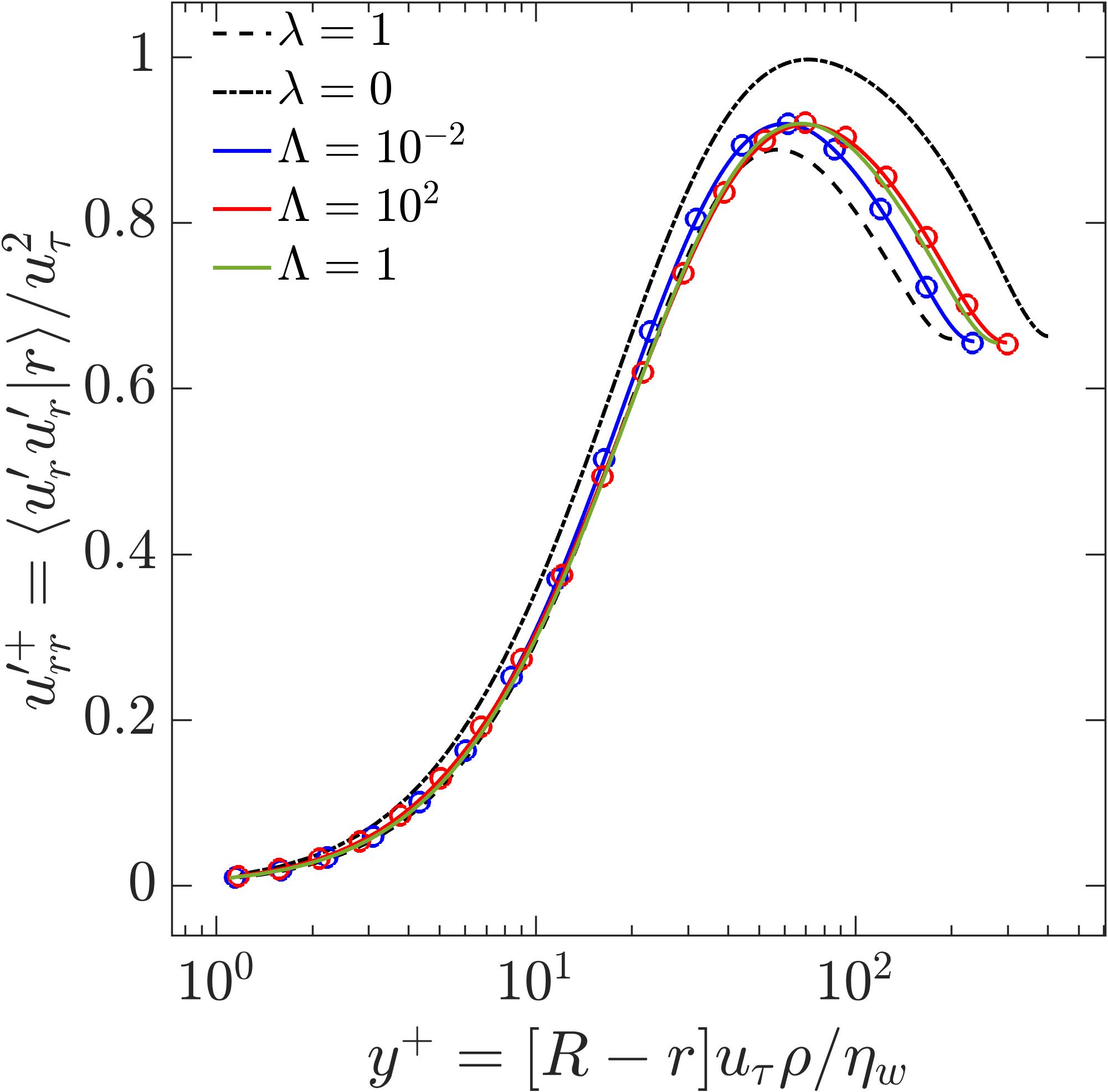}\\
(a) & (b)\\
\includegraphics[width=0.45\textwidth]{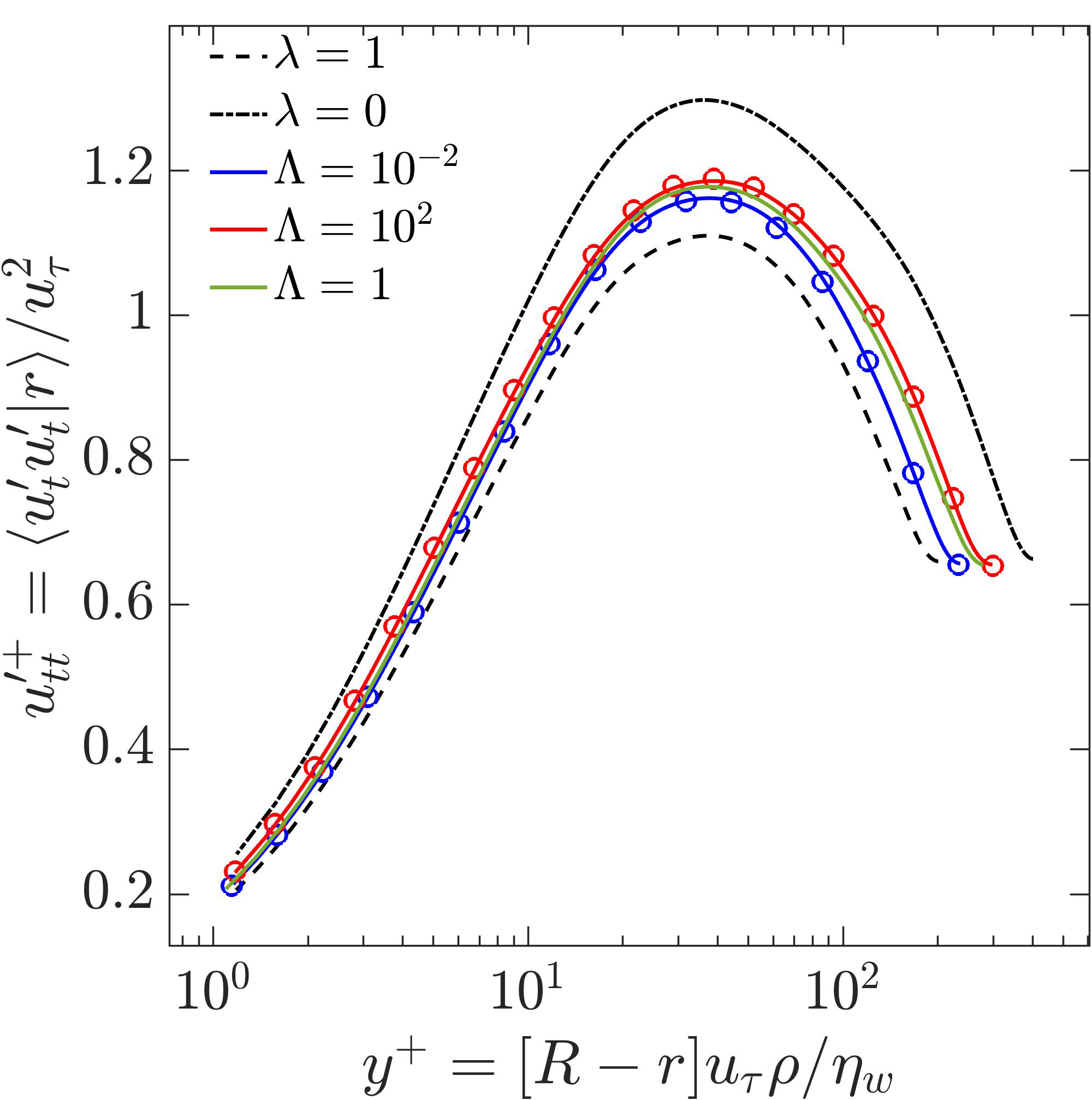}&
\includegraphics[width=0.45\textwidth]{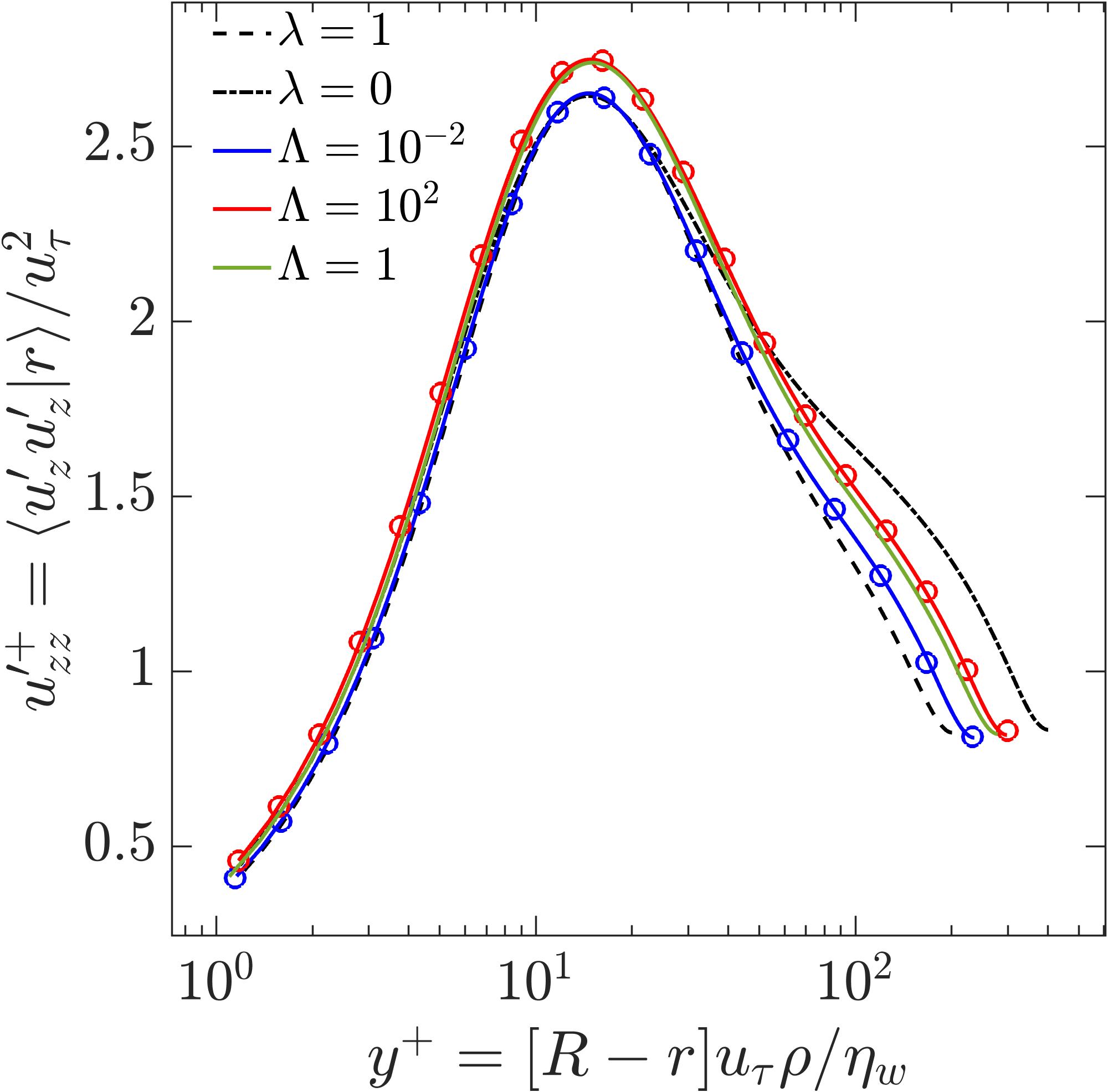}\\
(c) & (d)
\end{tabular}
\caption{Radial profiles of (a) radial/axial $u'_{rz}$, (b) radial $u'_{rr}$, (c) azimuthal $u'_{tt}$ and (d) axial $u'_{zz}$ Reynolds stresses, for thixotropic flows and Newtonian reference cases. Lines indicates results from DNS computations of thixotropic and Newtonian models and circles indicate results from the analytic closures (\ref{eq:lag_lambda_slow}) and (\ref{eq:lag_lambda_fast}).}
\label{fig:Mprofiles_2}
\end{figure}

The influence of viscosity on turbulence intensity is illustrated in Figure~\ref{fig:Mprofiles_2}. For the Newtonian reference cases, the profiles of all the turbulence intensities (except $u'_{zz}$) are higher for the unstructured case $\lambda=1$, with peaks for $u'_{rr}$ and $u'_{rz}$ moving further away from the walls, which is due to the increased $Re_G$ for the unstructured case. These observations are consistent with well-established trends of Newtonian turbulence ~\citep{Toonder1997,ElKhoury2013}. For the thixotropic cases, the azimuthal turbulence intensity $u'_{tt}$ and the $u'_{rz}$ Reynolds stress transition monotonically with increasing $\Lambda$ from the fully structured ($\lambda=1$) to the unstructured ($\lambda=0$) case, in terms of both turbulence intensity and peak shift.

The profiles of the axial turbulence intensity $u'_{zz}$ transition monotonically in terms of turbulence intensity, whereas, the profiles of the radial turbulence intensity $u'_{rr}$ transition monotonically with in terms of peak shift. 

\begin{figure}
\centering
\begin{tabular}{cc}
\includegraphics[width=0.45\textwidth]{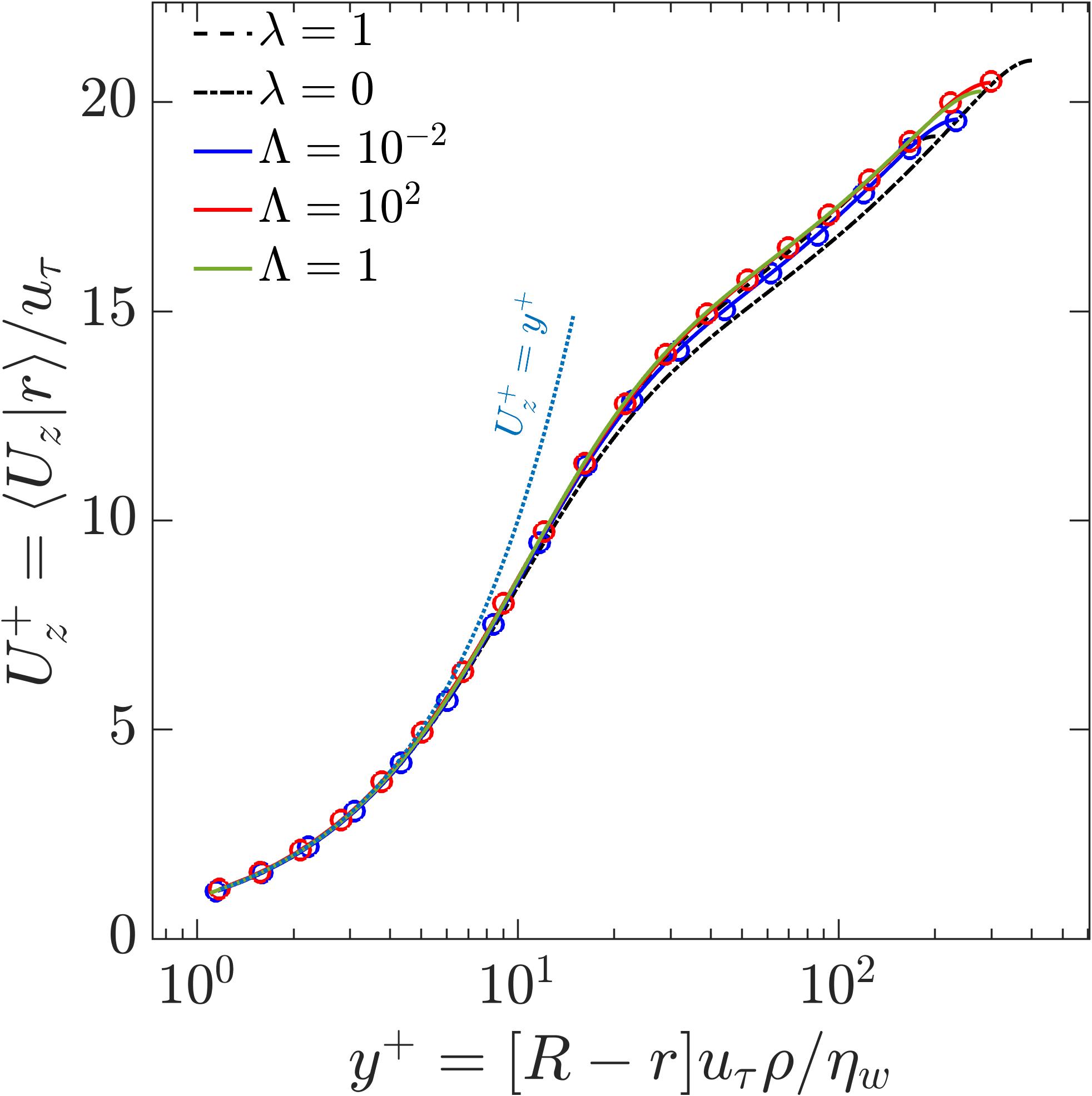}&
\includegraphics[width=0.45\textwidth]{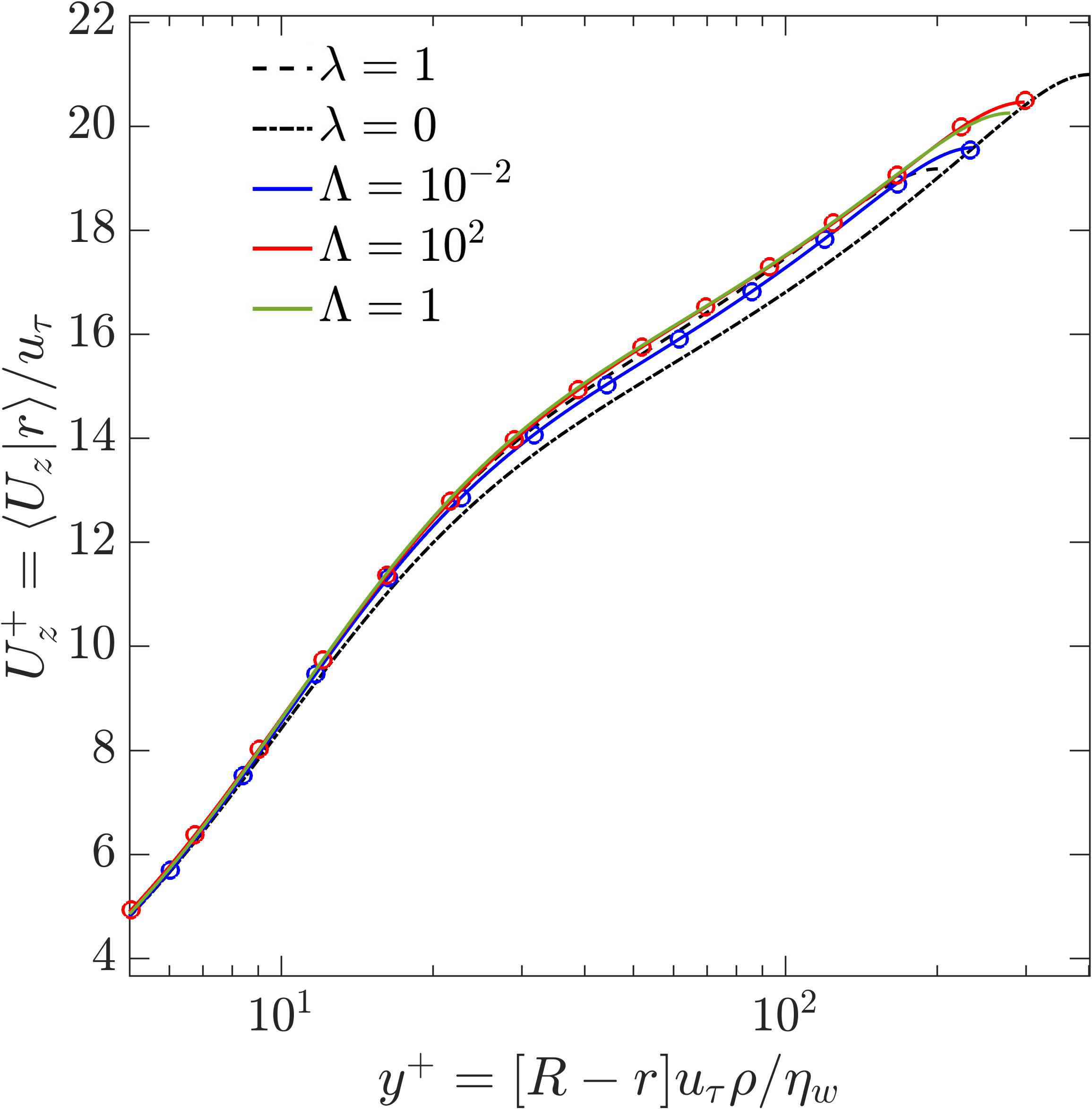}\\
(a) & (b)
\end{tabular}
\caption{Mean radial profiles of (a-b) axial velocity $U_z$ and 
for thixotropic flows and Newtonian reference cases. Lines indicates results from DNS computations of thixotropic and Newtonian models and circles indicate results from closures (\ref{eq:lag_lambda_slow}) and (\ref{eq:lag_lambda_fast}).}
\label{fig:Mprofiles_3}
\end{figure}

The mean axial velocity profiles shown in Figure~\ref{fig:Mprofiles_3}a demonstrate that all flow profiles roughly follow a linear relationship $U_z^+=y^+$ in the viscous sub-layer~\citep{Pope2001}. Figure~\ref{fig:Mprofiles_3}b provides a clearer view of these profiles,
with the deviation monotonically increases with increasing $\Lambda$. 
The peak deviation occurs in the buffer layer ($30 \leq y^+ \leq 2000$) for all the DNS cases due to increased turbulent intensities in that region (see Figure~\ref{fig:Mprofiles_2}). For the thixotropic cases, the area integral of the deviation over the pipe cross-section gives us the excess bulk flow rate (or drag reduction), thus drag reduction increases with $\Lambda$, which is consistent with previous studies~\citep{Escudier1996,Pereira2001}.

\section{Lagrangian Thixotropy}\label{sec:Lagrangian Thixotropy}

\subsection{Lagrangian Frame}\label{subsec:Lagrangian frame}

The results in \S\ref{sec:Eulerian Characteristics of Thixotropic Turbulence} suggest that the shear history along pathlines plays a critical role in organising thixotropic turbulence. As the ADRE (\ref{eq:transport_nD}) for $\lambda$ is advection-dominated ($\Pen\gg 1$), the Lagrangian frame is well-suited to study the feedback mechanisms that govern thixotropic turbulence, as this naturally encodes the shear history experienced by fluid elements. In this section we develop a Lagrangian model of thixotropy that provides insights into the feedback mechanisms that govern thixotropic turbulence and use these insights to develop analogue time-independent rheological models. 

We first consider evolution of the structural parameter $\lambda$ in the Lagrangian frame $\mathbf{X}$ as $\lambda(t;\mathbf{X},t_0)=\lambda(\mathbf{x}(t;\mathbf{X},t_0),t)$, where $\mathbf{x}(t;\mathbf{X},t_0)$ is the solution of the advection equation for a fluid particle initially at position $\mathbf{X}$ at time $t=t_0$:
\begin{equation} \label{eq:advect}
\frac{d\mathbf{x}(t;\mathbf{X},t_0)}{dt}=\mathbf{v}(x(t;\mathbf{X},t_0),t)
,\quad 
\mathbf{x}(t_0;\mathbf{X},t_0)=\mathbf{X}.
\end{equation}
In the limit of vanishing diffusivity ($Pe \rightarrow \infty$), the ADRE (\ref{eq:transport_nD}) can be transformed to the Lagrangian frame as 
\begin{equation} \label{eq:ADRE2}
 \frac{\partial\lambda(t;\mathbf{X},t_0)}{\partial t} = \Lambda\left([1-\lambda(t;\mathbf{X},t_0]-K\dot\gamma(t;\mathbf{X},t_0)\lambda(t;\mathbf{X},t_0))\right)
 ,\quad 
 \lambda(t_0;\mathbf{X},t_0)=\lambda_0.
\end{equation}
Solution of (\ref{eq:ADRE2}) shows that the structural parameter evolves with respect to Lagrangian shear history as
\begin{equation} \label{eq:memory}
	\begin{split}
		\lambda(t;\mathbf{X},t_0)&=\frac{1}{G'(t;\mathbf{X},t_0)}\left[\lambda_0+\Lambda\int_0^t G'(t^\prime;\mathbf{X},t_0)dt^\prime\right],\\
G'(t;\mathbf{X},t_0)&=\exp\left[\Lambda\int_{t_0}^t g(t^\prime,\mathbf{X},t_0)\,dt^\prime\right],
	\end{split}
\end{equation}
where $g(t,\mathbf{X},t_0)\equiv 1+K\dot\gamma(t,\mathbf{X},t_0)$. As we are interested in the long-time behavior of $\lambda(t)$, given by $t_0\rightarrow -\infty$ (or more accurately $t-t_0\gg 1/\Lambda$), the initial condition $\lambda_0$ has no impact on $\lambda$ and (\ref{eq:memory}) may be recast as
\begin{equation}\label{eq:memory2}
	\begin{split}
\lambda(t;\mathbf{X},t_0)&=\frac{\Lambda\int_0^\infty G(t-s;\mathbf{X},t_0)ds}{G(t;\mathbf{X},t_0)},\\
	G(t;\mathbf{X},t_0)&=\exp\left[-\Lambda\int_0^\infty g(t-s;\mathbf{X},t_0)\,ds\right].
	\end{split}
\end{equation}
Here $G$ in (\ref{eq:memory2}) represents a fading memory kernel $\hat{\mathcal{F}}$
\begin{equation}
G(t;\mathbf{X},t_0)=\hat{\mathcal{F}}\left[\dot\gamma(t-s;\mathbf{X},t_0)|_{s=0}^{\infty}\right],\label{eqn:memory3}
\end{equation}
that encodes the shear history from $s=0$ to $s\rightarrow-\infty$. Hence the most recent shear rates have the greatest impact, and the persistence of memory (analogous to relaxation time for viscoelastic fluids) is controlled by the thixoviscous number.

Equations (\ref{eq:memory2}), (\ref{eqn:memory3}) also shows that advective transport is an important mechanism in non-stationary thixotropic turbulent flow. From a stochastic perspective, the Lagrangian shear rate history $\dot\gamma(t-s;\mathbf{X},t_0)$ can be considered as a random process that is non-stationary in space and so is also dependent upon the history $s$ of Eulerian positions $\mathbf{x}(t-s;\mathbf{X},t_0)$ along pathlines. For fully-developed turbulent pipe flow the shear rate $\dot\gamma$ is non-stationary in the radial direction $r$, hence the functional (\ref{eqn:memory3}) simplifies to $G(r(t),t)=\hat{\mathcal{F}}[\dot\gamma(t-s;r(t-s))|_{s=0}^{\infty}]$, and the evolution equation  (\ref{eq:memory2}) needs to be supplemented by an advective transport model for evolution of radial position $r(t)$. This stochastic approach will be considered further in \S\S\ref{subsec:stochastic_model}.

Conversely, for statistically stationary turbulent flows (\ref{eqn:memory3}) simplifies to $G(t)=\hat{\mathcal{F}}[\dot\gamma(t-s)|_{s=0}^{\infty}$. If the Lagrangian shear rate follows a simple autocorrelation process with decorrelation time $\tau_{\dot\gamma}$
\begin{align}
    R_{\dot\gamma\dot\gamma}(|t-t^\prime|)\equiv\langle\dot\gamma^\prime(t;\mathbf{X},t_0)\dot\gamma^\prime(t^\prime;\mathbf{X},t_0)\rangle\sim\exp\left(-|t-t^\prime|/\tau_{\dot\gamma}\right),\label{eqn:shear_decorr}
\end{align}
then $\dot\gamma(t-s)$ may be modeled as a Markovian random process, leading to a fairly simple random walk model for the evolution of $\lambda$ via (\ref{eq:memory2}).

\begin{figure}
\centering
\begin{tabular}{ccc}
\includegraphics[width=0.33\textwidth]{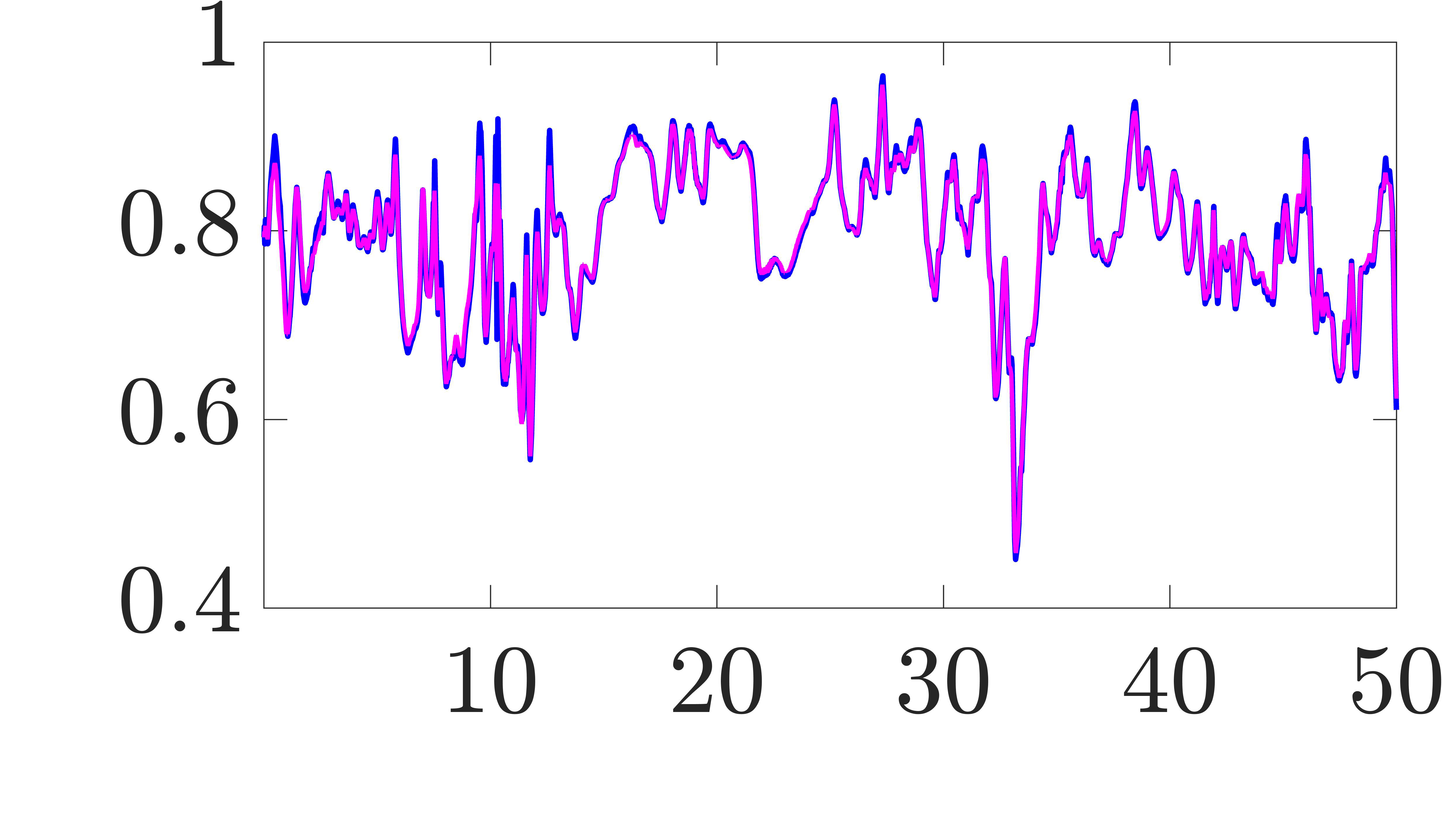}&
\includegraphics[width=0.33\textwidth]{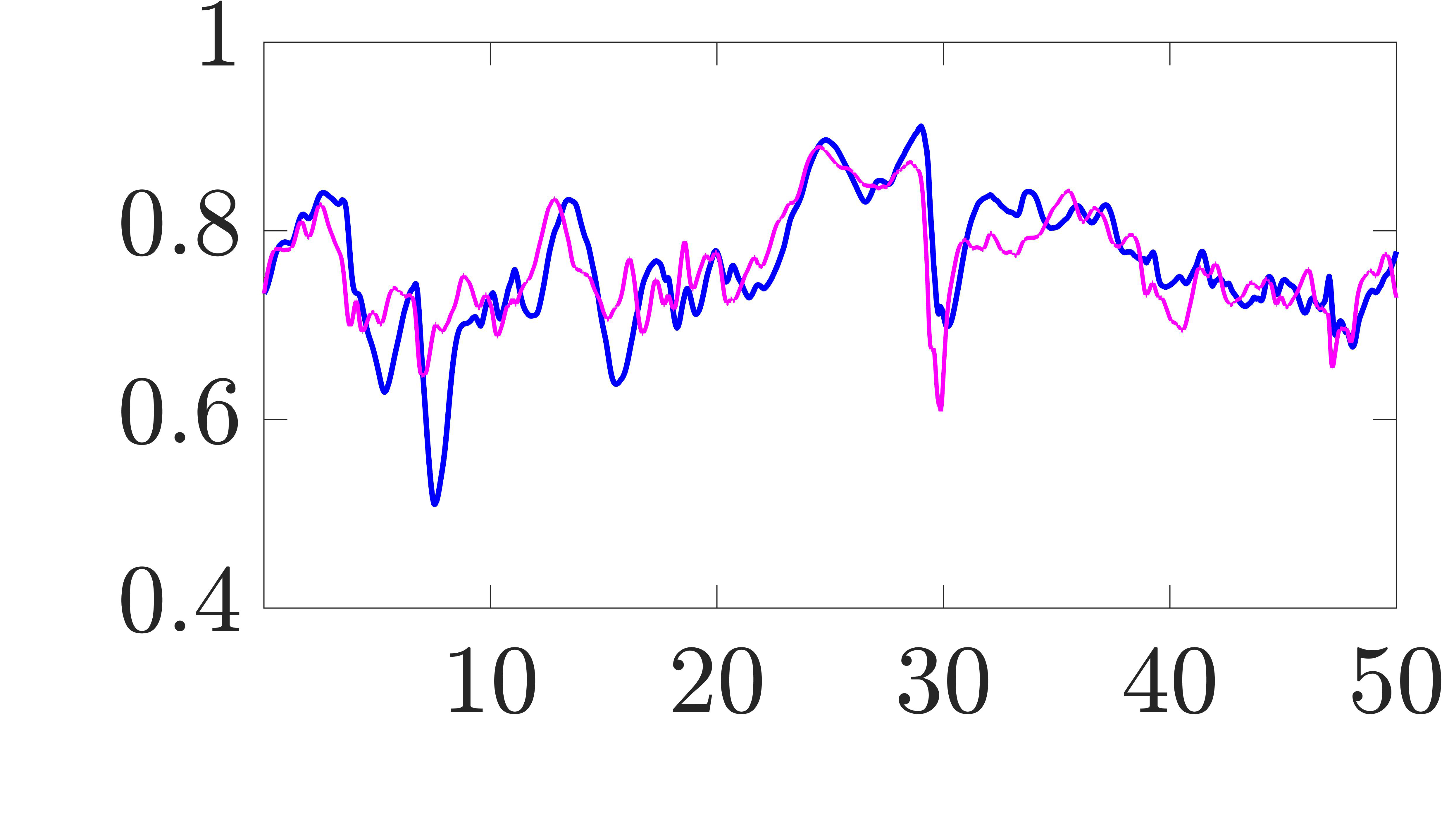}&
\includegraphics[width=0.33\textwidth]{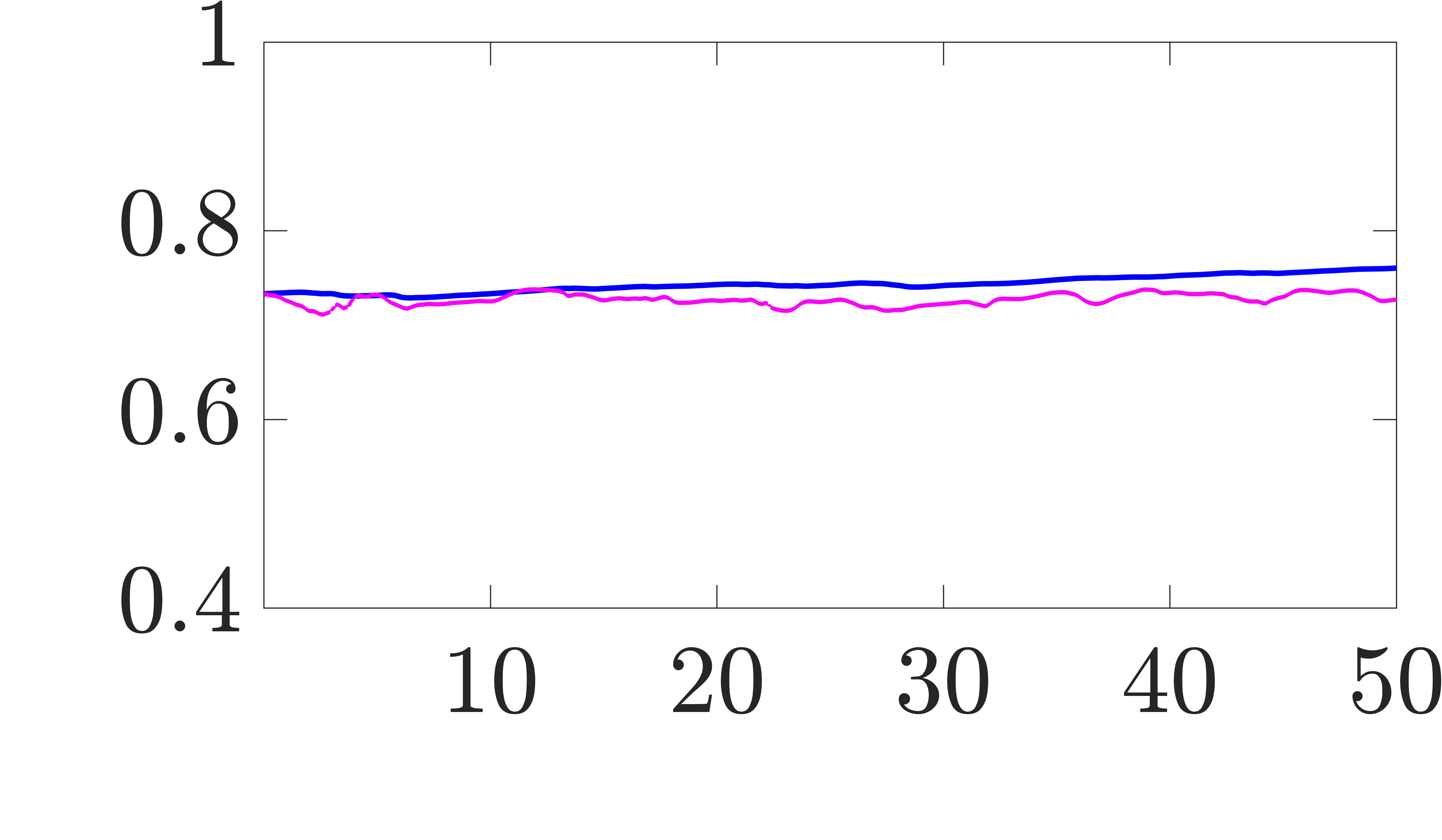}\\
\includegraphics[width=0.33\textwidth]{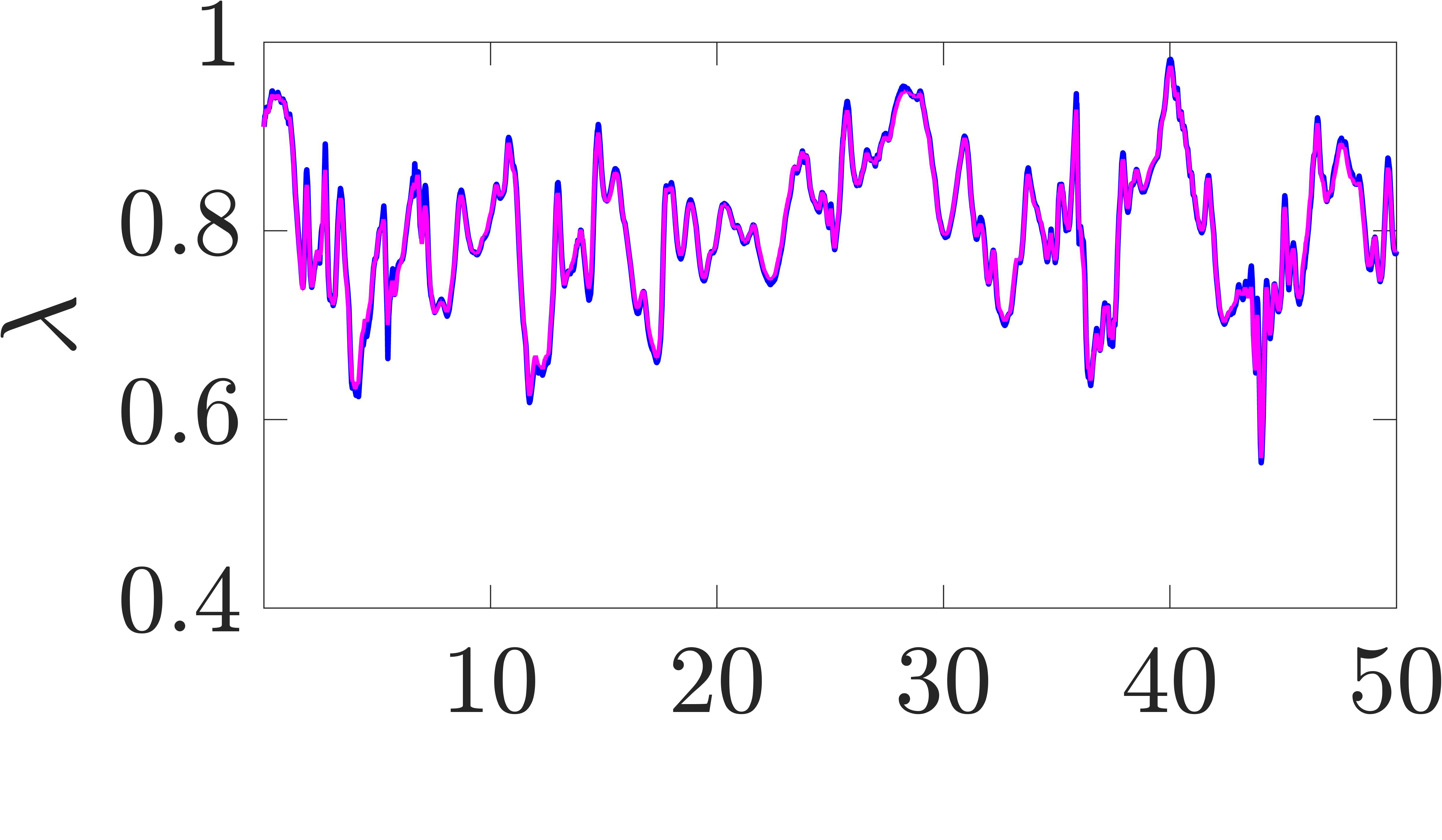}&
\includegraphics[width=0.33\textwidth]{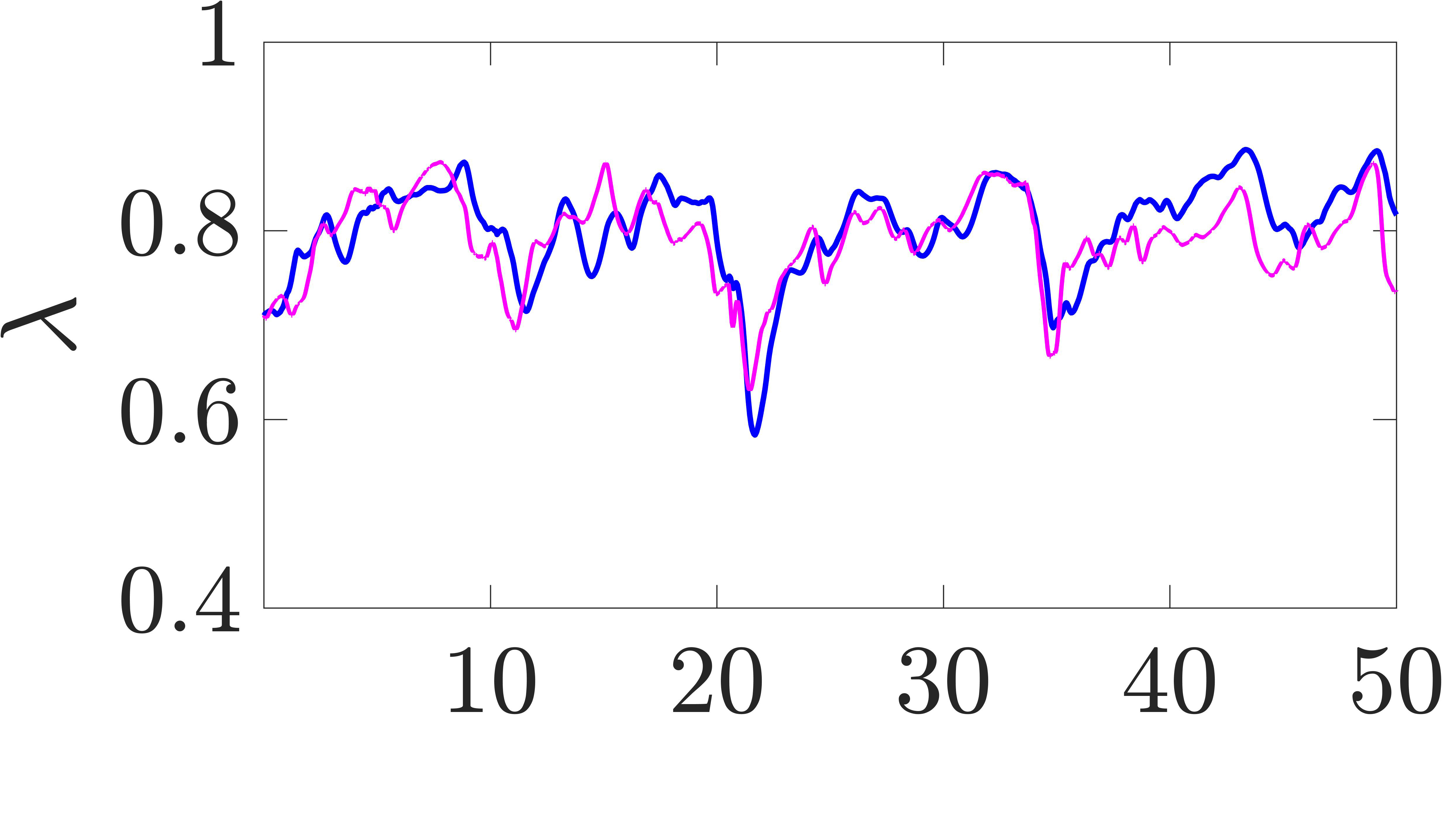}&
\includegraphics[width=0.33\textwidth]{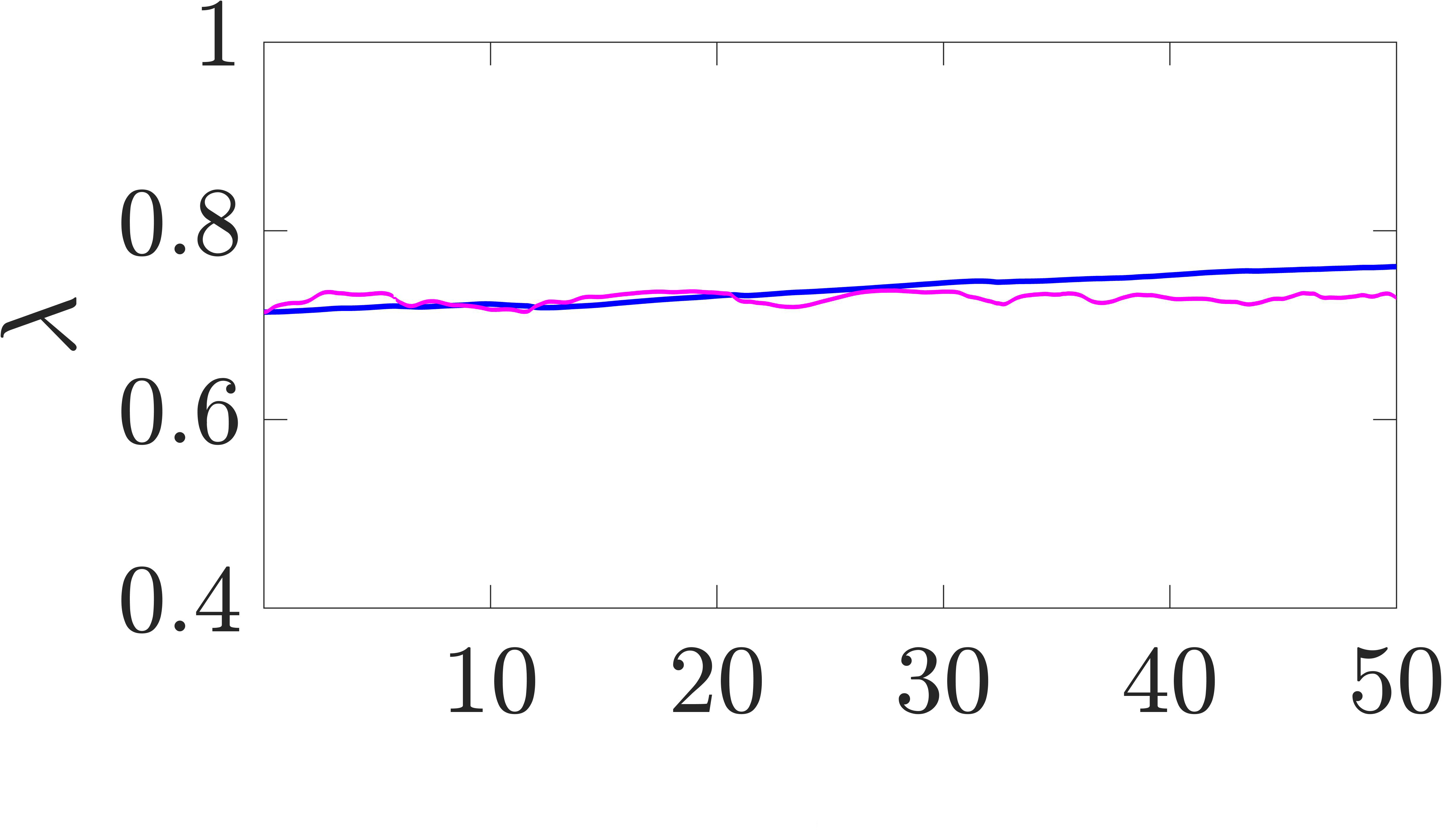}\\
\includegraphics[width=0.33\textwidth]{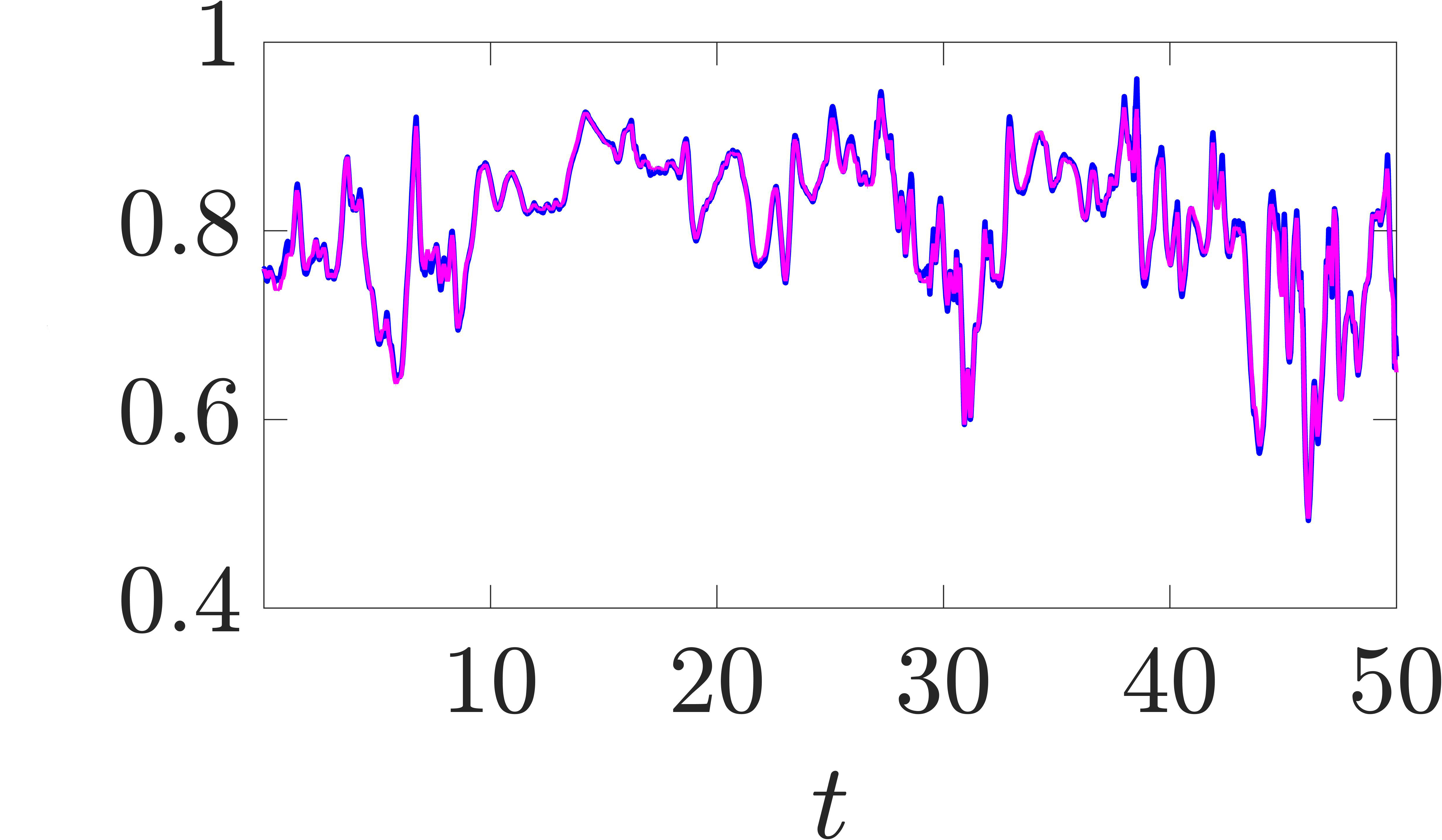}&
\includegraphics[width=0.33\textwidth]{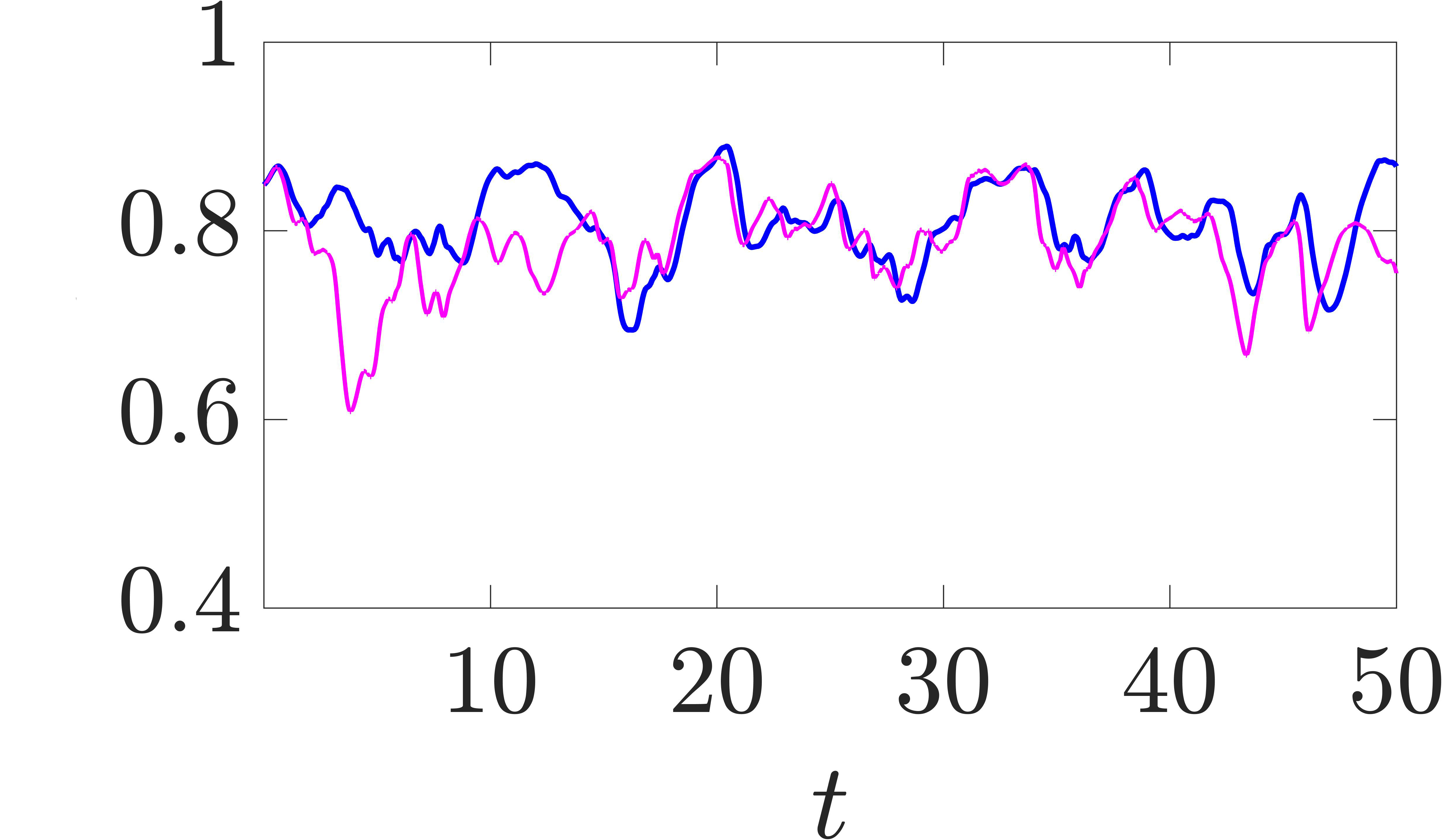}&
\includegraphics[width=0.33\textwidth]{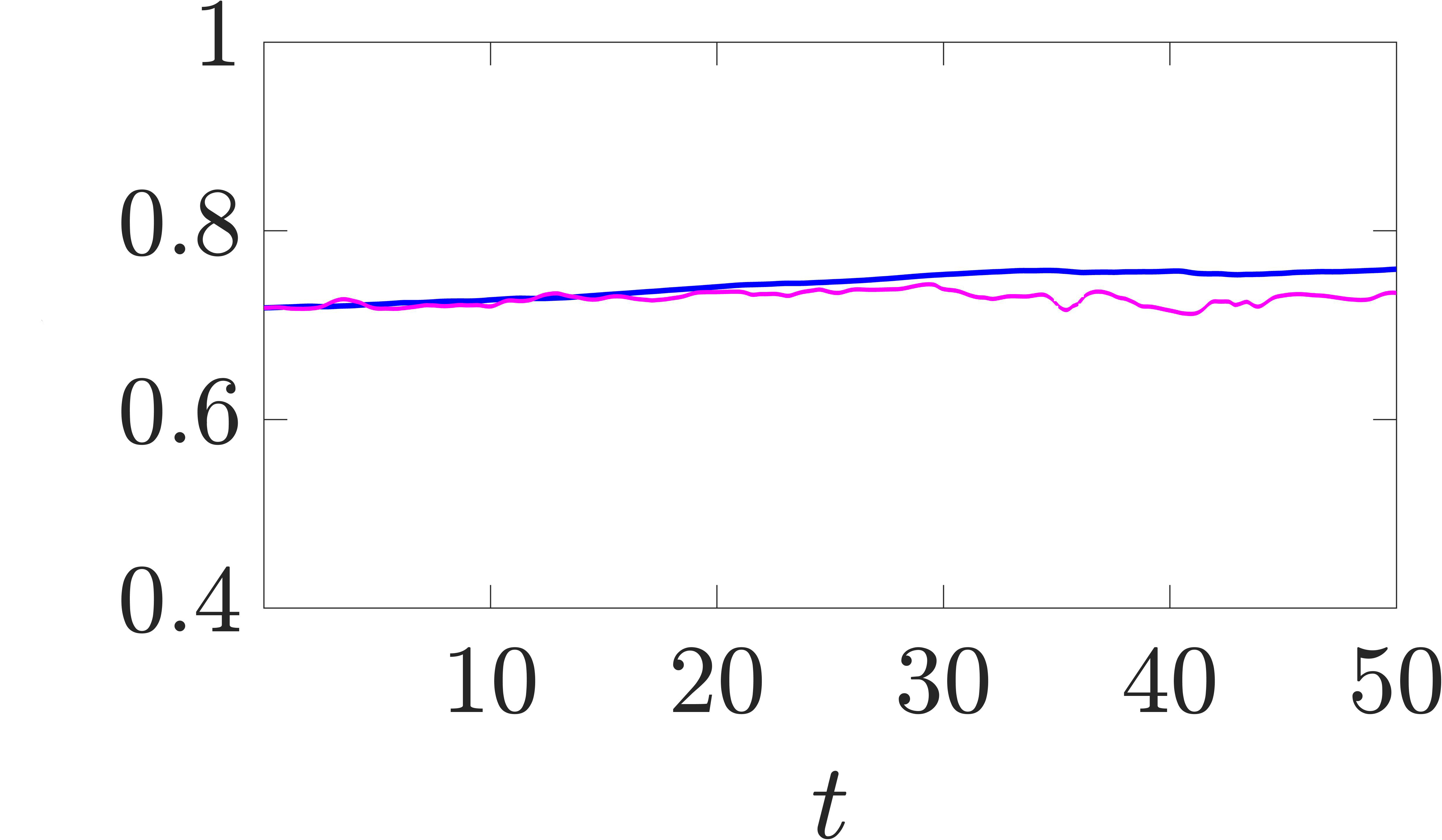}\\
(a) $\Lambda=10^{2}$ & (b) $\Lambda=1$ & (c) $\Lambda=10^{-2}$
\end{tabular}
\caption{Comparison of structural parameter $\lambda(t;\mathbf{X},t_0)$ along typical particle trajectories for thixotropic flows with (a) $\Lambda=10^{2}$, (b) $\Lambda=1$, and (c) $\Lambda=10^{-2}$, from (\textcolor{blue}{\textbf{--$\!$--}}) DNS results and (\textcolor{blue}{\textbf{--$\!$--}}) solution of the Lagrangian equation (\ref{eq:ADRE2}).}
\label{fig:FDM_all_compare}
\end{figure}

To test the Lagrangian assumption, the evolution of $\lambda$ along different pathlines is compared in Figure~\ref{fig:FDM_all_compare} between DNS results and computation of $\lambda$ via in (\ref{eq:ADRE2}). For the fast thixotropic kinetics ($\Lambda = 10^2$) with a high Damk\"{o}hler number ($Da = 10^5$), the DNS results closely match the Lagrangian solution (\ref{eq:ADRE2}), with a relative $L_2$ error of 0.18\% over $10^4$ trajectories. For intermediate kinetics ($\Lambda=1$), diffusion is significant although thixotropic kinetics still dominate ($Da = 10^3$), and the Lagrangian solution (\ref{eq:ADRE2}) has relative $L_2$ error of 1.79\%. For the slow kinetics ($\Lambda = 10^{-2}$), the timescales of diffusion approach those of the thixotropic kinetics ($Da = 10$), leading to noticeable discrepancies in pathlines and relative $L_2$ error of 4.03\%. The discrepancies in Figure~\ref{fig:FDM_all_compare}c are due to to the moderate P\'{e}clet number ($\Pen=10^3$) used for numerical stability, corresponding to $Da\sim 10$ for $\Lambda=10^{-2}$. As previously discussed, both $\Pen$ and $Da$ are much higher for complex fluids, hence the Lagrangian framework is broadly applicable to thixotropic flows. Although the Lagrangian assumption does not strictly hold in this study for $\Lambda\ll1$, we shall show in the following that this does not significantly alter the overall dynamics of the flow, and and accurate analytic closures for the limiting cases of $\Lambda$ can be developed.

\subsection{Fast Kinetics}\label{subsec:Fast Kinetics}

In the fast kinetic regime, $\Lambda \rightarrow \infty$, the thixotropic timescale $\tau_r$ is much shorter than the eddy turnover time $\tau_v$ that governs the Lagrangian shear rate decorrelation time, hence only the very recent shear history dictates $\lambda$. Following (\ref{eq:memory2}), we derive an expression for the structural parameter $\lambda$ in the limit $\Lambda \rightarrow \infty$ by first performing a series expansion of $\dot\gamma(t-s;\mathbf{X},t_0)$ about $t$ as
\begin{equation}
    \dot\gamma(t-s;\mathbf{X},t_0)=\dot\gamma(t;\mathbf{X})+\sum_{n=1}^\infty\dot\gamma_n(t;\mathbf{X})\left(\frac{-s}{\tau_{\dot\gamma}}\right)^n,
\end{equation}
where from the definition of $\tau_{\dot\gamma}$, the coefficients $\dot\gamma_n(t;\mathbf{X},t_0)\equiv\dot\gamma(t;\mathbf{X},t_0)^{(n)}\tau_{\dot\gamma}^n/n!\sim 1$, and $x^{(n)}(t)$ denotes the $n$-th derivative of $x$ with respect to $t$. Introducing the rescaled thixotropic time $t_r=t\Lambda$, the structural parameter evolves according to
\begin{equation} \label{eq:largeT}
\lim_{\Lambda\rightarrow\infty}\lambda(t;\mathbf{X},t_0)=\lim_{\Lambda\rightarrow\infty}\lambda(t_r/\Lambda;\mathbf{X},t_0)=\frac{\lim_{\Lambda\rightarrow\infty}\int_{t_0}^{r/\Lambda}G\left(t_r/\Lambda;\mathbf{X},t_0\right)dt_r}{\lim_{\Lambda\rightarrow\infty}G\left(t_r/\Lambda;\mathbf{X},t_0\right)}
\end{equation}
where the history function $G(t;\mathbf{X},t_0)$ in the limit $\Lambda\rightarrow\infty$ is then
\begin{equation}
\begin{split}
\lim_{\Lambda\rightarrow\infty}G\left(t_r/\Lambda;\mathbf{X},t_0\right)=&\lim_{\Lambda\rightarrow\infty}\exp\left(\int_0^\infty 1+K\dot\gamma\left(t_r/\Lambda-s_r/\Lambda;\mathbf{X},t_0\right)ds_r\right),\\
=&\lim_{s_r\rightarrow\infty}\exp\left(\left[1+K\dot\gamma\left(t_r/\Lambda;\mathbf{X},t_0\right)\right]s_r\right).
\end{split}
\end{equation}
Application of L'Hopital's rule to (\ref{eq:largeT}) yields
\begin{equation}\label{eq:lag_lambda_fast}
\begin{split}
\lim_{\Lambda\rightarrow\infty}\lambda(t;\mathbf{X},t_0)
=&\frac{\lim_{\Lambda\rightarrow\infty}G\left(t;\mathbf{X},t_0\right)}{\lim_{\Lambda\rightarrow\infty}\left(1+K\dot\gamma\left(t;\mathbf{X},t_0\right)\right)G\left(t;\mathbf{X},t_0\right)}.
\\=&\frac{1}{1+K\dot\gamma\left(t;\mathbf{X},t_0\right)}.
\end{split}
\end{equation}
Hence the structural parameter in Eulerian space for $\Lambda\rightarrow\infty$ is given by the equilibrium value
\begin{equation} \label{eq:eul_lambda_fast}
\lim_{\Lambda\rightarrow\infty}\lambda(\mathbf{x},t)=\frac{1}{1+K\dot\gamma(\mathbf{x},t)}.
\end{equation}
For $\Lambda \rightarrow \infty$ the structural parameter instantly equilibrates with respect to the local shear rate, hence the effective viscosity of the thixotropic psuedo-Newtonian fluid is a time-independent shear-thinning generalised Newtonian fluid~\citep{Scott1951,Barnes1997} that corresponds to a Cross fluid with unit index 
\begin{equation}\label{eq:eul_GN_fast}   \lim_{\Lambda\rightarrow\infty}\eta(\lambda)\equiv\eta_{\Lambda\rightarrow\infty}(\dot\gamma)=\eta_\infty+\frac{\eta_0-\eta_\infty}{1+k\dot\gamma}
    ,
\end{equation}
\begin{figure}
\centering
\begin{tabular}{cc}
\includegraphics[width=0.33\columnwidth]{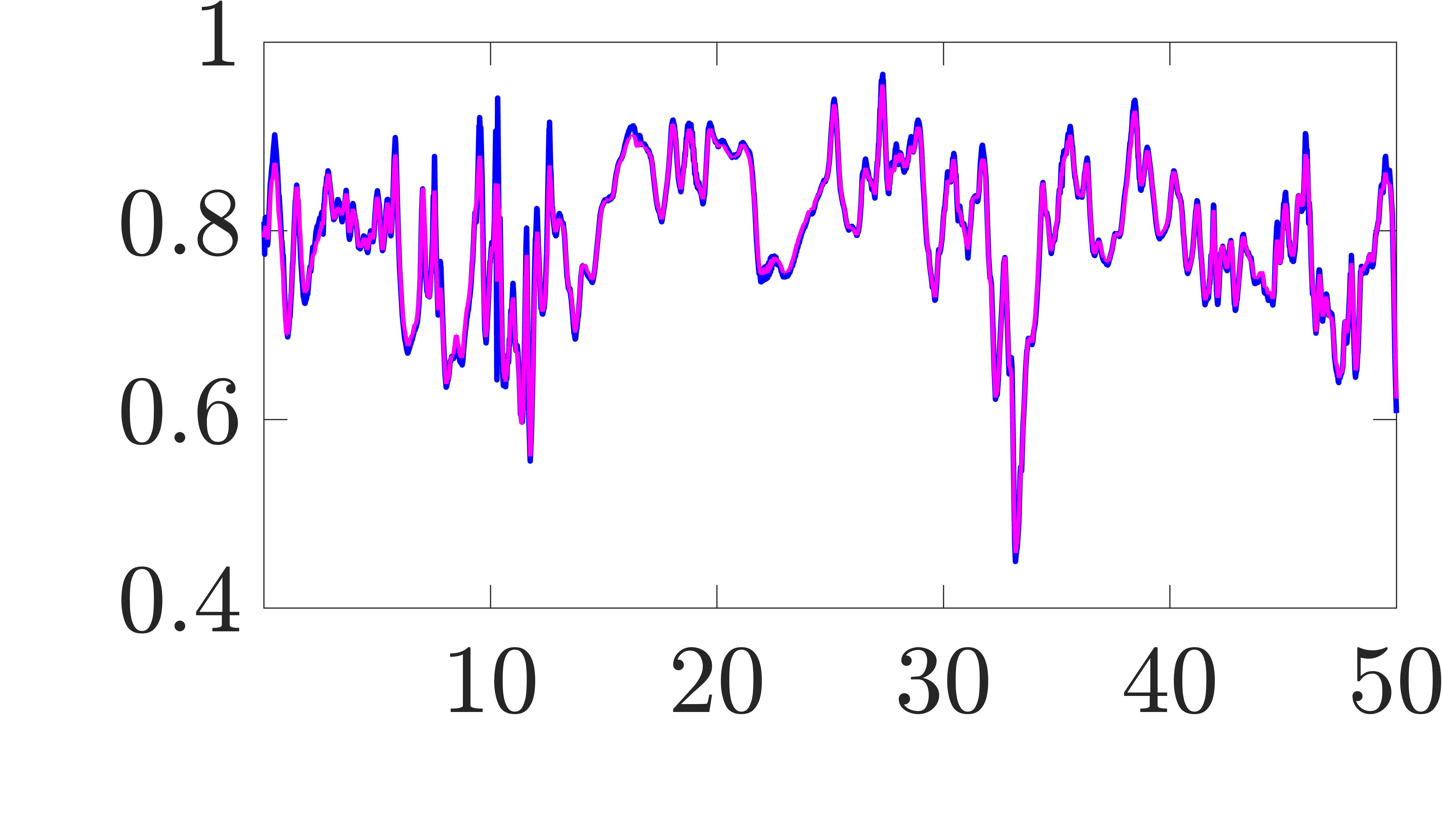} & \multirow{3}{*}[4.5em]{\includegraphics[width=0.5\columnwidth]{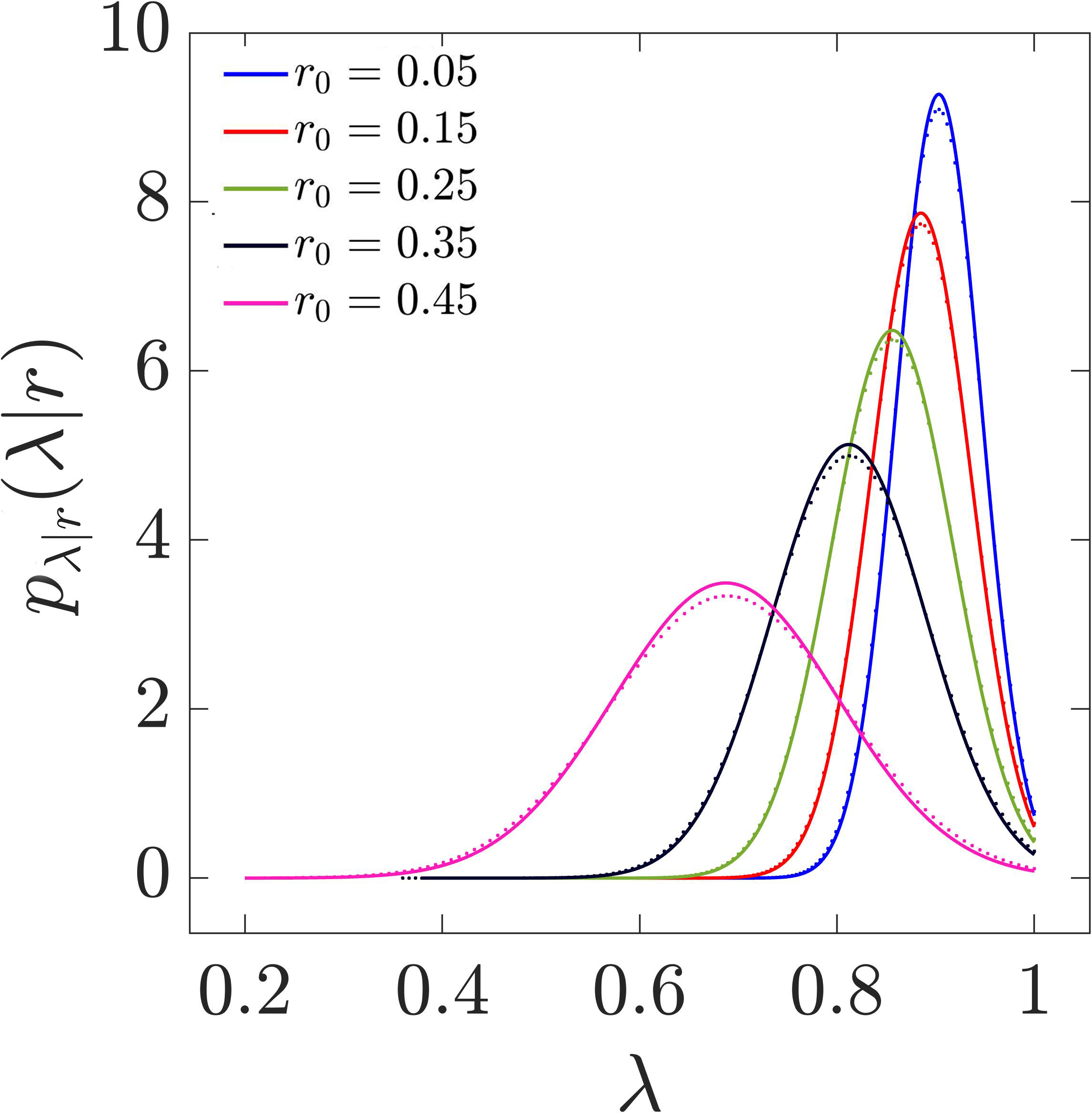}} 
\\
\includegraphics[width=0.33\columnwidth]{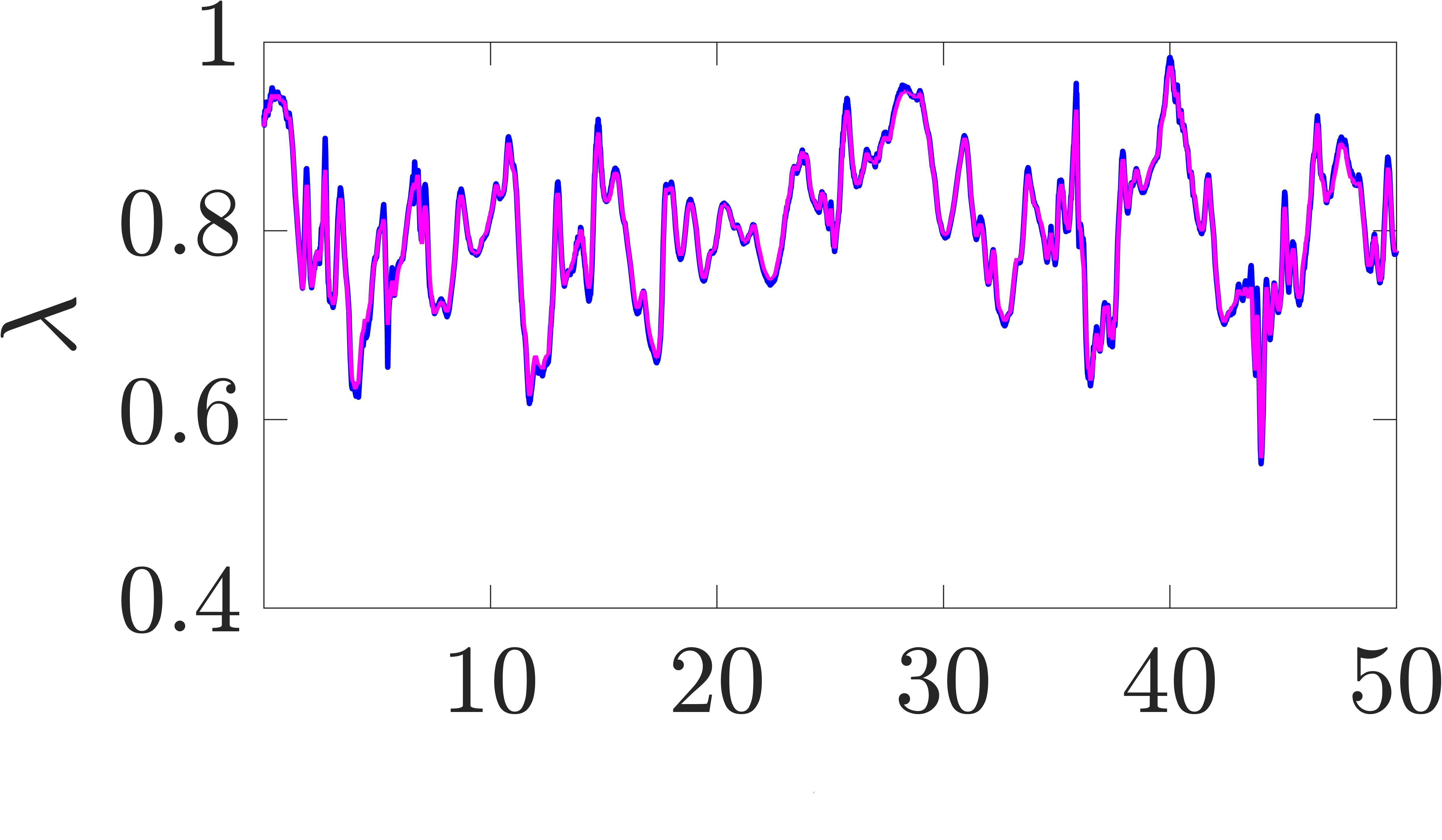} &                    
\\
\includegraphics[width=0.33\columnwidth]{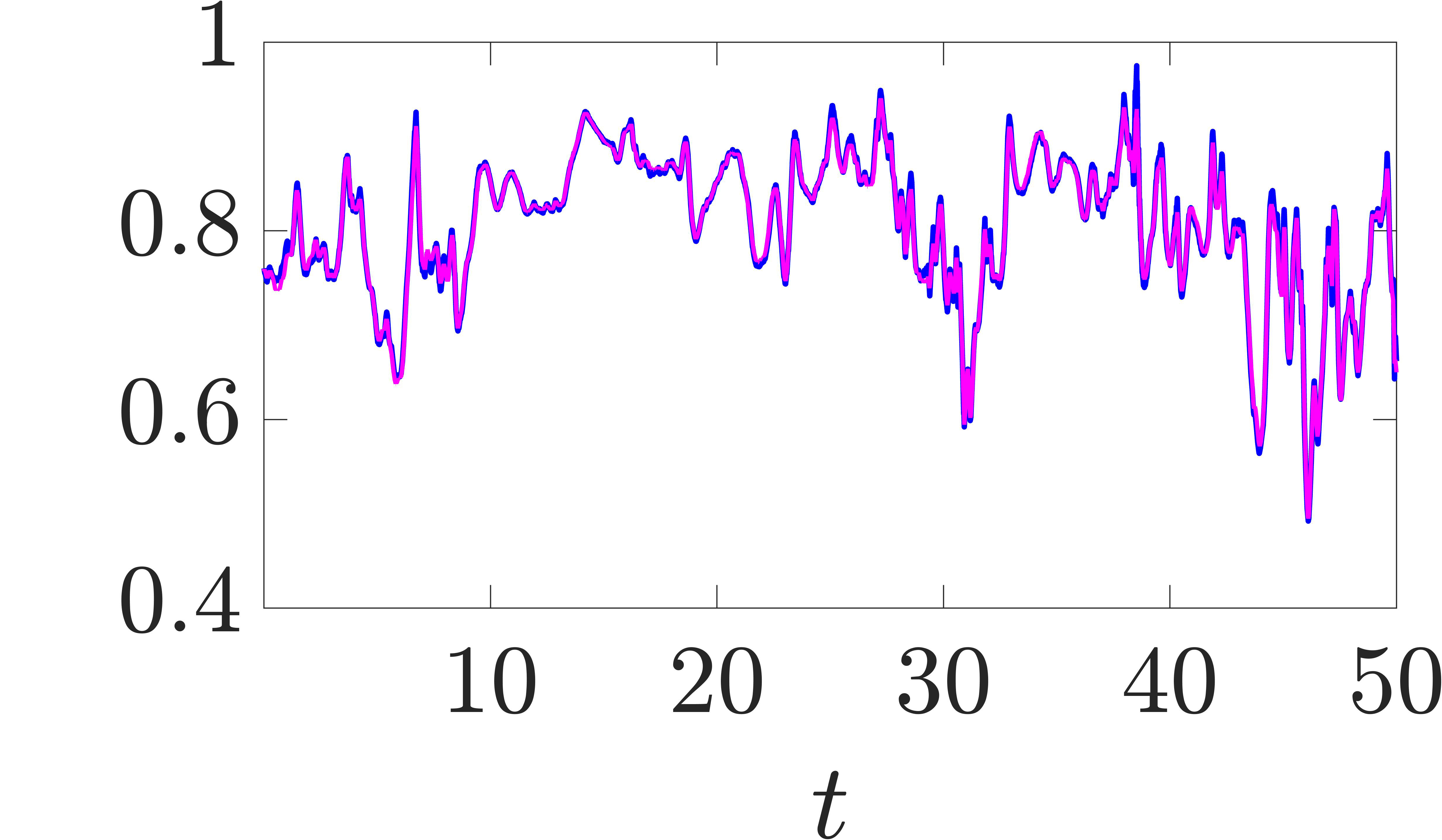} &                    
\\
(a) & \multicolumn{1}{c}{(b)}
\end{tabular}
\caption{(a) Comparison of structural parameter $\lambda(t;\mathbf{X},t_0)$ along typical particle trajectories for thixotropic flows with $\Lambda=10^{2}$, from (\textcolor{blue}{\textbf{--$\!$--}}) DNS results and (\textcolor{blue}{\textbf{--$\!$--}}) the analytic closure (\ref{eq:lag_lambda_fast}). (b) Comparison of conditional p.d.f. of structural parameter $p_\lambda(\lambda|r)$ along typical particle trajectories for thixotropic flows with $\Lambda=10^{2}$, from (solid lines) DNS results and (dotted lines) the analytic closure (\ref{eq:lag_lambda_fast}).}
\label{fig:compare_fast}
\end{figure}

Figure~\ref{fig:compare_fast} verifies (\ref{eq:lag_lambda_fast}) via comparison against DNS computations for $\Lambda = 10^2$. Figure~\ref{fig:compare_fast}a shows that the fast kinetics model (\ref{eq:lag_lambda_fast}) for evolution of $\lambda$ along sample trajectories matches DNS results very well, with a relative $L_2$ error of 0.1\% across $10^4$ Lagrangian trajectories. Figure~\ref{fig:compare_fast}b also shows that the conditional p.d.f. $p_{\lambda|r}(\lambda|r)$ of $\lambda$ at various radial positions from (\ref{eq:lag_lambda_fast}) also matches the DNS results very well, verifying both the Lagrangian assumption and the closure model (\ref{eq:lag_lambda_fast}).

To test the closure (\ref{eq:lag_lambda_fast}) and the corresponding effective viscosity model (\ref{eq:eul_GN_fast}), DNS computations using (\ref{eq:eul_GN_fast}) are compared with those using the thixotropic kinetics with $\Lambda=10^2$ in Figures~\ref{fig:Mprofiles_1} - \ref{fig:Mprofiles_3}. Computations based on the effective viscosity model (\ref{eq:eul_GN_fast}) show excellent agreement with the full thixotropic model, with relative errors of under 0.1\% for the mean structural parameter $\lambda$, viscosity $\eta$ and axial velocity $U_z$. The errors in turbulent statistics ($u'_{rr},u'_{tt},u'_{zz},u'_{rz}$) are also all less than 0.5\%. These results confirm that, as expected, for $\Lambda\gg 1$ thixotropic fluids exhibits turbulent fluid dynamics akin to that of shear-thinning fluids. Note that this result extends to a broad range of rheologies and thixotropic models described by (\ref{eqn:eta}), (\ref{eqn:thixo_homog}) respectively.

\subsection{Slow Kinetics}\label{subsec:Slow Kinetics}

To analyse thixotropic rheology in the limit of vanishingly slow thixotropic kinetics, $\Lambda \rightarrow 0$, we decompose the memory kernel into an ensemble average $\langle\dot\gamma\rangle$ that, due to ergodicity, corresponds to the Lagrangian average 
\begin{equation}
    \langle\dot\gamma\rangle\equiv\int_{-\infty}^\infty\dot\gamma p_{\dot\gamma}(\dot\gamma) d\dot\gamma = \lim_{T\rightarrow\infty}\frac{1}{T}\int_0^T\dot\gamma(t;\mathbf{X},t_0)dt,
\end{equation} 
and a fluctuating component $\dot\gamma'$ as
\begin{equation}   \dot\gamma(t;\mathbf{X},t_0)=\langle\dot\gamma\rangle+\dot\gamma'(t/\tau_{\dot\gamma};\mathbf{X},t_0),
\end{equation}
where the scaling $t/\tau_{\dot\gamma}$ ensures $\partial\dot\gamma'/\partial t\sim 1$. Using this decomposition, the memory kernel (\ref{eq:memory2}) is then
\begin{equation}
    \begin{split}
        G(t;\mathbf{X},t_0)&=\exp\left(\Lambda\int_{t_0}^t 1+K\langle\dot\gamma\rangle+K\dot\gamma'(t/\tau_{\dot\gamma};\mathbf{X},t_0) dt^\prime\right),\\
        &=\exp\left(\Lambda(1+K\langle\dot\gamma\rangle)(t-t_0)\right)\exp\left(\Lambda K\int_{t_0}^t\dot\gamma'(t/\tau_{\dot\gamma};\mathbf{X},t_0) dt^\prime\right).
    \end{split}
\end{equation}
and integration by parts yields
\begin{equation}
\begin{split}
        \Lambda\int_{t_0}^t G(t^\prime;\mathbf{X},t_0)dt^\prime
        &
    =\frac{G(t^\prime;\mathbf{X},t_0)-1}{1+K\langle\dot\gamma\rangle}- \Lambda \int_{t_0 \tau_{\dot\gamma}}^{t \tau_{\dot\gamma}}\frac{K\dot\gamma'(t_2^\prime;\mathbf{X},t_0)}{1+K\langle\dot\gamma\rangle}G(t_2^\prime \tau_{\dot\gamma};\mathbf{X},t_0)
    dt_2^\prime,
    \end{split}
\end{equation}
where $t_2=t \tau_{\dot\gamma}$. In the limits $t_0\rightarrow-\infty$, $\Lambda \rightarrow 0$, the structural parameter given by (\ref{eq:memory2}) then simplifies to
\begin{equation} \label{eq:lag_lambda_slow}
    \lim_{\Lambda\rightarrow 0}\lambda(t;\mathbf{X},t_0)=\frac{1}{1+K\langle\dot\gamma\rangle}=\lim_{\Lambda\rightarrow 0}\lambda(t,\mathbf{x}).
\end{equation}
Hence for $\Lambda \rightarrow 0$, the thixotropic kinetics are so slow compared to the fluctuation rate of the shear rate that the structural parameter effectively samples the ensemble of shear rates during convergence to equilibrium, and so the structural parameter is steady and is governed by the ensemble averaged shear rate $\langle\dot\gamma\rangle$. Hence the structural parameter in Eulerian space for $\Lambda\rightarrow0$ is given by the equilibrium value at the mean shear rate as:
\begin{equation} \label{eq:eul_lambda_slow}
 \lim_{\Lambda\rightarrow 0}\lambda(\mathbf{x},t)=\frac{1}{1+K \langle\dot{\gamma}\rangle}.
\end{equation}
and so the thixotropic viscosity for $\Lambda \rightarrow 0$ simplifies to the Newtonian viscosity
\begin{equation}\label{eq:eul_GN_slow}
    \lim_{\Lambda \rightarrow 0}\eta(\lambda)\equiv\eta_{\Lambda \rightarrow 0}=\eta_\infty+\frac{\eta_0-\eta_\infty}{1+K\langle\dot\gamma\rangle}
    ,\quad 
    \eta_\infty\leqslant\eta_{\Lambda \rightarrow 0}\leqslant\eta_0.
\end{equation}

\begin{figure}
\centering
\begin{tabular}{cc}
\includegraphics[width=0.33\columnwidth]{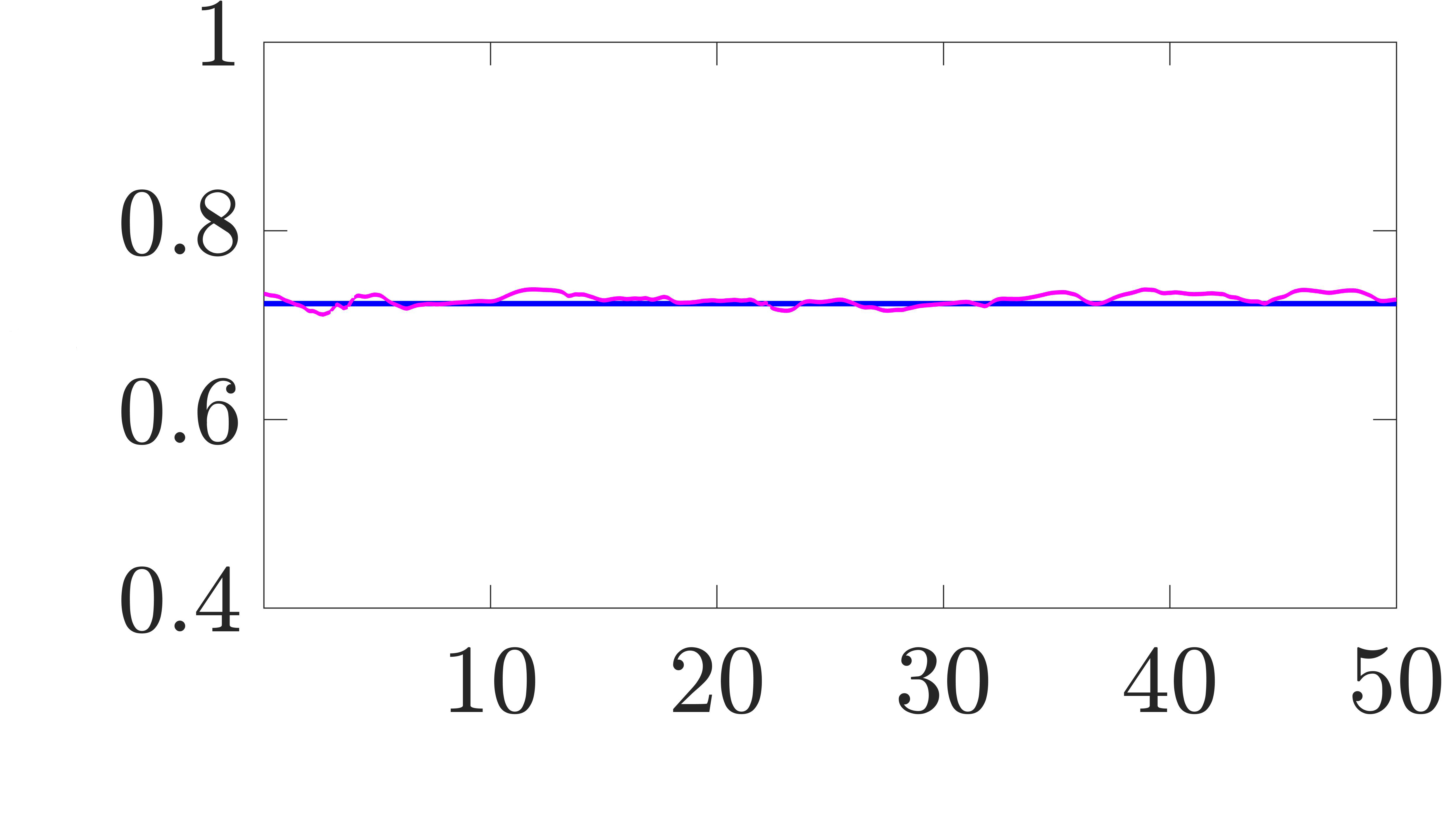} & \multirow{3}{*}[4.5em]{\includegraphics[width=0.5\columnwidth]{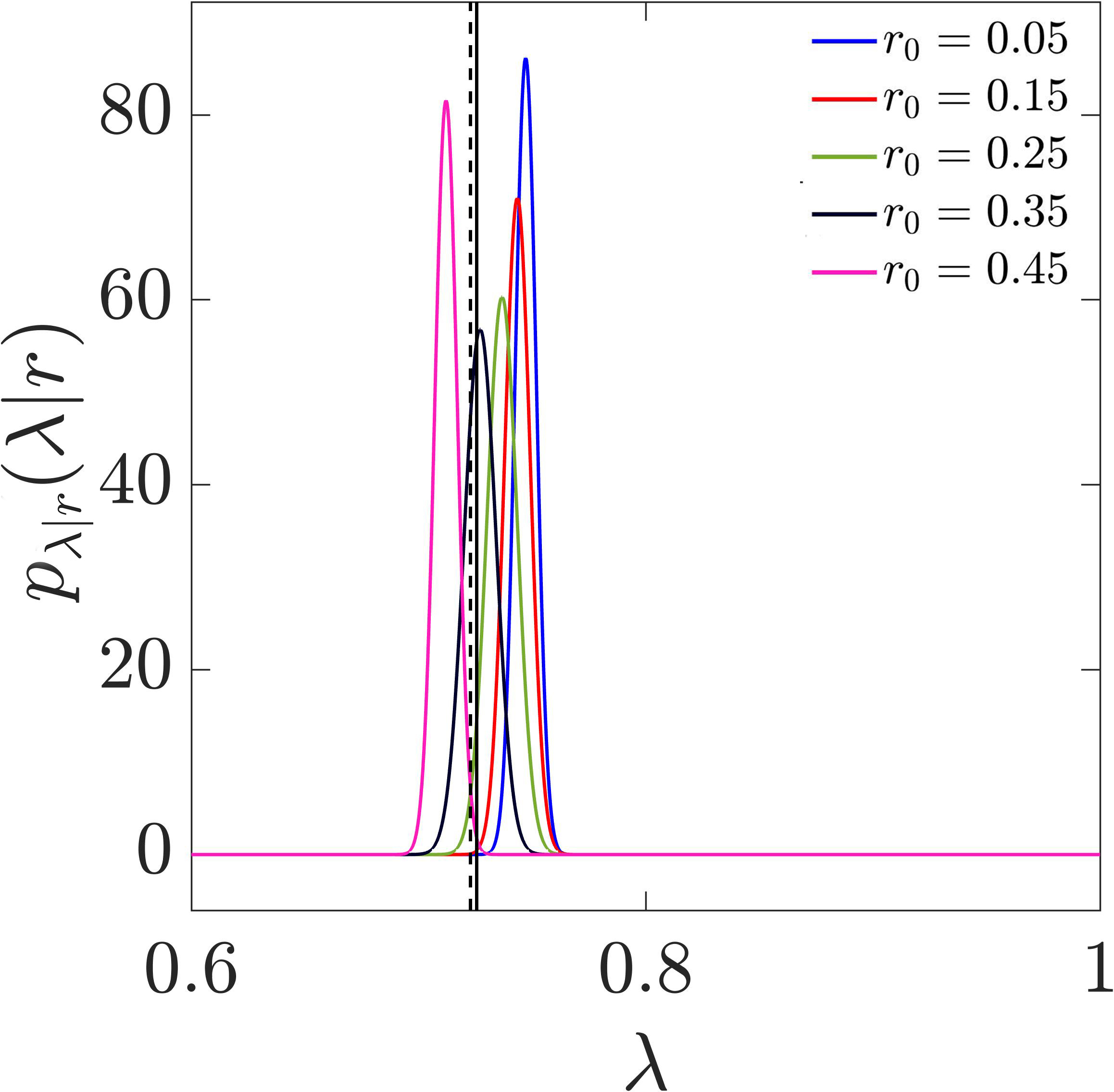}} 
\\
\includegraphics[width=0.33\columnwidth]{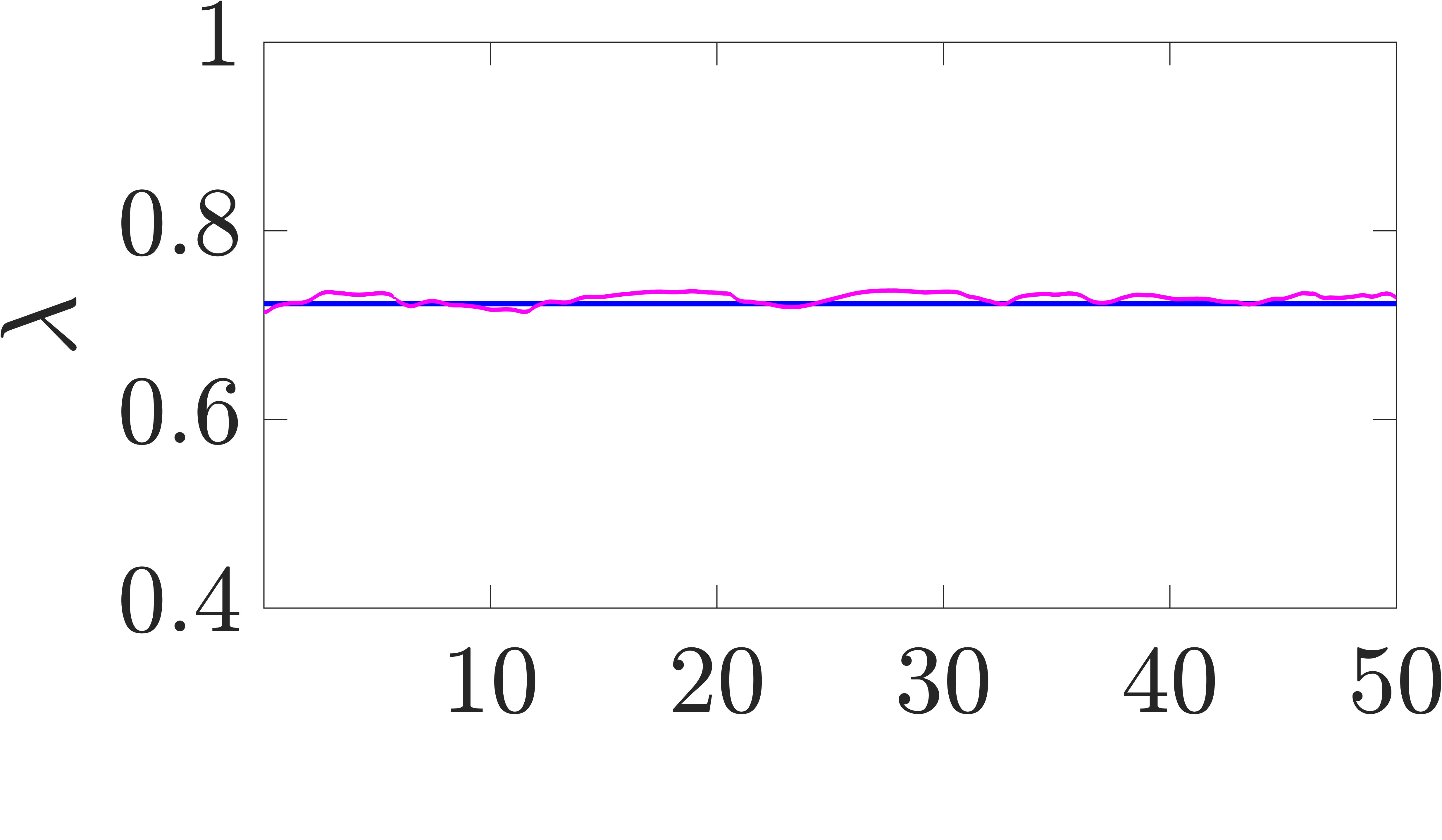} &                    
\\
\includegraphics[width=0.33\columnwidth]{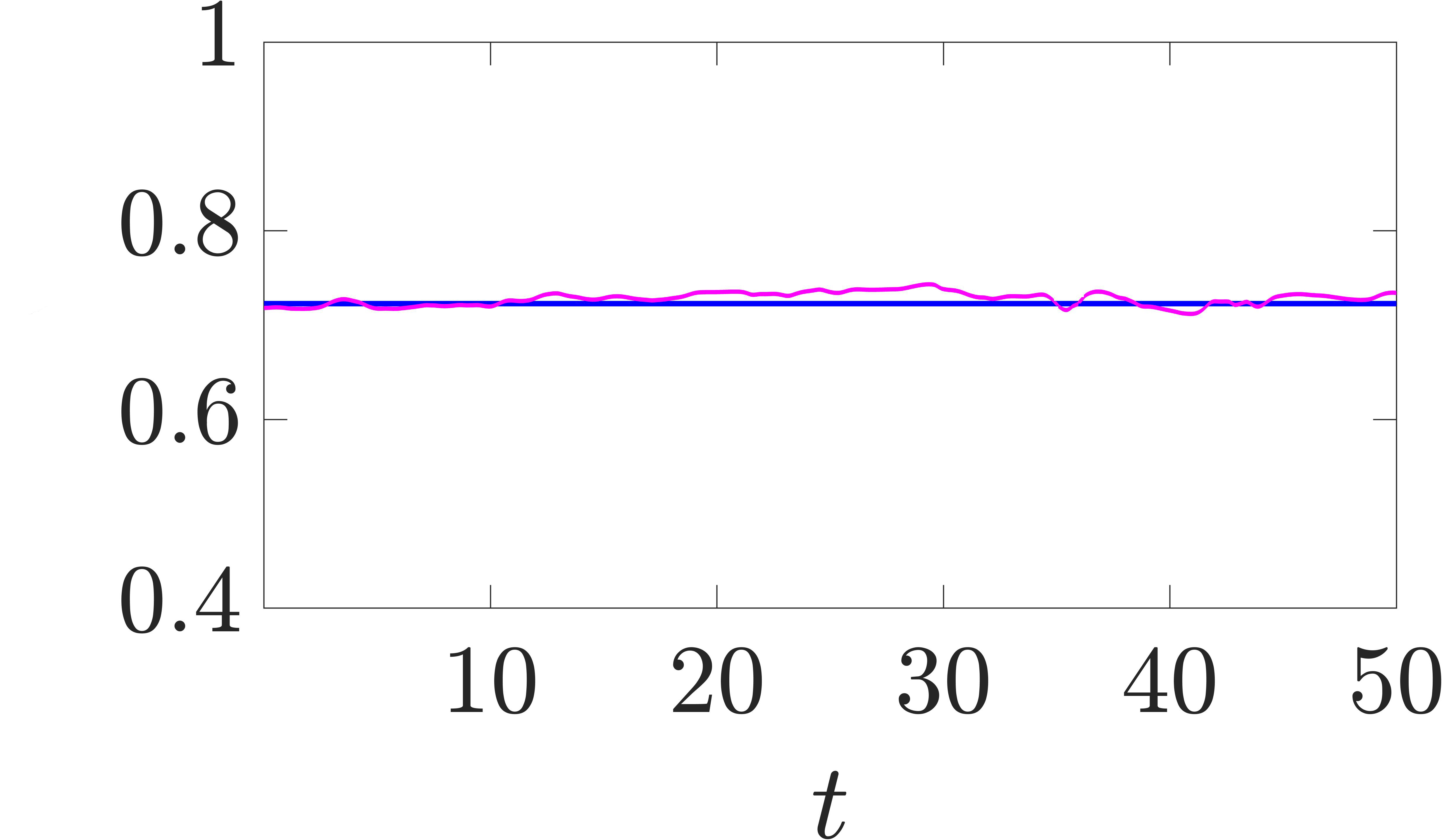} &                    
\\
(a) & \multicolumn{1}{c}{(b)}
\end{tabular}
\caption{(a) Comparison of structural parameter $\lambda(t;\mathbf{X},t_0)$ along typical particle trajectories for thixotropic flows with $\Lambda=10^{-2}$, from (\textcolor{blue}{\textbf{--$\!$--}}) DNS results and (\textcolor{blue}{\textbf{--$\!$--}}) the analytic closure (\ref{eq:lag_lambda_slow}). (b) Comparison of conditional p.d.f. of structural parameter $p_\lambda(\lambda|r)$ along typical particle trajectories for thixotropic flows with $\Lambda=10^{-2}$, from (solid lines) DNS results with (solid vertical line) its mean value, and (dotted line) the analytic closure (\ref{eq:lag_lambda_slow}).}
\label{fig:compare_slow}
\end{figure}

Figure~\ref{fig:compare_slow} verifies the slow kinetics analytic closure (\ref{eq:lag_lambda_slow}) by comparison with DNS computations for $\Lambda = 10^{-2}$. Despite that the Lagrangian framework does not strictly apply for this case, Figure~\ref{fig:compare_slow}a shows that time-series of $\lambda$ values along sample trajectories computed via DNS fluctuate close to the equilibrium value given by (\ref{eq:lag_lambda_slow}), with a relative $L_2$ error of 1.08\% across $10^4$ Lagrangian trajectories. Similarly, Figure~\ref{fig:compare_slow}b shows that the conditional p.d.f. $p_{\lambda|r}(\lambda|r)$ from the DNS computations at different radial positions $r$ all have small variance around the equilibrium value given by (\ref{eq:lag_lambda_slow}). These results show that despite breakdown of the Lagrangian approximation (\ref{eq:ADRE2}) for $\Lambda\ll 1$ due to significant diffusion of $\lambda$ (i.e. $Da=10$), the slow kinetics closure (\ref{eq:lag_lambda_slow}) is still accurate as the sampling of shear rates that governs convergence of $\dot\gamma(t;\mathbf{X},t_0)\rightarrow\langle\dot\gamma\rangle$ in (\ref{eq:lag_lambda_slow}) is not affected by the diffusivity of $\lambda$.

To test the closure (\ref{eq:lag_lambda_slow}) and the corresponding effective viscosity model (\ref{eq:eul_GN_slow}), DNS computations using (\ref{eq:eul_GN_slow}) are compared with those using the thixotropic kinetics with $\Lambda=10^{-2}$ in Figures~\ref{fig:Mprofiles_1} - \ref{fig:Mprofiles_3}. Computations based on the effective viscosity model (\ref{eq:eul_GN_slow}) show good agreement with the full thixotropic model, with relative errors of 2.24\% for the mean structural parameter $\lambda$, 0.95\% for the mean viscosity $\eta$ and 0.29\% for the axial velocity $U_z$. The slightly higher errors than for the fast kinetics closure may be partially due to breakdown of the Lagrangian approximation in the slow kinetics case. The errors in turbulent statistics ($u'_{rr},u'_{tt},u'_{zz},u'_{rz}$) are also all less than 0.7\%. These results confirm that, as expected, for $\Lambda\ll 1$ thixotropic fluids exhibits turbulent fluid dynamics akin to that of a generalised Newtonian fluids. Note that this result extends to a broad range of rheologies and thixotropic models described by (\ref{eqn:eta}), (\ref{eqn:thixo_homog}) respectively.

\section{Thixotropic Turbulence as a Non-Stationary Random Walk}
\label{sec:stochastic}

In \S\ref{sec:Lagrangian Thixotropy} we developed a Lagrangian framework for evolution of the structural parameter, and used this framework to establish convergence of the thixotropic rheology to the shear thinning (\ref{eq:eul_GN_fast}) and Newtonian (\ref{eq:eul_GN_slow}) models respectively in the limits of fast ($\Lambda\gg 1$) and slow ($\Lambda\ll 1$) thixotropic kinetics. In this section we extend the Lagrangian framework to the case of intermediate thixotropic kinetics ($\Lambda\sim 1$) via development of a stochastic model for the Lagrangian shear history experienced by fluid elements. This stochastic model yields a rheological model for the effective viscosity that is accurate for arbitrary thixoviscous numbers $\Lambda\in[0,\infty^+)$. 

\subsection{Stochastic Thixotropy as a Path Integral}
\label{subsec:path}

As discussed in \S\ref{sec:Governing Equations and Numerical Method}, fully-developed turbulent pipe flow is radially non-stationary, hence the random process governing the Lagrangian shear rate history $\dot\gamma(t-s;\mathbf{X},t_0)$ is also radially non-stationary. Specifically, the microstructural evolution equation (\ref{eq:memory2}) may be expressed as
\begin{equation}\label{eq:memoryr}
	\begin{split}
\lambda(t,r(t))&=\frac{\Lambda\int_0^\infty G(t-s,r(t-s))ds}{G(t,r(t))},\\
	G(t,r(t))&=\exp\left[\Lambda\int_0^\infty 1+K\dot\gamma(t-s,r(t-s))\,ds\right],
	\end{split}
\end{equation}
where the Lagrangian shear rate $\dot\gamma(t,r(t))$ is considered as a random variable that is non-stationary with respect to $r$ with conditional p.d.f. $p_{\dot\gamma|r}(\dot\gamma | r)$. Hence we require coupled stochastic models for both the radial position $r(t)$ and for the Lagrangian shear rate $\dot\gamma(t,r(t))$ which is conditional on $r(t)$. From (\ref{eq:memoryr}), the dependence of the structural parameter $\lambda$ at time $t$ and position $r(t)$ on the Lagrangian shear history can be encoded as
\begin{equation}
	\lambda(t,r(t))=\hat{\mathcal{F}}\left[\dot\gamma(t-s,r(t-s))\big|_{s=0}^{s\rightarrow\infty}\right],\label{eqn:Fr}
\end{equation}
and so the evolution of $\lambda$ follows a non-Markov process, while the evolution of $\dot\gamma(t,r(t))$ and $r(t)$ may be Markovian in an appropriate frame. Following the GN models developed in \S\ref{sec:Lagrangian Thixotropy}, we propose a simple viscosity model for arbitrary $\Lambda$ where, based on the radial non-stationarity of the flow, the local viscosity at any radial position $r$ is based on the conditionally averaged structural parameter $\langle\lambda | r\rangle$ as
\begin{equation}
\eta_\text{eff}(r,\dot\gamma)\equiv\eta(\langle\lambda | r\rangle,\dot\gamma),\label{eqn:eta_r}
\end{equation}
where 
\begin{equation}
\langle\lambda|r\rangle\equiv\int \lambda\,p_{\lambda|r}(\lambda|r)\,d\lambda,\label{eqn:lambda_r}
\end{equation}
and $p_{\lambda|r}(\lambda|r)$ is the conditional probability of $\lambda$ occurring at radial position $r$. This viscosity model (\ref{eqn:eta_r}) effectively assumes that the variation in $\lambda$ about its conditionally average has minimal impact on the flow, which shall be tested in due course. Discretizing the $s$-domain as $s_n=n\Delta s$, with $\Delta s\ll \tau_{\dot\gamma}$, the functional (\ref{eqn:Fr}) can be expressed as
\begin{equation}
\lambda(t,r(t))=\mathcal{F}[\dot{\boldsymbol\gamma}],
\end{equation}
where $\dot{\boldsymbol\gamma}$ is the vector $\dot{\boldsymbol\gamma}\equiv(\dot\gamma_0, \dot\gamma_1,\dots,\dot\gamma_q)$ of Lagrangian shear rates $\dot\gamma_n\equiv\dot\gamma(t-n\Delta s;\mathbf{X},t_0)$. As $G(t-s)\sim\exp(-\Lambda s)$, if $q\gg 1/(\Delta s\Lambda)$ then the shear history for $s>q\Delta s$ has no impact on $\lambda(t,r(t))$. Similarly, the discrete Lagrangian radial history can be encoded as $\mathbf{r}=\{r_0,r_1,\dots,r_q\}$, where $r_n\equiv r(t-n\Delta s)$ for $n=0:q$.

From (\ref{eqn:Fr}), the conditional average (\ref{eqn:lambda_r}) can be conceptualised as a \emph{path integral}~\citep{Kleinert2006,Wio2013} of $\mathcal{F}$ over all the shear $\dot{\boldsymbol\gamma}$ and radial $\mathbf{r}$ histories backwards in time from a particle at position $r(t)$ at current time $t$ to $r(t-q\Delta s)$ at time $t-q\Delta s$ as
\begin{equation}
\langle\lambda|r\rangle=\int_\mathbf{r}\int_{\dot{\boldsymbol\gamma}}\mathcal{F}[\dot{\boldsymbol\gamma}]\mathcal{P}[\dot{\boldsymbol\gamma},\mathbf{r}|r_0=r]d^q\dot{\boldsymbol\gamma}d^q\mathbf{r},\label{eqn:path}
\end{equation}
where $\mathcal{P}[\dot{\boldsymbol\gamma},\mathbf{r}|r_0=r]$ is the conditional probability of the shear history $\dot{\boldsymbol\gamma}$ and radial history $\mathbf{r}$ given current radial position $r$.  This general formalism describes evolution of the structural parameter over a broad range of thixotropic and rheological models.

Via Bayes' theorem, the conditional probability $\mathcal{P}[\dot{\boldsymbol\gamma},\mathbf{r}|r_0=r]$ is related to the joint probability $\mathcal{P}[\dot{\boldsymbol\gamma},\mathbf{r}]$ of shear rate history $\dot{\boldsymbol\gamma}$ and radial history as
\begin{equation}
\mathcal{P}[\dot{\boldsymbol\gamma},\mathbf{r}]=\mathcal{P}[\dot{\boldsymbol\gamma}|r]\mathcal{P}[\mathbf{r}|r_0=r],
\end{equation}
where $\mathcal{P}[\mathbf{r}|r_0=r]$ is the conditional probability of $\mathbf{r}$ given $r_0=r$. Under the assumption that the shear rate and radial position evolve in a Markovian manner in time, the joint probability $\mathcal{P}[\dot{\boldsymbol\gamma},\mathbf{r}]$ can then be expanded as
\begin{equation}
\begin{split}
\mathcal{P}[\dot{\boldsymbol\gamma},\mathbf{r}]=&\mathcal{P}[\{\dot\gamma_0,r_0\},\{\dot\gamma_1,r_1\},\dots,\{\dot\gamma_q,r_q\}],\\
=&P_{\Delta s}(\dot\gamma_{q},r_{q}|\dot\gamma_{q-1},r_{q-1})\dots P_{\Delta s}(\dot\gamma_2,r_2|\dot\gamma_1,r_1)P_{\Delta s}(\dot\gamma_1,r_1|\dot\gamma_0,r_0)p_{\dot\gamma,r}(\dot\gamma_0,r_0),
\end{split}
\end{equation}
where the \emph{backwards propagator} $P_{\Delta s}(\dot\gamma_{n+1},r_{n+1} | \dot\gamma_{n},r_{n})$ is the conditional probability of a material element being at position $r_{n+1}$ with shear rate $\dot\gamma_{n+1}$ at time $t-(n+1)\Delta s$ given it is at position $r_{n}$ with shear rate $\dot\gamma_{n}$ at time $t-n\Delta s$. Using the Chapman-Kolmogorov equation for Markov processes
\begin{equation}
P_{2\Delta s}(\dot\gamma_n,r_n|\dot\gamma_{n+2},r_{n+2})=\int_{-\infty}^\infty\int_{-\infty}^\infty P(\dot\gamma_n,r_n|\dot\gamma_{n+2},r_{n+2})P(\dot\gamma_n,r_n|\dot\gamma_{n+2},r_{n+2})d\dot\gamma_{n+1}dr_{n+1},\label{eqn:Chap_Kom}
\end{equation}
 the path integral (\ref{eqn:path}) can then be expressed explicitly as
\begin{equation}
\begin{split}
\langle\lambda | r\rangle=&\int_{-\infty}^\infty \dots \int_{-\infty}^\infty 
d\dot{\gamma}_0\dots d\dot{\gamma}_q dr_1\dots dr_q\\
&\mathcal{F}[\dot{\boldsymbol\gamma}]
P_{\Delta s}(\dot\gamma_{q},r_{q}|\gamma_{q-1},r_{q-1})\dots P_{\Delta s}(\dot\gamma_1,r_1 | \dot\gamma_0,r)p_{\dot\gamma|r}(\dot\gamma_0|r),\label{eqn:path2}
\end{split}
\end{equation}
where $p_{\dot\gamma|r}(\dot\gamma|r)$ is the conditional probability of $\dot\gamma$ given $r$. Hence the conditionally averaged structural parameter $\langle\lambda | r\rangle$ is dependent upon the thixotropic functional $\mathcal{F}$ and the stochastic process for $\dot\gamma(t,r(t))$ and $r(t)$ via the backwards propagator $P_{\Delta s}$. Typically, path integrals such as (\ref{eqn:path2}) are notoriously difficult to solve for all but simple linear systems~\citep{Kleinert2006}, however, in \S\S\ref{subsec:struct} we show that significant simplifications to (\ref{eqn:path}) arise for the simple stochastic models for shear and radial history that are developed as follows.

\subsection{Stochastic Models for Radial Position and Shear Rate History}
\label{subsec:stochastic_model}

In order to solve the path integral (\ref{eqn:path2}), we require stochastic models for Lagrangian shear rate and radial position. Although there exist more sophisticated models for radial transport in turbulent pipe flow~\citep{Bocksell2006,Mofakham2019}, we seek a simple model that yields analytic closure of (\ref{eqn:path2}). Following~\cite{Dentz2015}, a suitable model for $r(t)$ is developed as follows. In cylindrical coordinates, the advection-dispersion equation for the concentration $c(r,t)$ of an ensemble of passive tracer particles is given by the advection-dispersion equation (ADE)
\begin{equation}
    \frac{\partial c(r,t)}{\partial t}
    +\frac{1}{r}\frac{\partial}{\partial r}\left[r \bar{v}_r(r) c(r,t)\right]
    -\frac{1}{r}\frac{\partial}{\partial r}\left(r D_r(r)\frac{\partial c}{\partial r}\right)=0,\quad\frac{\partial c}{\partial r}\Big|_{r=0}=\frac{\partial c}{\partial r}\Big|_{r=1}=0,\label{eqn:ADE}
\end{equation}
where $\bar{v}_r(r)$ and $D_r(r)$ respectively are the Lagrangian radial mean transport velocity and dispersion coefficient. Here $\bar{v}_r(r)\equiv\langle v_r(\mathbf{x}(t;\mathbf{X},t_0),t)\rangle=0$ and $D_r(r)$ is related to the Lagrangian radial velocity fluctuations $v'_r(\mathbf{x}(t;\mathbf{X},t_0),t)\equiv v_r(\mathbf{x}(t;\mathbf{X},t_0),t)-\bar{v}_r(r)$ via the fluctuation-dissipation theorem as
\begin{equation}
    D_r(r)=\int_0^\infty \langle v_r^\prime(\mathbf{x}(t;\mathbf{X},t_0),t)v_r^\prime(\mathbf{x}(t+\tau;\mathbf{X},t_0),t+\tau)\rangle d\tau,\label{eqn:Dr}
\end{equation}
where the angled brackets denotes an average over all times $t$, azimuthal $\theta$ and axial $z$ coordinates. For any well-behaved $D_r(r)$, in the long-time limit $t\rightarrow\infty$ the particle concentration approaches the homogeneous state $c(r,t)\rightarrow c_0$. The particle concentration $c(r,t)$ is related to the probability $p_r(r,t)$ of finding a tracer particle at position $r$ at time $t$ as
\begin{equation}
    p_r(r,t)=2\pi r c(r,t),
\end{equation}
and so this probability is governed by the radial Fokker-Planck equation
\begin{equation}
\begin{split}
    \frac{\partial p_r(r,t)}{\partial t}+\frac{\partial}{\partial r}\left[\hat{v}_r(r)p_r(r,t)\right]-\frac{\partial^2}{\partial r^2}\left[D_r(r)p_r(r,t)\right]=0,\label{eqn:Fokker}
    \end{split}
\end{equation}
with insulating boundary conditions
\begin{equation}
\frac{\partial p_r}{\partial r}\Big|_{r=0}=\frac{\partial p_r}{\partial r}\Big|_{r=1/2}=0,\label{eqn:FPbcs}
\end{equation}
and
\begin{equation}
\hat{v}_r(r)\equiv\frac{D_r(r)}{r}+\frac{dD_r}{dr},
\end{equation}
is the effective radial velocity due to the radial dispersivity $D_r(r)$. In the long-time limit $t\rightarrow\infty$, the particle probability in (\ref{eqn:Fokker}) approaches the equilibrium distribution $p_r(r)=8r$. Following the It\^{o} interpretation of a stochastic process, the equivalent Langevin equation for the radial position $r(t)$ of a single tracer particle with $p_r(r,t)=\langle\delta(r-r(t))\rangle$ is then
\begin{equation}
    \frac{dr(t)}{dt}=\hat{v}_r[r(t)]+\sqrt{2D(r(t))}\xi_r(t),\quad r(t)\in[0,1/2],\label{eqn:Langevin}
\end{equation}
where $\xi_r(t)$ is a Gaussian white noise with zero mean and unit variance which is delta correlated in time; $\langle\xi_r(t)\xi_r(t^\prime)\rangle=\delta(t-t^\prime)$.

The above formulation makes several assumptions regarding the radial position and radial velocity processes. First, it is assumed that the Lagrangian radial velocity fluctuation $v_r^\prime(\mathbf{x}(t;\mathbf{X},t_0),t)$ is a stochastic Markov process in that it decorrelates on a timescale $\tau_v$ that is much faster than the radial position decorrelation time $\tau_r$, i.e. $\tau_v\ll\tau_r$, which validates the assumption of delta-correlated Gaussian noise $\xi_r(t)$ in (\ref{eqn:Langevin}). The other assumption is that the decorrelation processes for radial velocity and position are radially-independent. These assumptions are tested as follows.

\begin{figure}
\centering
\begin{tabular}{cc}
\includegraphics[width=0.41\columnwidth]{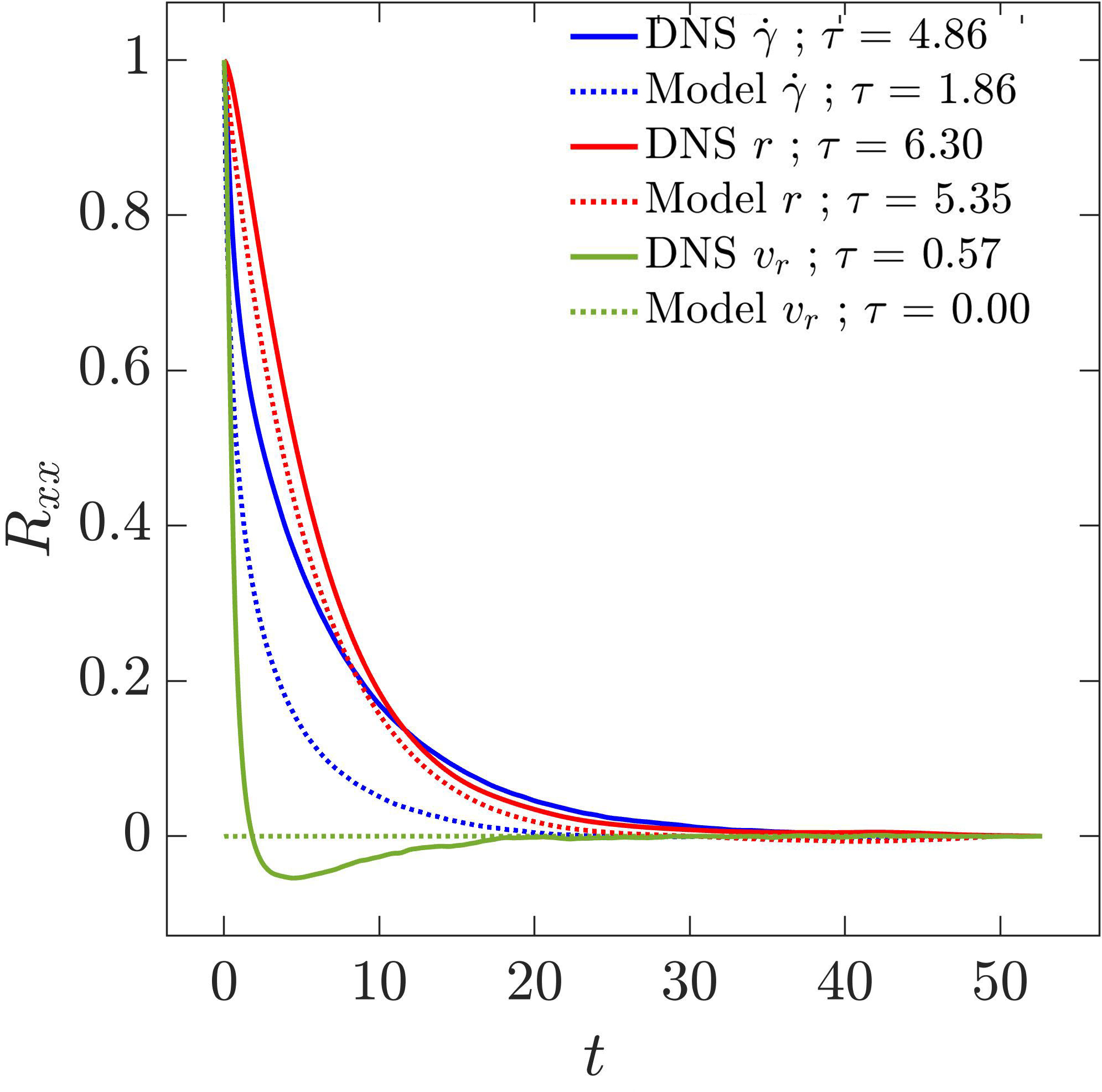}&
\includegraphics[width=0.41\columnwidth]{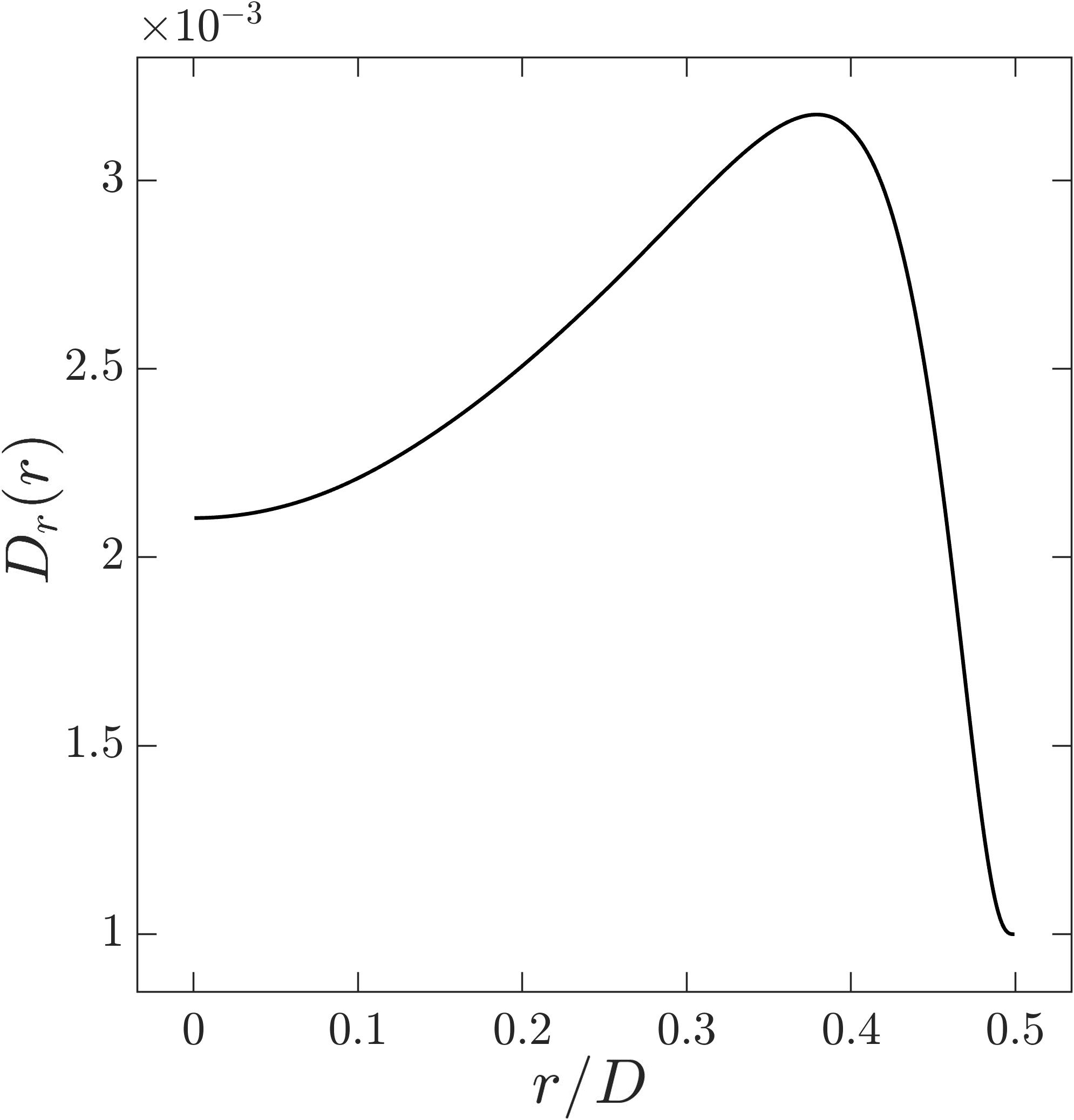}\\
(a) & (b)
\end{tabular}
\caption{(a) Comparison of Lagrangian autocorrelation functions $R_{xx}$ for radial velocity fluctuation $x=v_r'$, radial position $x=r'$ and shear rate $x=\dot\gamma'$ for thixotropic flows with $\Lambda=1$, from (solid lines) DNS results and (dotted lines) the stochastic model (\ref{eqn:Langevin}).(b) Radial dispersivity $D_r(r)$ from the DNS computations of the case $\Lambda=1$.}
\label{fig:autocorr}
\end{figure}

\begin{figure}
\centering
\begin{tabular}{cc}
\includegraphics[width=0.46\columnwidth]{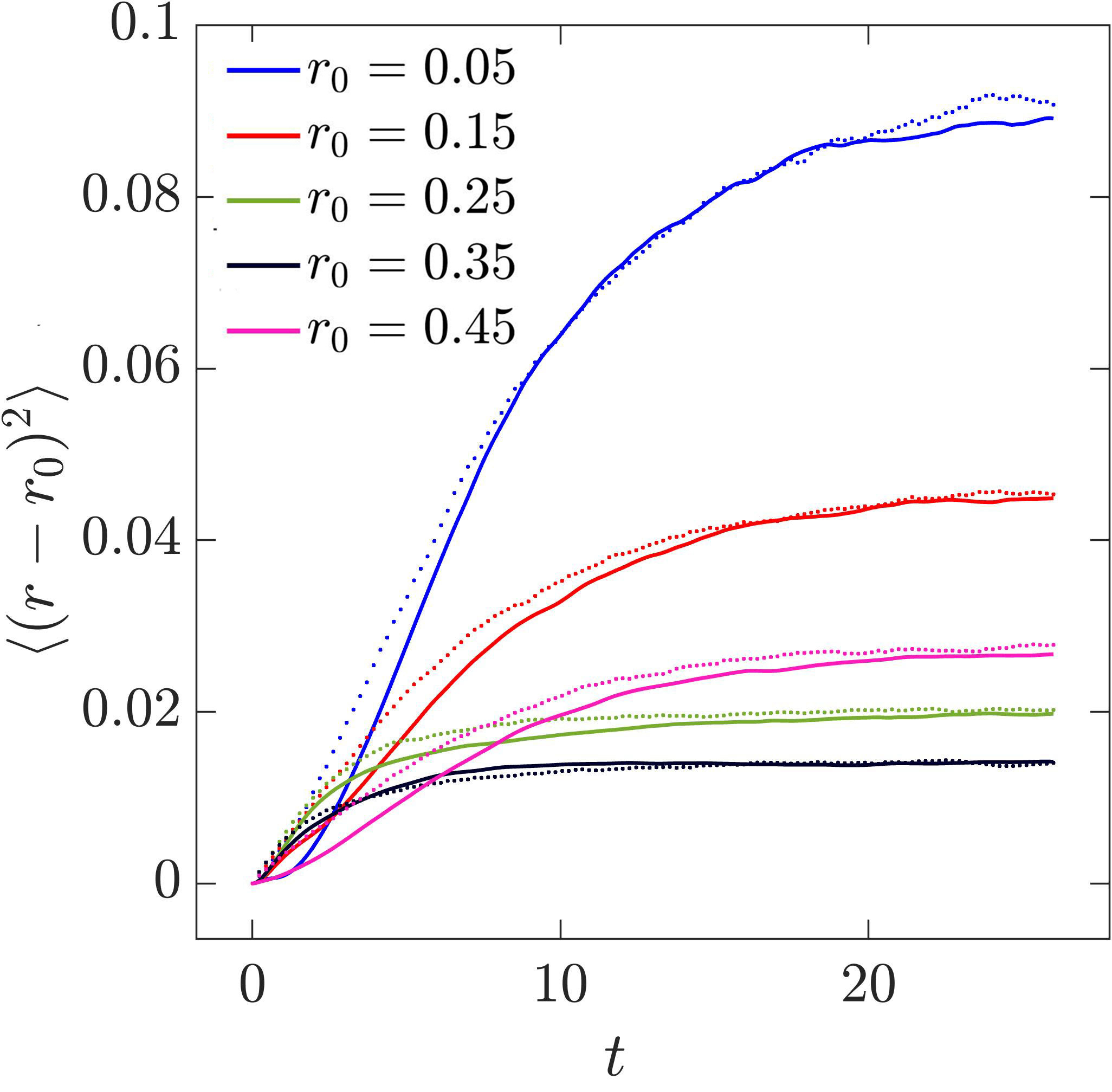}&
\includegraphics[width=0.46\columnwidth]{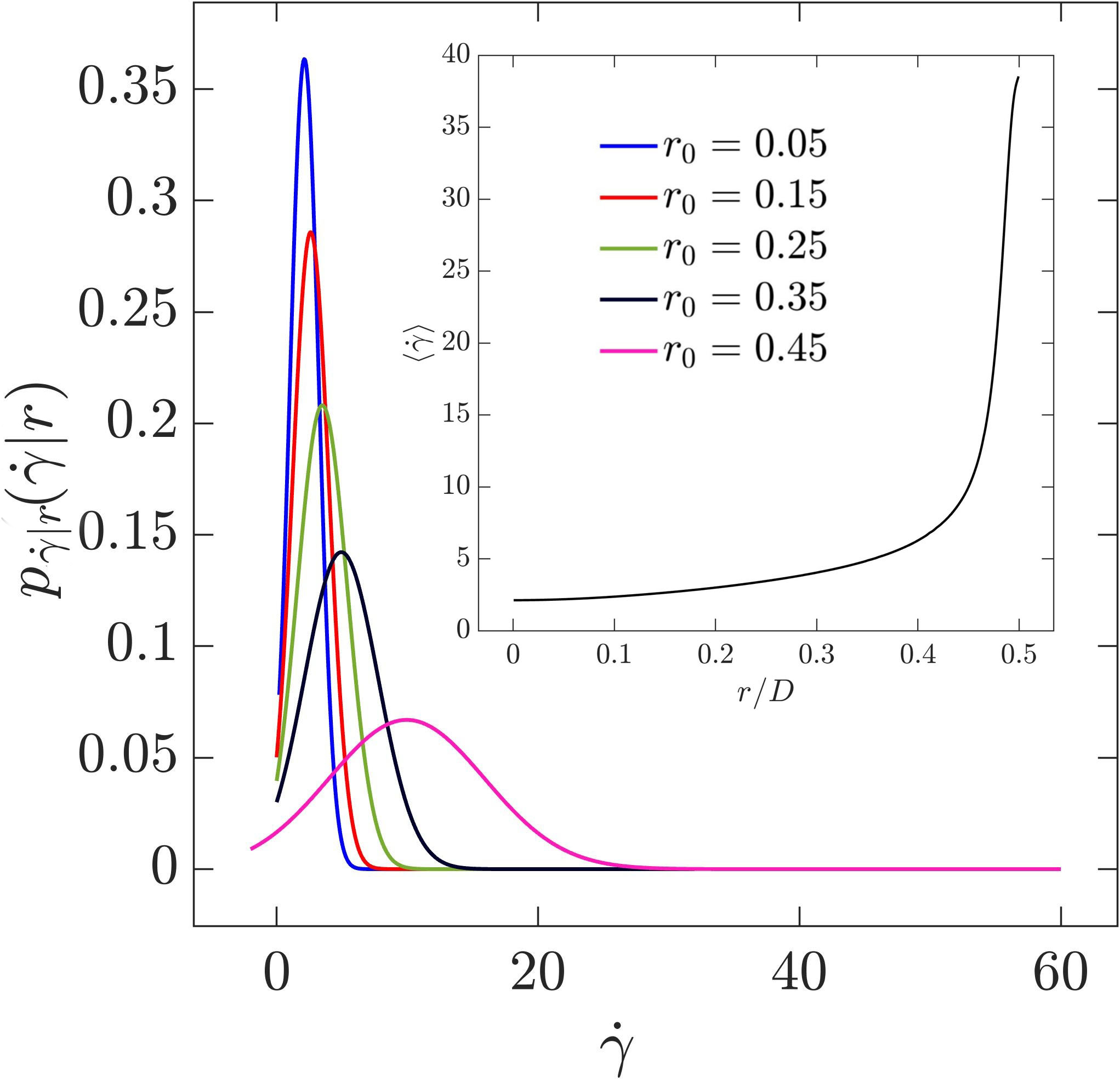}\\
(a) & (b)
\end{tabular}
\caption{(a) Comparison of mean square displacement $\langle (r-r_0)^2\rangle$ along typical particle trajectories for thixotropic flows with $\Lambda=1$, from (solid lines) DNS results and (dotted lines) the stochastic model (\ref{eqn:Langevin}).(b) Conditional p.d.f. of shear rate at various radial locations $p_{\dot\gamma|r}(\dot\gamma|r)$ from the DNS computations of the case $\Lambda=1$. Inset: distribution of conditionally averaged shear rate $\langle\dot\gamma|r\rangle$ from the DNS computations of $\Lambda=1$.}
\label{fig:radial_ac}
\end{figure}

Figure~\ref{fig:autocorr}a shows the normalised Lagrangian autocorrelation functions for the radial velocity ($R_{v_r'v_r'}$), radial position ($R_{r'r'}$) and shear rate ($R_{\dot\gamma'\dot\gamma'}$) fluctuations from the DNS simulations for $\Lambda=1$ as
\begin{align}
    R_{xx}(\tau)&\equiv\frac{1}{\sigma^2_{x}}\langle x(\mathbf{x}(t;\mathbf{X},t_0),t)x(\mathbf{x}(t+\tau;\mathbf{X},t_0),t+\tau)\rangle,
\end{align}
with $x=v_r'$, $x=r'$, $x=\dot\gamma'$. From this data, the decorrelation times for the radial velocity, radial position and shear rate are computed as $\tau_v\approx 0.57$, $\tau_r\approx 6.30$, $\tau_{\dot\gamma}\approx 4.86$. Hence the assumption that $\tau_v\ll\tau_r$ is approximately true but not strictly valid. Under the assumption that the radial velocity fluctuation follows a Markov process with the exponentially decaying autocorrelation function
\begin{equation}
    R_{v_r^\prime v_r^\prime}(\tau)=\exp(-\tau/\tau_v),
\end{equation}
the radial dispersivity $D_r(r)$ in (\ref{eqn:Dr}) is
\begin{equation}
    D_r(r)=\tau_v\sigma^2_{v_r}(r),\label{eqn:Dr2}
\end{equation}
where $\sigma^2_{v_r}(r)$ is the conditional variance $\sigma^2_{v_r}(r)=\langle v_r^\prime(\mathbf{x}|_{r},t)v_r^\prime(\mathbf{x}|_{r},t)\rangle$. To compare results of the stochastic model with the DNS solutions (which include finite diffusivity of $\lambda$ for numerical stability), the scalar diffusivity $1/\Pen=10^{-3}$ is also added to (\ref{eqn:Dr2}). However we note for many complex fluids such as suspensions and emulsions, $1/\Pen$ is negligibly small and so can be neglected. The resultant radial dispersivity is shown in Figure~\ref{fig:autocorr}b, indicating that as expected, dispersion is moderate in the pipe core, reaches a maximum in the buffer layer before decaying to the scalar diffusivity $\alpha$ at the pipe wall. Application of this radial dispersivity to the random walk model (\ref{eqn:Langevin}) yields the mean square displacement data shown in Figure~\ref{fig:radial_ac}a, which agrees well with the DNS results for $\Lambda=1$, and provides a close match to the radial position decorrelation curve shown in Figure~\ref{fig:autocorr}a.

As shown in Figure~\ref{fig:autocorr}a, the Lagrangian shear rate and radial position fluctuations appear to decorrelate on similar timescales $\tau_{\dot\gamma}\sim\tau_r$, suggesting that shear rate fluctuations are dominated by fluctuations in radial position rather than temporal fluctuations at fixed radial position. As such, we propose a very simple deterministic relationship between the Lagrangian shear rate $\dot\gamma(t,r(t))$ and radial position based on the conditional mean as
\begin{equation}
\dot\gamma(t,r(t))\equiv\dot\gamma(t,\mathbf{X};t_0)\approx\langle \dot\gamma|r(t)\rangle,\label{eqn:shear_model}
\end{equation}
hence all fluctuations in Lagrangian shear rate are driven by fluctuations in radial position. The conditional shear rate p.d.f. is shown in Figure~\ref{fig:radial_ac}b, indicating that the shear rate is fairly peaked in the pipe bulk but broadens near the wall, and the inset of this figure shows that the average shear rate monotonically decreases from the pipe bulk to the wall. Figure~\ref{fig:autocorr}a indicates that this stochastic model yields a significantly faster decorrelation rate ($\tau_{\dot\gamma}=1.86$) than that observed in the DNS simulations ($\tau_{\dot\gamma}=4.86$), which may be attributed to approximating the p.d.f.s in Figure~\ref{fig:radial_ac}b with delta functions. Although approximate, this stochastic model for Lagrangian shear rate history closes the path integral (\ref{eqn:path2}).

From (\ref{eqn:path2}), the conditional average $\langle\lambda|r\rangle$ involves an ensemble average over all trajectories that arrive at position $r$ at current time $t$. Hence we consider the adjoint problem that describes the evolution of Lagrangian radial position or tracer particles that are currently at position $r$ at time $t$ backwards in time. If we write the radial position probability as the probability of tracer particle being at position $r$ at time $t$ given it as originally at position $r_0$ at time $t-s<t$ as
$p_r(r,t)=p_r(r,t|r_s,t-s)$, then the backward Fokker-Planck equation is given by the adjoint of (\ref{eqn:Fokker}) as
\begin{equation}
    \frac{\partial p_r(r,t|r_s,t-s)}{\partial s}-\hat{v}_r(r_s)\frac{\partial}{\partial r_s}p_r(r,t|r_s,t-s)-D_r(r_s)\frac{\partial^2}{\partial r_s^2}p_r(r,t|r_s,t-s)=0,\label{eqn:Fokker_back}
\end{equation}
with insulating boundary conditions (\ref{eqn:FPbcs}) and initial condition $p_r(r,t|r_0,t)=\delta(r_0-r)$ at $s=0$. Using (\ref{eqn:shear_model}) and (\ref{eqn:Fokker_back}), the path integral (\ref{eqn:path2}) then simplifies to
\begin{equation}
\langle\lambda | r\rangle=\int_{-\infty}^\infty \dots \int_{-\infty}^\infty dr_1\dots dr_q\mathcal{F}[\langle\dot\gamma|\mathbf{r}\rangle]
P_{\Delta s}(r_q|r_{q-1})\dots P_{\Delta s}(r_1|,r),\label{eqn:path2_simple}
\end{equation}
where the backwards propagator for $r$ is given as $P_{\Delta s}(r_{n+1}|r_n)\equiv p_r(r,t|r_{\Delta s},t-\Delta s)$, which is a Green's function for the backward Fokker-Planck equation.

\subsection{Structural Parameter}
\label{subsec:struct}

An expression for the structural parameter is given by discretising (\ref{eq:memoryr}) with respect to $s$ as
\begin{equation}
    \lambda(t,r(t))=\mathcal{F}[\dot{\boldsymbol\gamma}]=\frac{\Lambda\,\Delta s\sum_{n=1}^\infty G_n}{G_0},
\end{equation}
where $G_n\equiv G(t-n\Delta s;\mathbf{X},t_0)$ is given as
\begin{equation}
    G_n=\exp\left(\Lambda\,\Delta s\sum_{i=n}^\infty g_i\right)=\prod_{i=n}^\infty h_i,
\end{equation}
where $h_n\equiv\exp[\Lambda\,\Delta s(1+K\dot\gamma_n)]>1$. From these relations, the structural parameter is then
\begin{equation}
\begin{split}
    \lambda(t,r(t))=\mathcal{F}[\dot{\boldsymbol\gamma}]=\Lambda\,\Delta s\sum_{n=1}^\infty\frac{1}{\prod_{i=0}^n h_i}=&\Lambda\,\Delta s\left(\frac{1}{h_0}+\frac{1}{h_0h_1}+\frac{1}{h_0h_1h_2}+\dots\right).
    \end{split}\label{eqn:lambdan}
\end{equation}
Hence the functional $\mathcal{F}[\dot{\boldsymbol\gamma}]$ that governs the structural parameter $\lambda$ depends upon the shear rate history, and convergence of the sum in (\ref{eqn:lambdan}) is governed by the thixoviscous number $\Lambda$. Averaging  of (\ref{eqn:lambdan}) yields
\begin{equation}
\begin{split}
    \langle\lambda|r\rangle=&\Lambda\,\Delta s\left(\left\langle\frac{1}{h_0}\right\rangle+\left\langle\frac{1}{h_0h_1}\right\rangle+\left\langle\frac{1}{h_0h_1h_2}\right\rangle+\dots\right),
    \end{split}\label{eqn:lambda_av_gen}
\end{equation}
where evaluation of the ensemble averaged terms $\langle\cdot\rangle$ corresponds to solution of the path integral (\ref{eqn:path2}). Note that this expression is completely general and not contingent upon the stochastic model developed in \S\S\ref{subsec:stochastic_model}.

For this stochastic model, the Lagrangian shear rate $\dot\gamma(t,r(t))$ is given by the conditional average $\langle\dot\gamma|r(t)\rangle$, hence insertion of (\ref{eqn:lambdan}) into (\ref{eqn:path2_simple}) yields 
\begin{equation}
\begin{split}
    \langle\lambda|r\rangle=&\frac{\Lambda\,\Delta s}{h(r)} \left[1+\left\langle \frac{1}{h(r_1)}\right\rangle+\left\langle\frac{1}{h(r_1)h(r_2)}\right\rangle+\dots\right],
    \end{split}\label{eqn:lambda_av}
\end{equation}
where 
\begin{equation}
h(r)\equiv\exp\left(\Lambda\,\Delta s[1+K\langle\dot\gamma|r\rangle]\right).
\end{equation}
The averages in (\ref{eqn:lambda_av}) are then given by the backwards Fokker-Planck propagator
\begin{align}
f_1(r)&\equiv\left\langle \frac{1}{h(r_1)}\right\rangle=\int \frac{1}{h(r_{\Delta s})}P_{\Delta s}(r_{\Delta s}|r)dr_{\Delta s},\\
f_n(r)&\equiv\left\langle \frac{1}{\prod_{i=1}^n h(r_i)}\right\rangle=\int \frac{1}{h(r_{\Delta s})}P_{\Delta s}(r_{\Delta s}|r)f_{n-1}(r_{\Delta s})dr_{\Delta s},
\end{align}
and so
\begin{equation}
\begin{split}
    \langle\lambda|r\rangle=&\frac{\Lambda\,\Delta s}{h(r)} \left[1+\sum_{n=1}^\infty f_n(r)\right].
    \end{split}\label{eqn:lambda_av2}
\end{equation}
Note that the length of the relevant shear history depends strongly upon the thixoviscous parameter $\Lambda$.\\

In the limit of fast kinetics with $\Lambda\gg 1$, we set $\Delta s\lll 1$ very small such that $\Lambda\,\Delta s\ll 1$ and so for $s=n\Delta s$, $n=0:q$ the radial position $r(t-s)$ is still centred about $r(t)$, hence
\begin{align}
f_n(r)=\int \frac{1}{h(r_{\Delta s})}\delta(r-r_{\Delta s})f_{n-1}(r_{\Delta s})dr_{\Delta s}=\frac{1}{h(r)}f_{n-1}(r_{\Delta s})=\frac{1}{h(r)^n},
\end{align}
and so we recover the conditionally averaged shear rate analogue to the shear thinning viscosity (\ref{eq:eul_GN_fast}) as
\begin{equation}
    \lim_{\Lambda\rightarrow\infty}\langle\lambda|r\rangle= \frac{\Lambda\,\Delta s}{1-h(r)}=\frac{1}{1+K\langle\dot\gamma|r\rangle}.
\end{equation}
Note that if we relax the assumption (\ref{eqn:shear_model}), a similar derivation based on (\ref{eqn:lambda_av_gen}) exactly recovers the shear thinning viscosity (\ref{eq:eul_GN_fast}). In the slow kinetics regime, the terms in (\ref{eqn:lambda_av2}) converge very slowly as $\Lambda\ll 1$ and the conditional probability converges to
\begin{equation}
\lim_{s\rightarrow\infty}p_r(r,t|r_s,t-s)=p_{r}(r)=8r,
\end{equation}
then for $\Lambda\,\Delta s\ll1$, the ensemble averages are
\begin{equation}
f_n=\int\frac{1}{h(r_{\Delta s})}p_{r}(r_{\Delta s})\, f_{n-1}\,dr_{\Delta s}=[1-\Lambda\,\Delta s(1+K\langle\dot\gamma\rangle)]f_{n-1},
\end{equation}
and so
\begin{equation}
\sum_{n=0}^\infty f_n=\sum_{n=0}^\infty\left[1-\Lambda\,\Delta s(1+K\langle\dot\gamma\rangle)\right]^n=\frac{1}{\Lambda\,\Delta s(1+K\langle\dot\gamma\rangle)},
\end{equation}
and so we recover the Newtonian viscosity (\ref{eq:eul_GN_slow}) as
\begin{equation}
    \lim_{\Lambda\rightarrow 0}\langle\lambda|r\rangle=\frac{1}{1+K\langle\dot\gamma\rangle}.
\end{equation}
Hence the stochastic model for the structural parameter can be applied to the entire range of thixoviscous numbers $\Lambda\in[0,\infty^+)$, and recovers the limiting viscosity models for fast and slow kinetics derived in \S\ref{sec:Lagrangian Thixotropy}. As such, the effective viscosity for the entire spectrum of the thixotropic kinetic parameter $\Lambda\in[0,\infty^+)$ is given by the radially-dependent viscosity
\begin{equation}
    \eta_{\text{eff}}(r,\dot\gamma)=\eta(\langle\lambda|r\rangle,\dot\gamma)=\mu_\infty+(\mu_0-\mu_\infty)\langle\lambda|r\rangle,\label{eqn:visc_eff}
\end{equation}
the accuracy of which shall be tested as follows.

\subsection{Numerical Testing of Stochastic Model}
\label{subsec:numerical_stoch_model}

To test the stochastic model developed in the previous section, we compare model predictions for the structural parameter $\langle\lambda|r\rangle$ with DNS computations of $\Lambda=1$. As the expression (\ref{eqn:lambda_av2}) does not yield closed-form solutions for the averaged structural parameter, we compute $\langle\lambda|r\rangle$ via random walk simulations of the Langevin equation (\ref{eqn:Langevin}), combined with the path integral (\ref{eqn:lambda_av2}). We then test the effective viscosity model $\eta_{\text{eff}}(r,\dot\gamma)$ by comparing DNS simulations using this model with the full thixotropic model described in \S\ref{sec:Governing Equations and Numerical Method}. 

\begin{figure}
\centering
\begin{tabular}{cc}
\includegraphics[width=0.33\columnwidth]{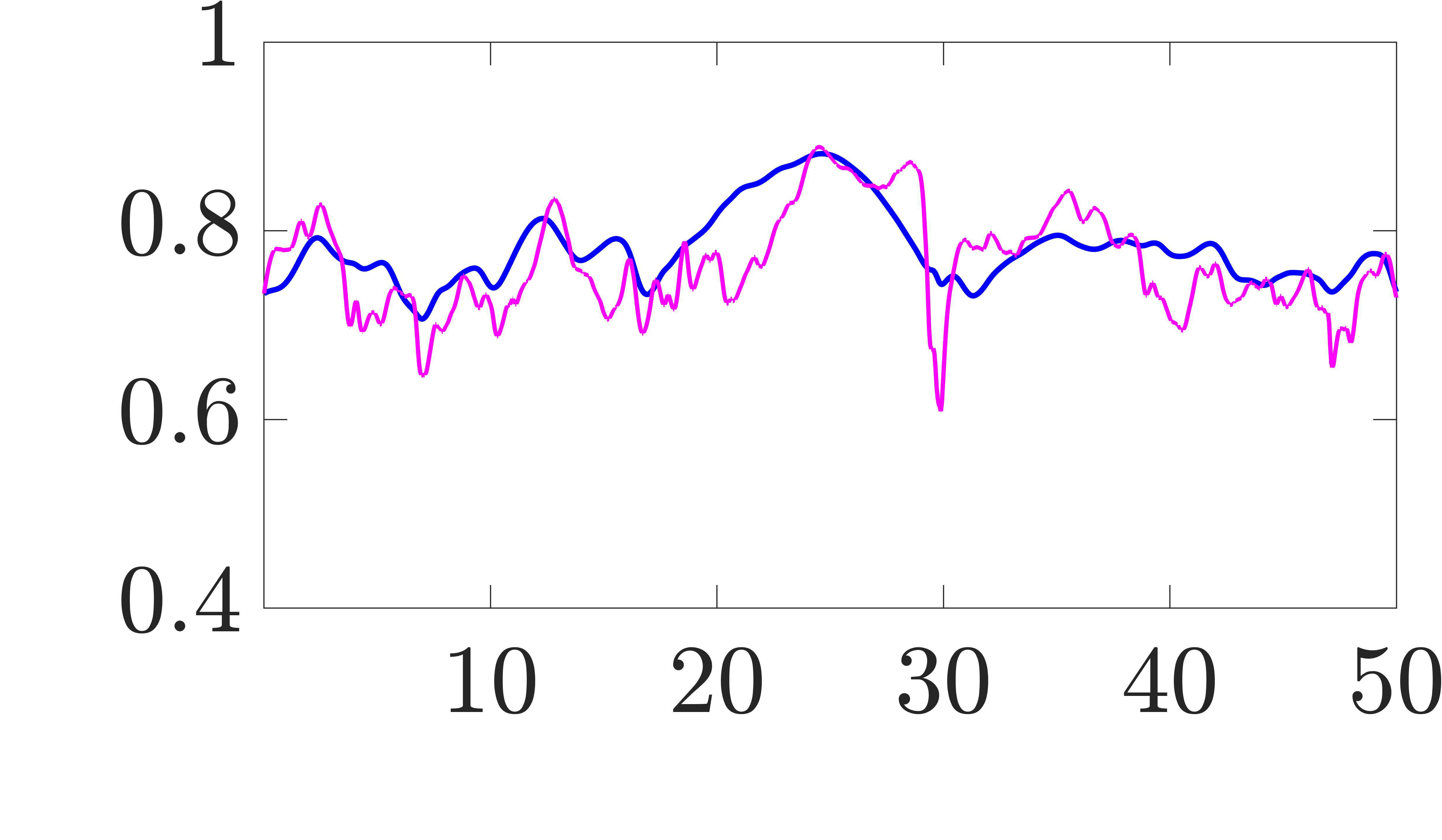} & \multirow{3}{*}[4.5em]{\includegraphics[width=0.5\columnwidth]{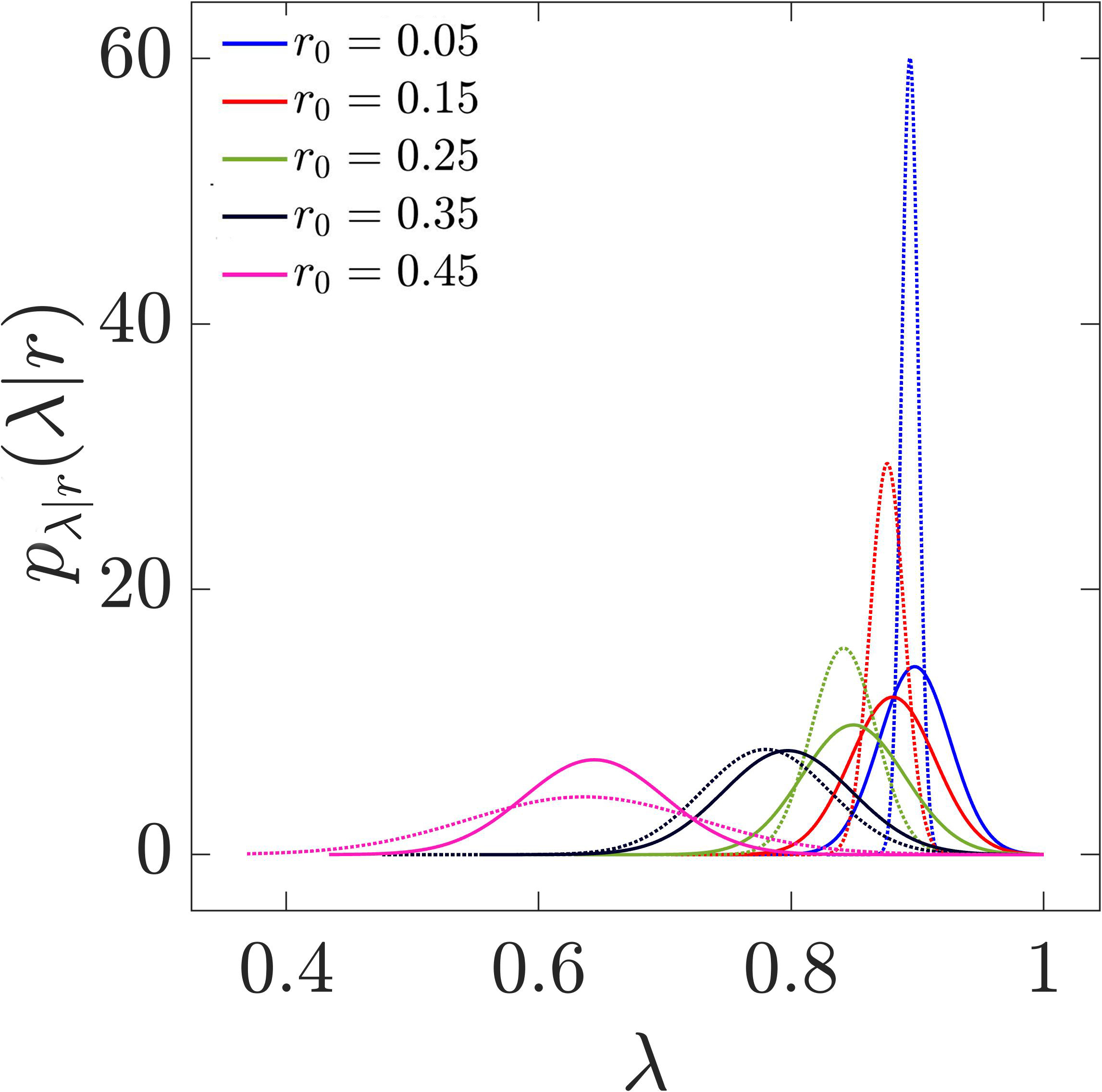}} 
\\
\includegraphics[width=0.33\columnwidth]{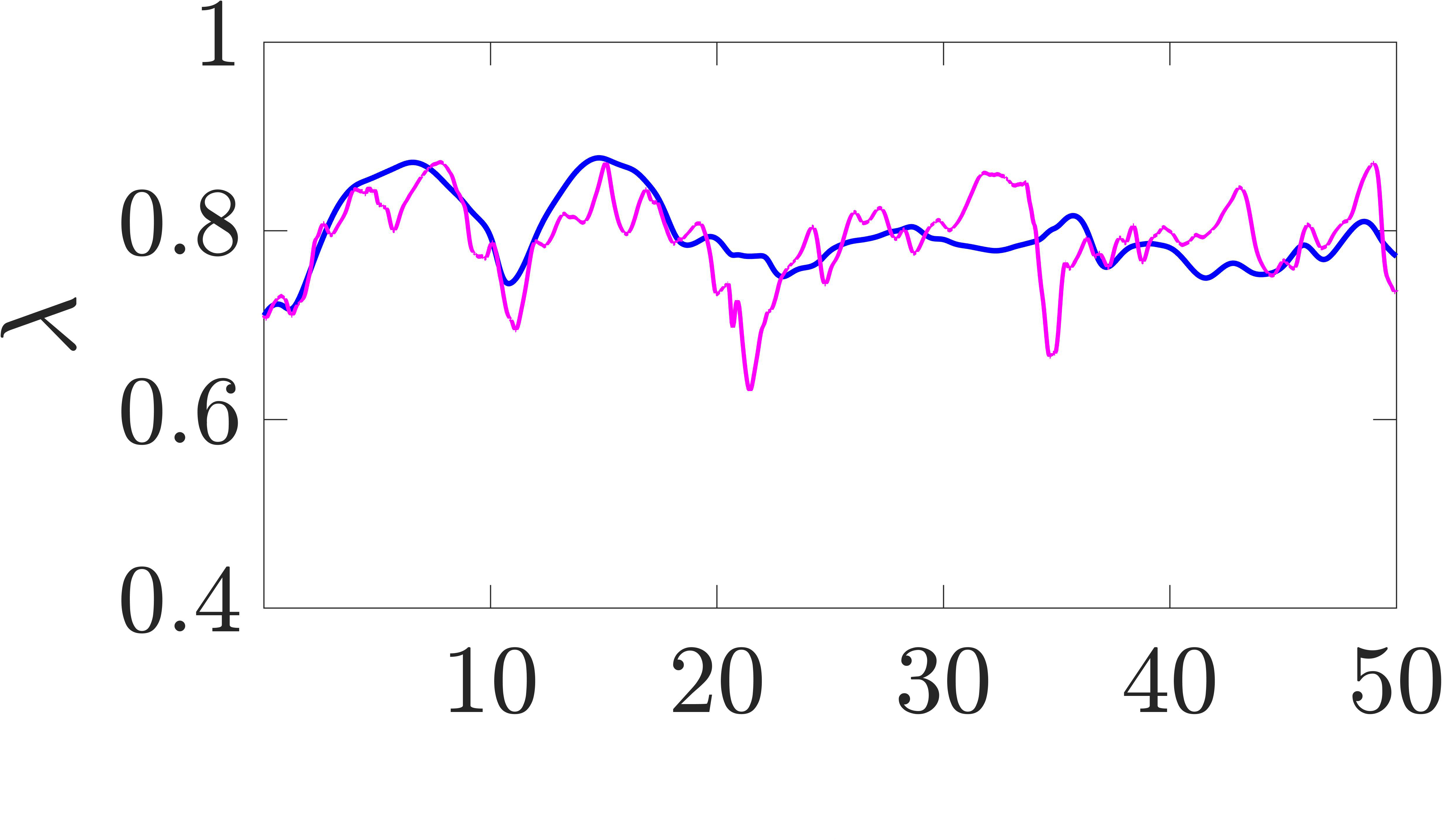} &                    
\\
\includegraphics[width=0.33\columnwidth]{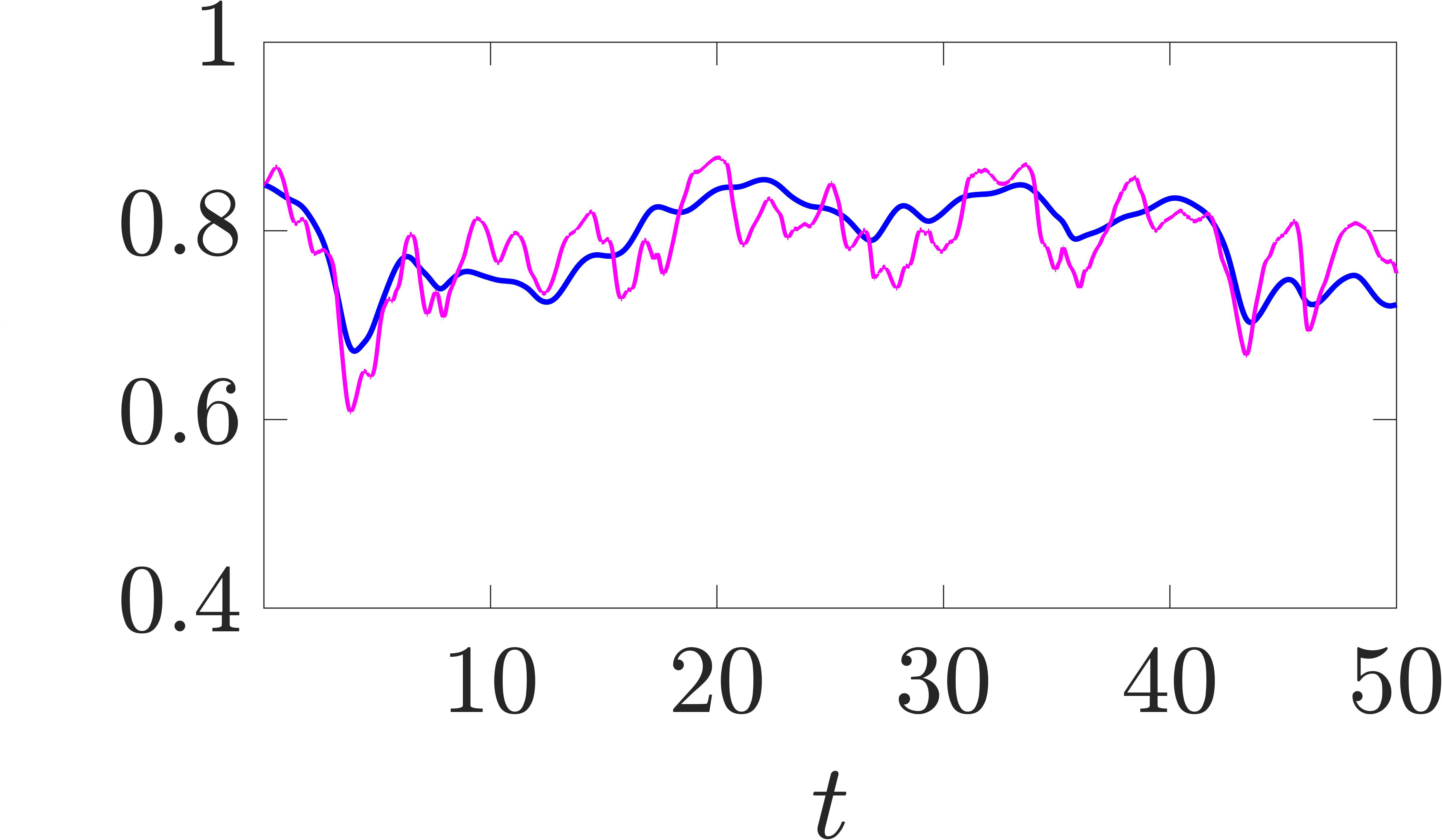} &                    
\\
(a) & \multicolumn{1}{c}{(b)}
\end{tabular}
\caption{(a) Comparison of structural parameter $\lambda(t;\mathbf{X},t_0)$ along typical particle trajectories for thixotropic flows with $\Lambda=1$, from (\textcolor{blue}{\textbf{--$\!$--}}) DNS results and (\textcolor{blue}{\textbf{--$\!$--}}) solution of Lagrangian equation (\ref{eq:ADRE2}) with $\langle \dot\gamma|r(t)\rangle$. (b) Comparison of conditional p.d.f. of structural parameter $p_\lambda(\lambda|r)$ along typical particle trajectories for thixotropic flows with $\Lambda=1$, from (solid lines) DNS results and (dotted lines) the stochastic model (\ref{eqn:Langevin}).}
\label{fig:compare_interm}
\end{figure}

Figure~\ref{fig:compare_interm}a shows that for for $\Lambda=1$, the Lagrangian structural parameter computed using Lagrangian shear rate $\dot\gamma(t;\mathbf{X},t_0)$ data (\ref{eq:ADRE2}) agrees fairly well with that given directly from the DNS simulations due to the large Damkh\"{o}ler number ($Da=10^3$), as previously discussed in \S\ref{sec:Lagrangian Thixotropy}. Conversely, the structural parameter computed via (\ref{eq:ADRE2}) using the conditionally averaged shear rate $\langle \dot\gamma|r(t)\rangle$ does not agree as well, but it does shadow the DNS results, indicating that this closure may still accurately predict $\langle\lambda|r\rangle$. Figure~\ref{fig:compare_interm}b compares the p.d.f. $p_\lambda(\lambda|r)$ for $\Lambda$ from DNS computations with those given by the stochastic model described in \S\S\ref{subsec:stochastic_model}. The stochastic model predicts significantly less variance of $\lambda$ at the core, however the distributions agree quite well near the pipe wall and the mean $\lambda$ agrees fairly well throughout, with the relative error in mean as much as 5\%.

\begin{figure}
\centering
\begin{tabular}{cc}
\includegraphics[width=0.45\columnwidth]{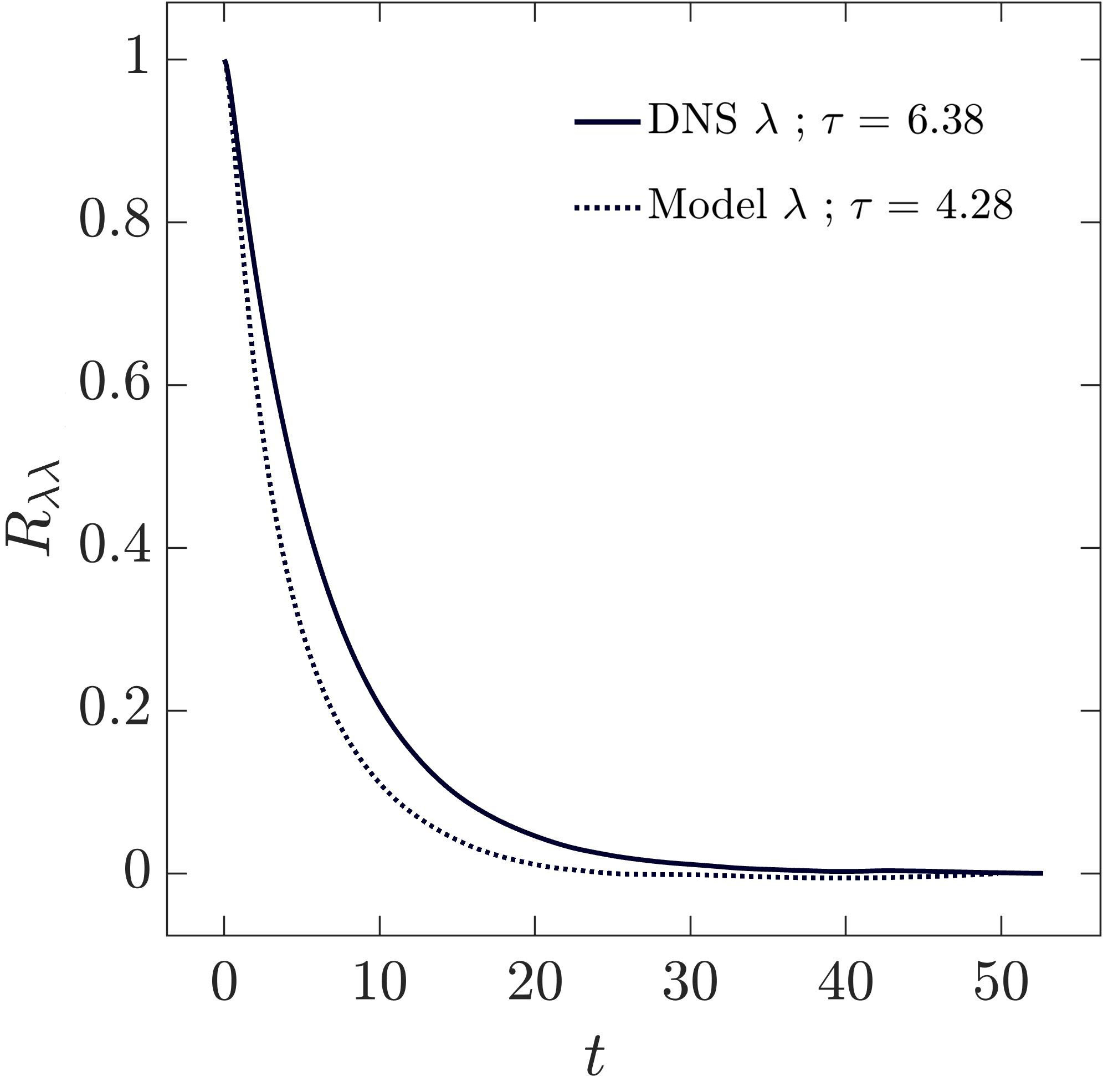}&
\includegraphics[width=0.45\columnwidth]{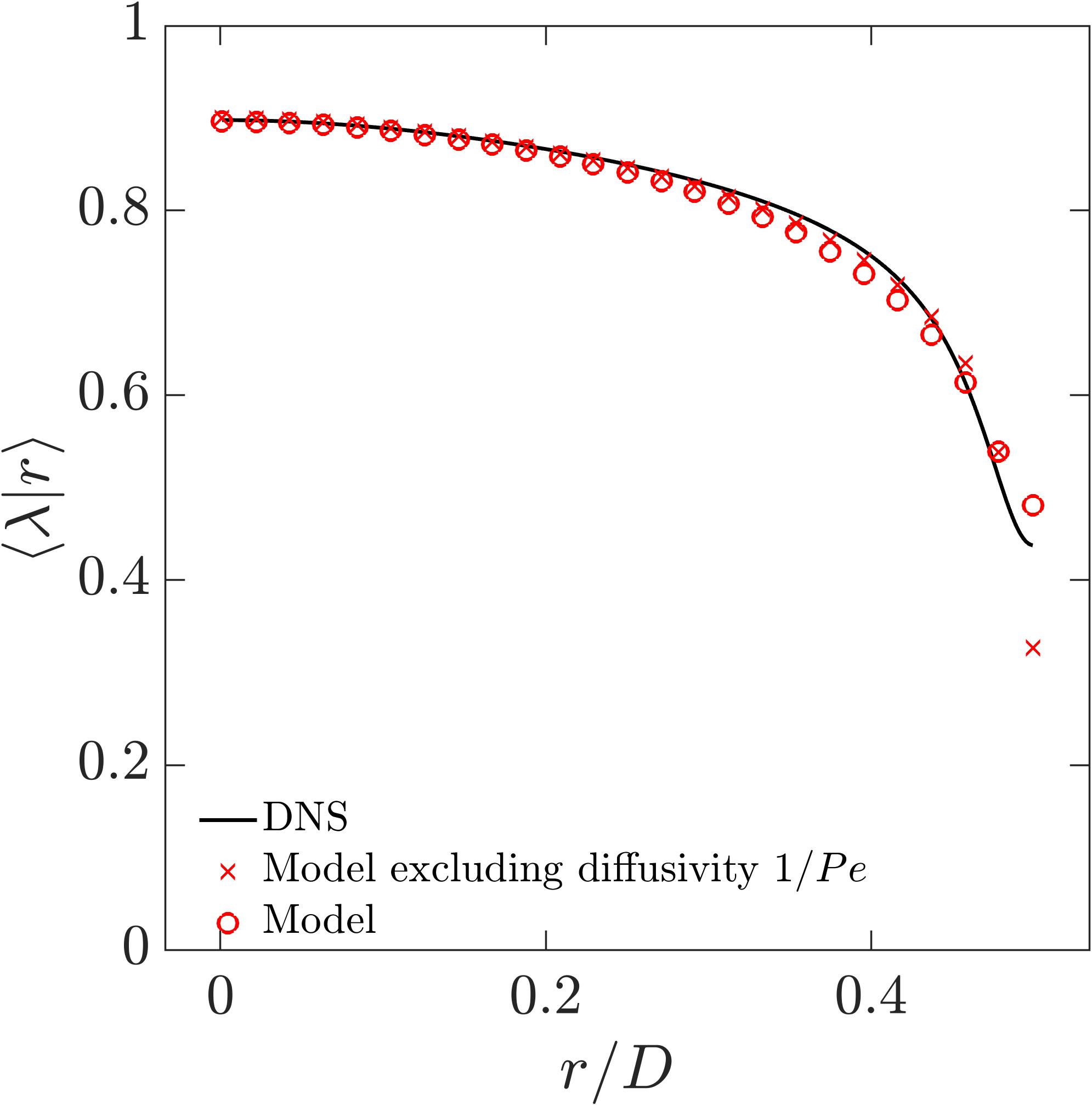}\\
(a) & (b)
\end{tabular}
\caption{(a) Comparison of Lagrangian autocorrelation functions for structural parameter $R_{\lambda \lambda}$ for thixotropic flows with $\Lambda=1$, from (solid lines) DNS results and (dotted lines) the stochastic model (\ref{eqn:Langevin}). (b) Comparison of conditionally averaged structural parameter $\langle\lambda|r\rangle$ for thixotropic flows with $\Lambda=1$, from (solid lines) DNS results, (\textcolor{red}{$\boldsymbol{\circ}$}) the stochastic model (\ref{eqn:Langevin}) and (\textcolor{red}{$\boldsymbol{\times}$}) the stochastic model (\ref{eqn:Langevin}) excluding diffusivity $1/Pe$.}
\label{fig:lambda_ac}
\end{figure}

Figure~\ref{fig:lambda_ac}a shows that for $\Lambda=1$, the structural parameter decorrelates slightly faster for the stochastic model than from the DNS simulations, possibly due to the shear model (\ref{eqn:shear_model}). Figure~\ref{fig:lambda_ac}b shows that despite these differences, $\langle\lambda|r\rangle$ is accurately predicted by the stochastic model when the numerical diffusivity $\alpha$ is included in the radial dispersivity $D_r(r)$, with relative $L_2$ error of 1.86\% . Conversely, exclusion of $\alpha$ leads to significant underestimation of $\lambda$ at the pipe wall as $D_r(r)$ is then zero, leading to trapping of particles for arbitrarily long periods in this high shear region.

\begin{figure}
  \centering
\begin{tabular}{cc}
\includegraphics[width=0.44\textwidth]{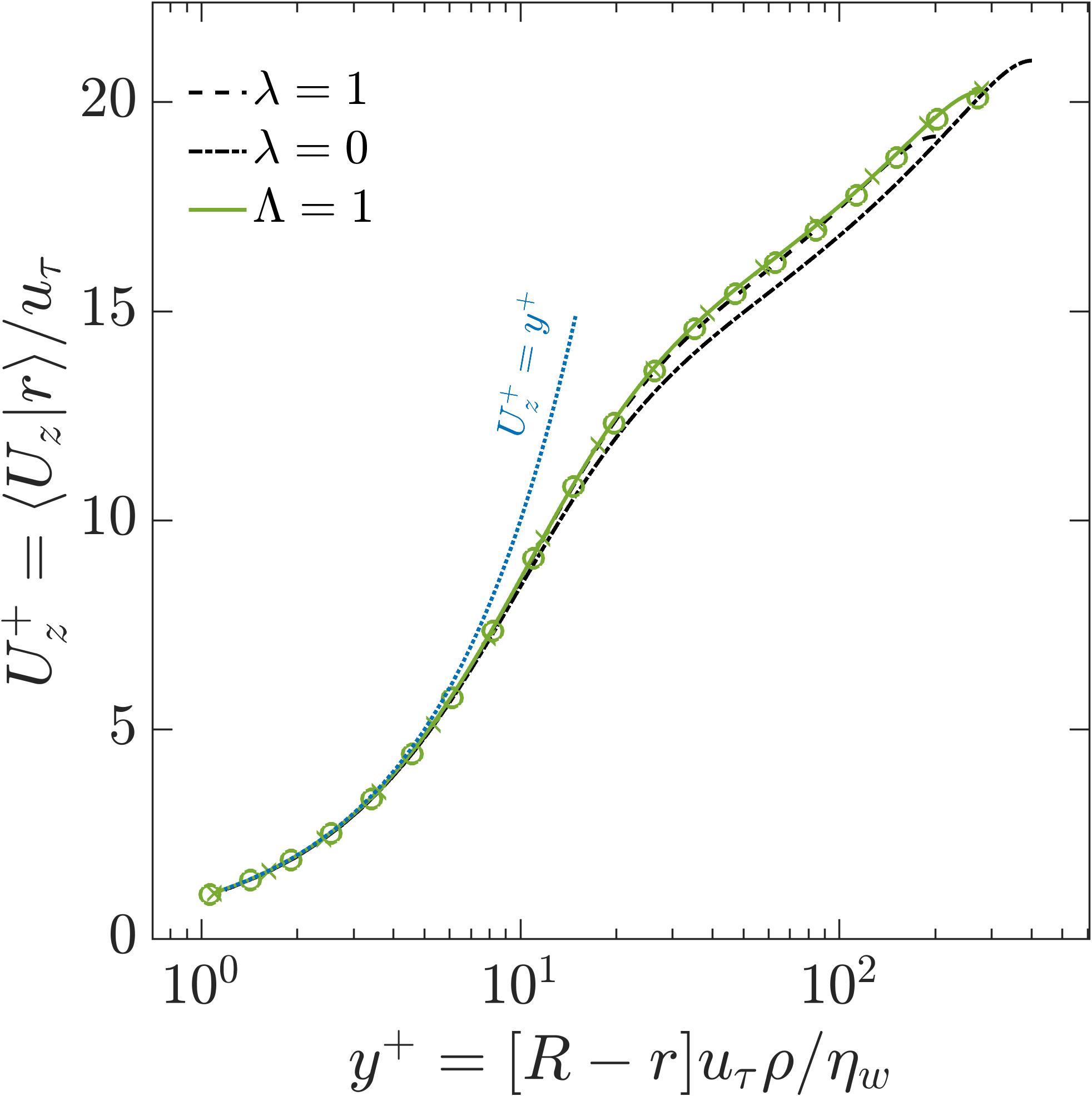}&
\includegraphics[width=0.44\textwidth]{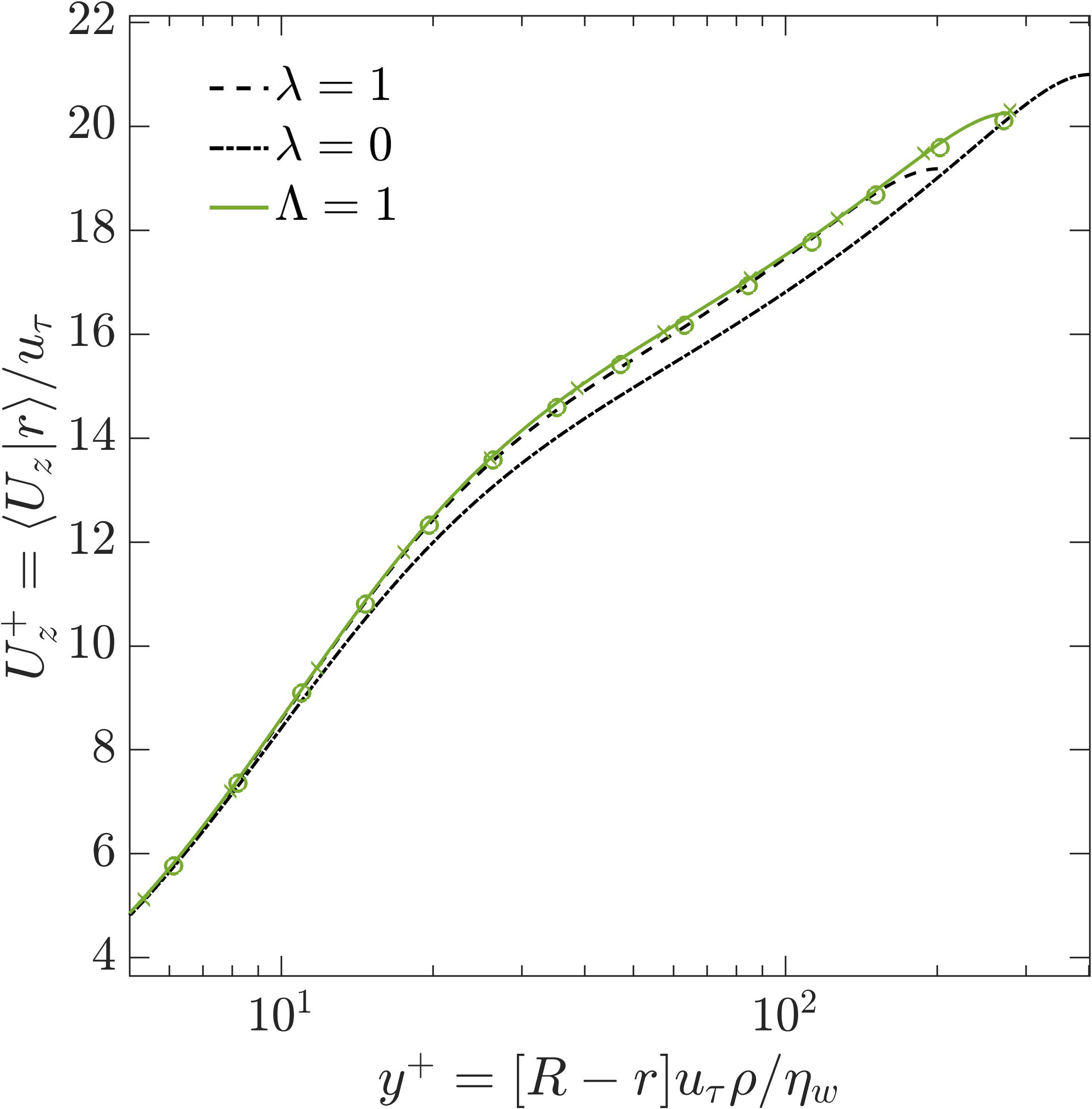}\\
(a) & (b)\\
\includegraphics[width=0.44\textwidth]{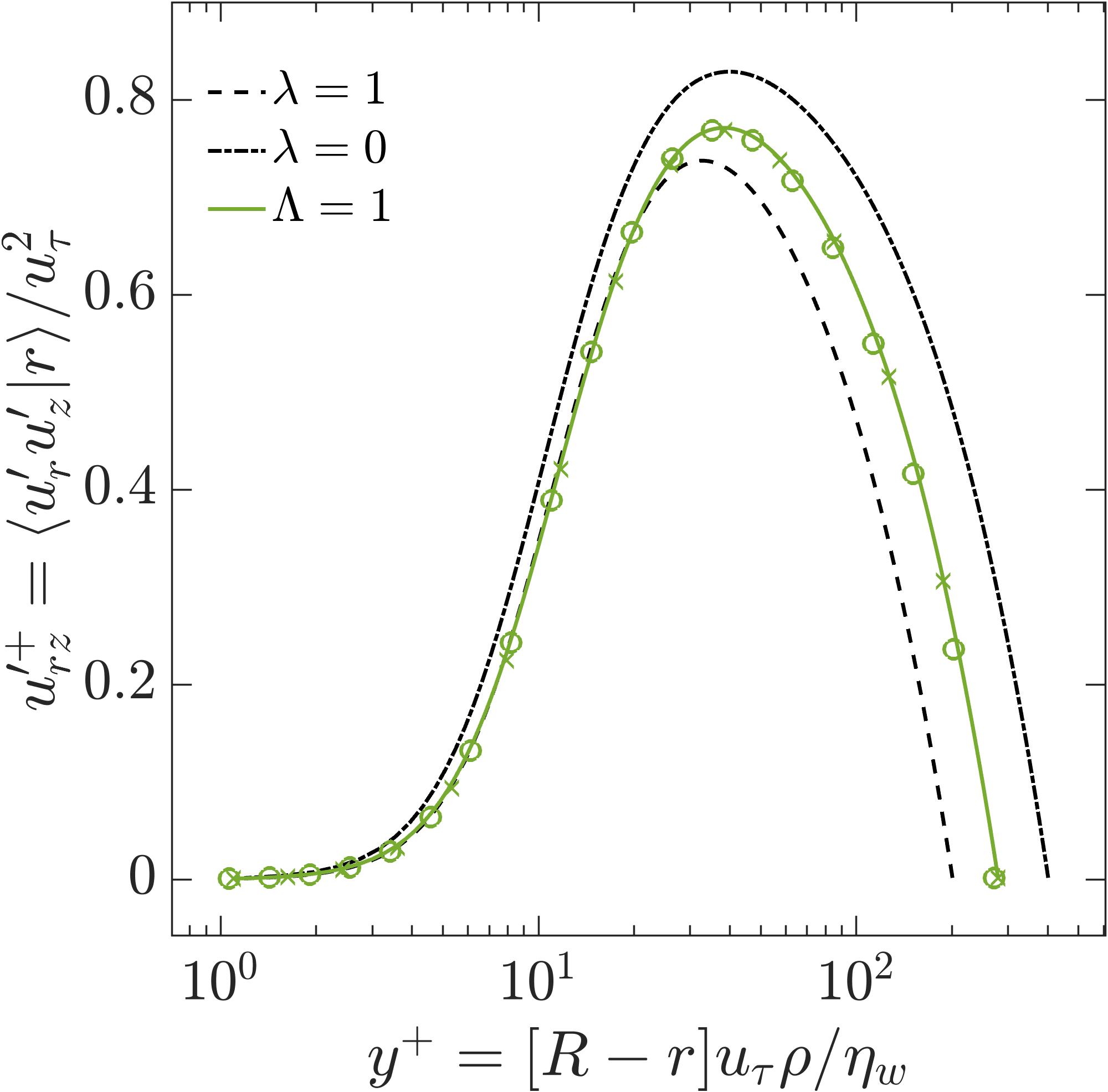}&
\includegraphics[width=0.44\textwidth]{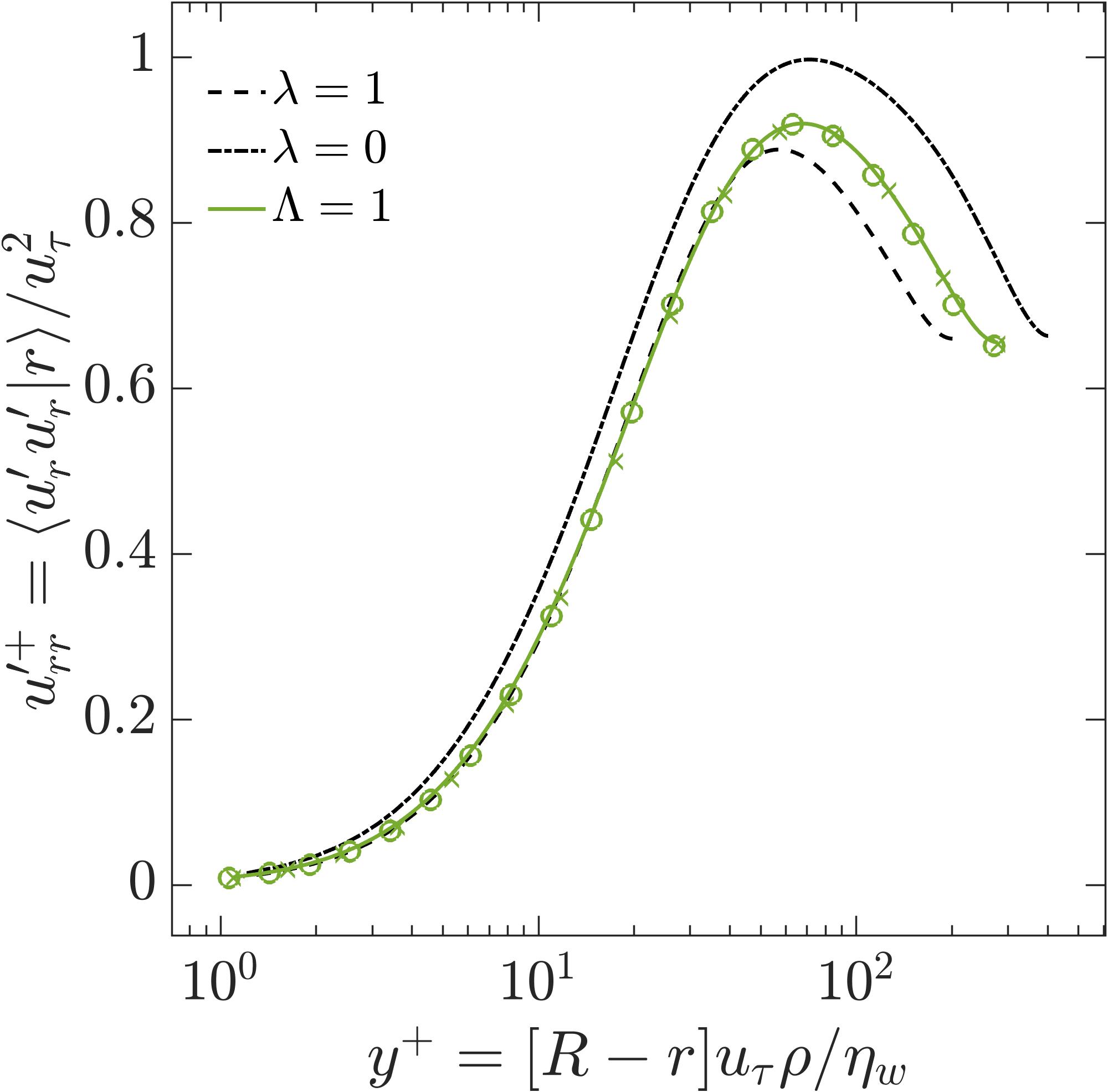}\\
(c) & (d)\\
\includegraphics[width=0.44\textwidth]{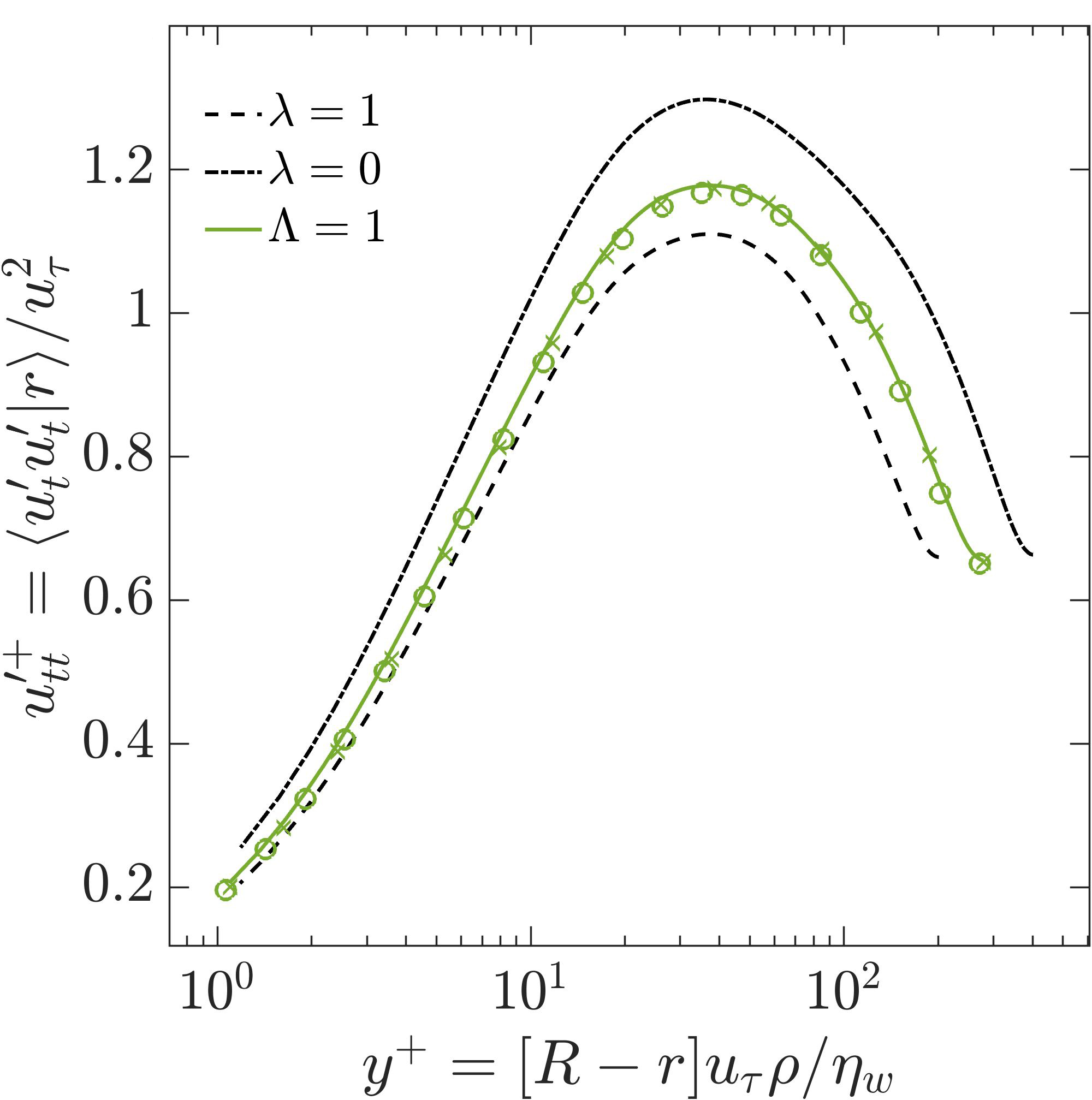}&
\includegraphics[width=0.44\textwidth]{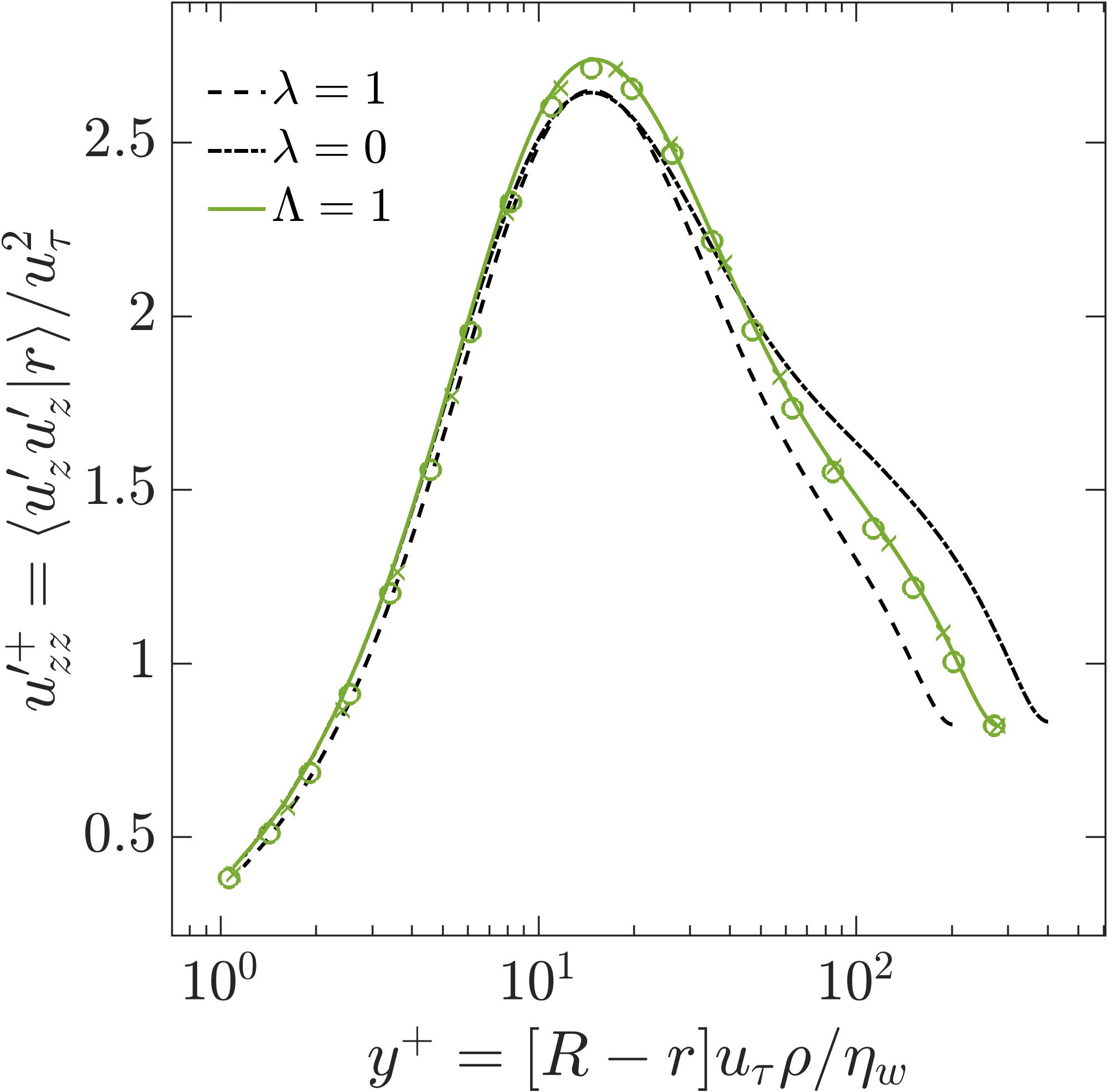}\\
(e) & (f)
\end{tabular}
\caption{Mean velocity radial profiles (a-b) and Reynolds stresses (c-f) for thixotropic flows and Newtonian reference cases. The plots are from (solid lines) DNS results, (\textcolor{black}{$\boldsymbol{\times}$}) the effective viscosity model based on the conditionally averaged structural parameter $\langle\lambda | r\rangle$ and (\textcolor{black}{$\boldsymbol{\circ}$}) the stochastic model (\ref{eqn:Langevin}).}
\label{fig:Mprofiles_stoch}
\end{figure}

Figure~\ref{fig:Mprofiles_stoch} compares results from DNS computations for $\Lambda = 1$ using the full thixotropic model (\ref{eq:transport_nD}) are compared with results from DNS computations using the effective viscosity model (\ref{eqn:visc_eff}) computed from  (i) the stochastic model for $\langle\lambda|r\rangle$ outlined above, and (ii) direct computation of $\langle\lambda|r\rangle$ from DNS results for thixotropic flow. The mean profiles shown in Figure~\ref{fig:Mprofiles_stoch}a and b computed from both the stochastic model and direct calculation of $\langle\lambda|r\rangle$ show excellent agreement with those obtained using the thixotropic model, with errors in the mean viscosity $\eta$ (0.84\% and $10^{-5}$\%) and axial velocity $U_z$ profiles (1\% and 0.16\%) are low. Figure~\ref{fig:Mprofiles_stoch}c-f also shows that the Reynolds stress profiles ($u'_{rr},u'_{tt},u'_{zz},u'_{rz}$) from the DNS results using the stochastic model and from direct computation of $\langle\lambda|r\rangle$ both agree very well with the those using the full thixotropic model, with errors as much as 2.4\% and 1.3\% respectively. 

These results verify both the stochastic model for the averaged structural parameter $\langle\lambda|r\rangle$ and the effective viscosity model $\mu_{\text{eff}}(r,\dot\gamma)$ for turbulent pipe flow. They also show that turbulent flow of time-dependent thixotropic fluids can be accurately approximated as time-independent generalised Newtonian fluids via the effective viscosity $\mu_{\text{eff}}(r,\dot\gamma)$. Although it is somewhat surprising that such a simple stochastic model for $\lambda$ and simple effective viscosity model $\eta_{\text{eff}}$ can accurately capture the turbulent dynamics of a time-dependent fluid, it is important to note that unlike e.g. viscoelastic turbulence, purely thixotropic flow is still an essentially a viscous flow, albeit one with a non-local viscosity.

Indeed, in the limit of fast $\Lambda\gg 1$ and slow $\Lambda\ll 1$ thixotropic kinetics, the rheology of these fluids is time-independent (generalised Newtonian), while for intermediate kinetics $\Lambda\sim 1$, the effective viscosity is well-approximated by the conditional average $\langle\eta|r\rangle$, indicating that fluctuations of $\eta$ around this mean are not important. As $\eta$ scales linearly with $\lambda$, Figure~\ref{fig:compare_interm}b indicates that the variance of $\eta$ around $\langle\eta|r\rangle$ is significant for all $r$. However, the results above indicate that these variations do not significantly impact the structure of turbulent pipe flow.

\subsection{Nonlinear Rheology and Thixotropy}
\label{subsec:nonlinear}

While we have considered simple linear rheological (\ref{eq:rheology}) and thixotropic (\ref{eq:transport}) models in this study, the rheology and thixotropic kinetics of most complex fluids is highly nonlinear. Here we briefly consider the extension of the findings of this study to nonlinear rheology (\ref{eqn:eta}) and thixotropic models (\ref{eq:kinetics}). These nonlinear models admit an equilibrium structural parameter $\lambda_{\text{eq}}(\dot\gamma)$ given by (\ref{eqn:thixo_homog}), leading to the following viscosity model in the fast kinetic limit $(\Lambda\gg 1)$ as
\begin{equation}\label{eqn:nonlin_fast}
\lim_{\Lambda\rightarrow\infty}\eta(\lambda,\dot\gamma)=\eta(\lambda_{\text{eq}}(\dot\gamma),\dot\gamma),
\end{equation}
whereas in the slow kinetic limit $(\Lambda\ll 1)$ we recover 
\begin{equation}\label{eqn:nonlin_slow}
\lim_{\Lambda\rightarrow 0}\eta(\lambda,\dot\gamma)=\eta(\lambda_{\text{eq}}(\langle\dot\gamma\rangle),\dot\gamma).
\end{equation}
For intermediate kinetics, if we again assume that the effective viscosity can be represented in terms of the radially averaged structural parameter, then 
\begin{equation}\label{eqn:nonlin_int}
\eta(\lambda,\dot\gamma)=\eta(\langle\lambda|r\rangle,\dot\gamma),
\end{equation}
which recovers (\ref{eqn:nonlin_fast}), (\ref{eqn:nonlin_slow}) respectively in the limits $\Lambda\rightarrow\infty$, $\Lambda\rightarrow 0$.

Although structural parameter models of the form (\ref{eq:kinetics}) do not possess analytic solutions in general, their solution is still of the form of the shear history functional $\hat{\mathcal{F}}$ in (\ref{eqn:Fr}). However, many nonlinear structural parameter models \citep{Mewis2009} can be expressed in non-dimensional form as
\begin{equation}
\frac{d\lambda}{dt}=\Lambda\left[\dot\gamma^{n_2}(1-\lambda)-K\dot\gamma^{n_1}\lambda\right],
\end{equation}
where the index $n_2=0$ corresponds to Brownian rebuild, and $n_2>0$ corresponds to shear-induced rebuild. These models have equilibrium solution
\begin{equation}
\lambda_{\text{eq}}(\dot\gamma)=\frac{1}{1+K\dot\gamma^{n_1-n_2}},
\end{equation}
and so are shear thinning for $n_1>n_2$ and shear thickening for $n_1<n_2$. This class of structural parameter model has explicit solution
\begin{equation}\label{eq:memoryr_nonlin}
	\begin{split}
\lambda(t,r(t))&=\frac{\Lambda\int_0^\infty G(t-s,r(t-s))ds}{G(t,r(t))},\\
	G(t,r(t))&=\exp\left[\Lambda\int_0^\infty \dot\gamma(t-s,r(t-s))^{n_2}+K\dot\gamma(t-s,r(t-s))^{n_1}\,ds\right],
	\end{split}
\end{equation}
and so the results derived in \S\S\ref{subsec:struct} can be directly applied to these nonlinear models with $\langle\lambda|r\rangle$ given by (\ref{eqn:lambda_av2}) and
\begin{equation}
h(r)=\exp\left[\Lambda\,\Delta s(\langle\dot\gamma|r\rangle^{n_2}+K\langle\dot\gamma|r\rangle^{n_1})\right].
\end{equation}
Clearly, further research is required to justify the assumptions associated with the stochastic model for Lagrangian shear rate and effective viscosity model (\ref{eqn:eta_r}) for nonlinear viscosity $\eta(\lambda,\dot\gamma)$ and thixotropy models.

\section{Conclusions}\label{sec:Conclusions}

In this study we have considered fully developed turbulent pipe flow of a time-dependent complex fluid, modeled via the simplest non-trivial thixotropic rheology model; a purely viscous Moore fluid~\citep{Moore1959} with structural parameter $\lambda$. Despite this simplicity, these results are relevant to a broad class of applications involving inelastic thixotropic flow of materials with otherwise generalised Newtonian (GN) viscous rheology, and so are relevant to a wide class of industrial flows (pipelining, heat transfer, mixing, etc) of complex materials including suspensions and emulsions, slurries, pastes and biological materials. The feedback between turbulence structure, rheology and microstructural state in these complex flows is not well understood.

To address this shortcoming, we use direct numerical simulation (DNS) to simulate fully developed moderately turbulent flow ($Re_G\approx 6,000-14,000$) over a broad range of thixotropic kinetics from slow to fast (with respect to the advective timescale), as reflected by the thixoviscous numbers $\Lambda\in[10^{-2},10^{2}]$. We consider transport of $\lambda$ in the advection-dominated regime ($\Pen=10^3$), and note that for most complex fluids, self-diffusion of $\lambda$ is negligible ($\Pen\sim10^{12}$).

As expected, in the limit of fast thixotropic kinetics ($\Lambda\rightarrow\infty^+$), the viscosity of thixotropic fluids converges to that given by the equilibrium structural parameter $\lambda_{\text{eq}}(\dot\gamma(\mathbf{x},t))$ (\ref{eq:equil}), and so the viscosity converges to the shear thinning case $\eta(\lambda,\dot\gamma)=\eta(\lambda_{\text{eq}}(\dot\gamma,\dot\gamma)$. In this limit, the turbulence structure of these flows is the same as that of a shear-thinning GN fluid.

Conversely, in the limit of slow thixotropic kinetics ($\Lambda\rightarrow 0$), the viscosity of thixotropic fluids converges to that given by the average structural parameter $\lambda_{\text{eq}}(\langle\dot\gamma\rangle)$ (\ref{eqn:lambda_av}) due to ergodic sampling of the shear rate along pathlines, and so the viscosity converges to $\eta(\lambda,\dot\gamma)=\eta(\lambda_{\text{eq}}(\langle\dot\gamma\rangle,\dot\gamma)$. In this limit, the turbulence structure of the Moore fluid is the same as that of a Newtonian fluid. In general, thixotropic fluid behave as GN fluids in this limit.

Hence, in the limits of fast and slow thixotropic kinetics, turbulent flow of time-dependent inelastic thixotropic fluids is identical to that of turbulent flow of purely viscous time-independent fluids. For intermediate thixotropic kinetics, the picture is more complex as the local (in space and time) structural parameter $\lambda$ and hence viscosity $\eta$ depends upon the Lagrangian history of shear rates.

From the thixotropic kinetic equation, we derive an expression for local $\lambda$ as a ``fading memory'' functional $\mathcal{F}$ of the Lagrangian shear history where $\Lambda$ governs the persistence of memory in this functional. 
As fully developed pipe flow is non-stationary in the radial coordinate, we propose that turbulent thixotropic pipe flow with intermediate thixotropic kinetics ($\Lambda\sim 1$) may be approximated as a purely viscous (time-independent) flow with radially-variable effective viscosity given by the conditionally averaged structural parameter as $\eta(\lambda,\dot\gamma)=\eta_{\text{eff}}(r,\dot\gamma)\equiv\eta(\langle\lambda|r\rangle,\dot\gamma)$ (\ref{eqn:visc_eff}). This path integral formulation is completely general and applies to all $\Lambda\in[0,\infty^+)$, recovering the previously derived closures for the fast and slow thixotropic kinetics.

A stochastic model for $\eta_{\text{eff}}$ can then be generated via a stochastic model for $\langle\lambda|r\rangle$, which represents a \emph{path integral} of the functional $\mathcal{F}$ backwards in time over the Lagrangian shear histories that arrive at $r$. We develop a simple non-stationary stochastic model for the Lagrangian shear history based based on a Fokker-Planck equation for radial position (\ref{eqn:Fokker_back}) under the assumption that the instantaneous Lagrangian shear rate is given by the conditional average $\dot\gamma=\langle\dot\gamma|r\rangle$. This simple stochastic model provides accurate estimates of the conditionally averaged structural parameter $\langle\lambda|r\rangle$, and DNS computations based on this model for $\Lambda=1$ agree with those given by the full thixotropic model to around 1\% in terms of Reynlds stresses and mean axial velocity profile.

Similarly, DNS computations based on direct computation of the conditionally averaged structural parameter $\langle\lambda|r\rangle$ from the full thixotropic DNS computations agree with these latter computations to a high degree of accuracy ($\sim$2.4\% error for axial velocity and Reynolds stress profiles). These results establish that the turbulent flow of time-dependent thixotropic fluids is very similar to that of time-independent purely viscous (generalised Newtonian) fluids, for all ranges of thixotropic kinetics over the range $\Lambda\in[0,\infty^+)$. For non-stationary flows such as fully developed turbulent pipe, this manifests as a radially-dependent effective viscosity $\eta_{\text{eff}}(r,\dot\gamma)$, whereas in general the effective viscosity is a function of every non-stationary direction of the flow. Similarly, homogeneous isotropic turbulent flow of thixotropic fluids are expected to behave in a similar manner to generalised Newtonian analogues; i.e. $\eta_{\text{eff}}=\eta_{\text{eff}}(\dot\gamma)$.

Although further research is required, these results suggest that they extend more broadly to a wide range of inelastic thixotropic fluids with nonlinear viscosity $\eta(\lambda,\dot\gamma)$ and nonlinear thixotropic kinetics, as the basic mechanisms that govern the effective viscosity in these flows persist. The observation that under turbulent flow conditions, time-dependent thixotropic fluids behave as time-independent purely viscous analogues is a significant simplification that impacts a broad range of applications. This correspondence is  attributed to the fact that thixotropic flows are viscous flows in terms of their stress-strain relationship (as opposed to e.g. visco-elastic flow), albeit ones with a non-local (Lagrangian history) viscosity dependence. The ergodic nature of turbulent flow appears to play an important role in rending these flows similar to time-independent purely viscous analogues.

\backsection[Funding]{The authors acknowledge financial support from the Australian Research Council, Melbourne Water, and Water Corporation, granted through ARC-Linkage 180100869. This research was also supported by the NCI Adapter Scheme, with computational resources provided by NCI Australia, an NCRIS-enabled facility supported by the Australian Government.}
\backsection[Declaration of interests]{The authors report no conflict of interest.}

\appendix

\section{Numerical Method for Thixotropic Kinetics}\label{app:Numericals}

The spectral element code~\citep{Blackburn2019} employs a second-order temporal integration method~\citep{Karniadakis1991} based on the operator splitting technique. The advective non-linear terms in the Navier-Stokes (\ref{eq:gvn_nD} and thixotropy \ref{eq:transport_nD}) equations are treated explicitly, while the diffusion terms are handled implicitly to ensure numerical stability and convergence, as described in previous studies~\citep{Rudman2006}. 
 To mitigate numerical instabilities, an implicit treatment of the reaction kinetics in (\ref{eq:transport_nD}) was implemented. The discretized reaction kinetics, derived from (\ref{eq:memory}), are expressed
\begin{equation} \label{eq:implicitFDM}
\lambda_{n+1} = \frac{\lambda_n}{g_n(\Delta t)} + \frac{\Lambda}{\Lambda\left[1 + K \dot\gamma_n \right]} \left[1 - \frac{1}{g_n(\Delta t)} \right]
,\quad 
g_n(\Delta t) = \exp \left[\Lambda\Delta t \left(1+K\dot\gamma_n \right) \right],
\end{equation}
where the subscript $n$ denotes quantities at time step $n$. This formulation assumes that the local shear rate remains constant over a time step $\Delta t$, i.e. $\dot\gamma(t) = \dot\gamma_n$. This implicit treatment improves numerical stability and allows the code to handle the stiff terms effectively, even for high $\Lambda$ cases.

\bibliographystyle{jfm}
\bibliography{jfm}

\end{document}